%
%

\input amstex
\documentstyle{amsppt}

\magnification=1200
\pagewidth{16.5truecm}
\pageheight{22.6truecm}

\TagsOnRight
\CenteredTagsOnSplits
{\normalbaselineskip15pt}
\NoBlackBoxes

\font\headerfont=cmcsc8

\def\half{\hbox{$\textstyle{1\over 2}$}}

\def\Ext{\operatorname{Ext}}
\def\Int{\operatorname{Int}}
\def\supp{\operatorname{supp}}
\def\dist{\operatorname{dist}}
\def\diam{\operatorname{diam}}
\def\free{{\text{free}}}

\rightline{To appear in J. Stat. Phys.}
\bigskip
\bigskip
\topmatter

\title
Surface induced Finite Size Effects\\
for First Order Phase Transitions
\endtitle

\leftheadtext\nofrills
{\headerfont C. Borgs, R.Koteck\'y}
\rightheadtext\nofrills
{\headerfont Surface induced Finite
 Size Effects}

\author
 C. Borgs\footnotemark"${}^\dag$"
and R. Koteck\'y\footnotemark"${}^\ddag$"
\endauthor

\footnotetext"${}^\dag$"{Heisenberg Fellow}
\footnotetext"${}^\ddag$"{Partly supported by
the grants GA\v CR 202/93/0499 and GAUK 376}

\affil
Department of Mathematics,
University of
California at Los Angeles,\\
and Center for Theoretical Study,
Charles University,
Prague  \endaffil
\address
Christian Borgs
\hfill\newline
Department of Mathematics
\hfill\newline
University of
California at Los Angeles
\hfill\newline
Los Angles, CA 90023
\endaddress
\email
borgs\@math.ucla.edu
\endemail

\address
Roman Koteck\'y
\hfill\newline
Center for Theoretical Study, Charles
University,\hfill\newline
 T\'aboritsk\'a 23, 130 87 Praha 3,
Czech Republic
\hfill\newline
\phantom{18.}and
\hfill\newline
Department of
Theoretical Physics, Charles University,
\hfill\newline
 V~Hole\v sovi\v
ck\'ach~2, 180~00~Praha~8, Czech Republic
\endaddress
\email
kotecky\@aci.cvut.cz
\endemail

\comment
\endcomment

\abstract
We consider
classical
lattice models
describing first-order phase transitions,
and study the finite-size scaling
of the magnetization and  susceptibility.
In order to model the effects
of an actual surface in systems
like small magnetic clusters,
 we consider models with free boundary
conditions.

For a field driven transition
with two coexisting phases  at the
infinite volume transition point
$h=h_t$, we prove that
the low temperature finite
volume magnetization $m_{\free}(L,h)$
per site in a cubic volume
of size  $L^d$  behaves like
$$
m_\free(L,h)=\frac{m_++m_-}2 + \frac{m_+-m_-}2
\tanh  \bigl(\frac{m_+-m_-}2\,L^d\,
(h-h_\chi(L))\bigr)+O(1/L),
$$
where $h_\chi(L)$ is the position
of the maximum of the (finite volume) susceptibility
and $m_\pm$  are the infinite
 volume magnetizations
at $h=h_t+0$ and $h=h_t-0$,
respectively.
We show that $h_\chi(L)$ is shifted by
an amound proportional
to $1/L$ with respect to
the infinite volume transitions point
$h_t$ provided the surface
free energies of the two phases
at the transition point are different.
This should be compared with the shift
for periodic boundary conditons,
 which for an asymmetric transition
with two coexisting phases
is proportional only to $1/L^{2d}$.

One can consider also other definitions
of finite volume transition points,
as, for example, the position $h_U(L)$ of the
maximum of the so called Binder cummulant
$U_\free(L,h)$. While
it is again
shifted by an amount proportional to $1/L$
with respect to the infinite volume
transition point $h_t$,
its shift with respect to $h_\chi(L)$ is of
 the much smaller
order $1/L^{2d}$. We give explicit
formulas for the  proportionality
factors, and show that, in the leading
$1/L^{2d}$ term, the relative shift
is the same as that for periodic
boundary conditions.

\endabstract
\endtopmatter

\document

\head{1. Introduction}
\endhead

In the last twenty years, the study of
finite size (FS) effects near
first and second
order phase transitions has gained
increasing interest. While the study
of FS effects
for the second order phase transitions
goes back to the work of Fisher and
coworkers in the early
seventies \cite{FB72, FF69, Fi71},
finite-size effects for first order
phase transitions were first
considered by Imry \cite{I80} and,
in the sequel,
by Fisher and Berker \cite{FB82},
Bl\"ote and Nightingale \cite{BN81},
Binder and coworkers \cite{Bi81, BL84, CLB86},
Privman and Fisher
\cite{PF83}, and others.

Recently, these studies have been
systematized in a rigorous framework by
Borgs and Koteck\'y \cite{BK90}
(see also \cite{BK92, BKM91}),
and by Borgs
and Imbrie \cite{BI92a, BI92b, Bo92}.
Their results
cover both finite size effects in
cubic volumes and long cylinders,
both field and temperature driven
transitions, but were always limited to
periodic boundary conditions.
While the periodic boundary conditions
are natural for the description of
computer experiments that are used
to study the bulk properties of
a system (note that
periodic boundary conditions are
used in these computer
experiments because they minimize
the unwanted finite size effects)
they do not allow for the description
 of FS effects in
actual physical systems like, e.g.,
 small magnetic clusters,
where surface effects are of
 major importance.

In this paper we start a rigorous
study of such surface effects.
We consider spin systems in
 a finite box $\Lambda=\{1,\ldots,L\}^d$,
imposing free or so called
 ``weak'' boundary
conditions (see Section 2 below)
instead of the periodic boundary
 conditions used in our
previous work.

In order to explain our main ideas,
 let us first review the
FSS for a system in a periodic box
\cite{BK90, BK92, BKM91}.
For a system describing the coexistence of
two phases, say an Ising magnet
 at low temperatures, the partition
function with periodic
boundary conditions can be approximated by
$$
Z_{\text{per}}(L,h)\cong Z_+(L,h) + Z_-(L,h),
\tag 1.1
$$
where $Z_\pm$ contain small
perturbations of the ground state
configurations
$\sigma_\Lambda\equiv +1$ and
$\sigma_\Lambda\equiv -1$,
respectively.
The error terms coming
from the tunneling configurations
can be bounded by
$O(L^de^{-L/L_0}) e^{-f(h)}$,
where $f(h)$ is the free energy
of the system and $L_0$ is
a constant of the order
of the infinite volume correlation
length.

In the asymptotic (large volume)
behavior of
$\log Z_\pm$ there should appear,
in principle,
volume, surface, ..., and corner
terms. A periodic box, however,
has neither surface, ...,
nor edges or corners, and one obtains
$$
Z_{{\text{per}}}(L,h)\cong
e^{-f_+(h)L^d}+e^{-f_-(h)L^d}
=
\,2\cosh\left(\frac{f_+(h)-f_-(h)}
2 L^d\right)
e^{-\frac{f_+(h)+f_-(h)}2 L^d},
\tag 1.2
$$
where $f_+(h)$ and $f_-(h)$ are the
(meta-stable) free energies
of the phase plus
and minus. Taylor expanding
$f_\pm(h)$ around the transition
point $h_t$, and introducing
the spontaneous magnetizations
$m_\pm$ of the phase plus and minus
at $h_t$,
one obtains the FSS of the magnetization
$m_{{\text{per}}}(L,h)=
L^{-d}d\log Z_{{\text{per}}}(L,h)/dh$ in the form
$$
m_{{\text{per}}}(L,h)\cong \frac{m_++m_-}2
+ \frac{m_+-m_-}2\tanh\left(
\frac{m_+-m_-}2(h-h_t)L^d\right).
\tag 1.3
$$
It describes the rounding of the
infinite volume
transition in a region of width
$$
\Delta h \sim L^{-d}
\tag 1.4
$$
with a shift $h_t(L)-h_t$
that vanishes in the approximation
(1.3). A more accurate calculation
shows that, in fact, for a
system describing the coexistence
of two low temperature
phases at the infinite volume
transition point $h_t$ and
with infinite volume susceptibilities
$\chi_\pm$, one has
$$
h_\chi(L)-h_t= \frac{6(\chi_+-\chi_-)}
{(m_+-m_-)^{3}}L^{-2d}+O(L^{-3d})
\tag 1.5
$$
if $h_\chi(L)$ is defined as the
position of the
maximum of the susceptibility
 in the volume $L^d$.

Turning to free boundary condition,
 we again expand
$\log Z_\pm(L,h)$ into
volume-surface-...-corner
terms. This time, however,
the volume $\Lambda$ has a boundary,
 and
the expansion yields
$$
-\log Z_\pm(L,h)=f_\pm^{(d)}(h)L^d +
f^{(d-1)}_\pm(h) 2dL^{d-1} + O(L^{d-2}),
\tag 1.6
$$
where $f^{(d)}_\pm(h)=f_\pm(h)$ are the
(meta-stable) bulk free energies,
while $f^{(d-1)}_\pm(h)$ are the
(meta-stable)
surface free energies of the phase
plus and minus, respectively.
As a consequence, (1.2) gets replaced by
$$
\align
Z_{\free}(L,h)&\cong
\exp\biggl(
-\frac{f_+(h)+f_-(h)}2 L^d -
\frac{f^{(d-1)}_+(h) + f^{(d-1)}_-(h)}
2 2dL^{d-1}
\biggr)\cr
&\times2\cosh\biggl(
\frac{f_+(h)-f_-(h)}2 L^d +
\frac{f^{(d-1)}_+(h)-f^{(d-1)}_-(h)}
2 2dL^{d-1}
\biggr).
\tag 1.7
\endalign
$$
At this point, one major difference
with respect to (1.2)
appears: while the free energies
$f_+$ and $f_-$ are
equal at the transition point $h_t$,
the surface free energies
are typically different at $h_t$
(obviously, there are systems
for which
$\tau_+:=f^{(d-1)}_+(h_t)$
and
$\tau_-:=f^{(d-1)}_-(h_t)$
are equal, as e.g. in the symmetric
Ising model where $\tau_+=\tau_-$
by symmetry,
but for asymmetric first order
transitions, this is typically not
the case). The leading terms
in the expansion around $h_t$ then
lead to the formula
$$
m_{\free}(L,h)\cong \frac{m_++m_-}2
+ \frac{m_+-m_-}2
\tanh
\left(
\frac{m_+-m_-}2(h-h_\chi(L))L^d
\right).
\tag 1.8
$$
Here
$$
h_\chi(L)=h_t + \frac{\tau_+-\tau_-}
{(m_+-m_-)}
\frac{2d}{L} + O(1/L^2)
\tag 1.9
$$
which, for $\tau_-\neq \tau_+$,
is now proportional to $1/L$,
while the width $\Delta h$
of the transition is still
proportional to $L^{-d}$.

In fact, a formula of the form
(1.8) has already been given in
\cite{PR90}, with heuristic arguments
very similar to those presented
above.
Here, our goal is twofold: first,
 we want to make the  arguments
leading to (1.8) rigorous,
deriving at the same time precise
error bounds on the subleading terms
(in fact, our method
allows to calculate
in a systematic way
the corrections to (1.8) in terms of an
infinite asymptotic series
in powers of $1/L$).
Second, we want to generalize
these results to a wider
class
of situations, including, in particular,
the finite-size scaling of
expectation values of arbitrary local observables.

It will turn out that the more precise
analysis of the subleading terms
reveals an interesting fact:
if one considers other
standard definitions
of the
finite  volume transition points,
as e.g. the position $h_U(L)$ of the
maximum of the so called Binder cummulant
$U_\free(L,h)$,
one finds that all of them
are
shifted,
with respect to the infinite volume
transition point $h_t$,
by an amount proportional to $1/L$.
Their mutual shifts,
however, are of the much smaller
order $1/L^{2d}$,
with  proportionality
factors that are the same
as those for the corresponding
shifts with periodic
boundary conditions,
see Section 2 for the precise statements.

The finite-size scaling of local observables,
on the other hand, will lead to the
construction of certain "meta-stable"
states
$\langle\cdot\rangle^{h}_\pm$
and their finite-volume analogues
$\langle\cdot\rangle^{L,h}_\pm$,
such that
$$
\langle A \rangle_{\free}^{L,h}
\cong \frac{A_+(L)+A_-(L)}{2}
+\frac{A_+(L)-A_-(L)}{2}
\tanh\Bigl\{\frac{m_+-m_-}{2}(h-h_\chi(L))
L^d\Bigr\}.
\tag 1.10
$$
Here
$A_\pm(L)
=\langle A \rangle^{L,h_t}_\pm$
differ from the corresponding infinite
volume expectation values
$A_\pm=\langle A \rangle^{h_t}_\pm$
by an amount which is exponentially
small in the distance
$\dist(\supp A,\partial\Lambda)$,
see Theorem 3.2 in Section 3.4
for the precise
statement in the more general
context of $N$ phase coexistence.
Note that the argument of the hyperbolic
tangent in (1.10) is the same as in (1.9),
and is independent of the particular choice for
$A$.
Thus the finite-size scaling of all
local observables is synchronized in the
sense that, after subtracting
the ``offset'' $\frac{A_+(L)-A_-(L)}{2}$,
the functions $\langle A \rangle_{\free}^{L,h}$
asymptotically only differ by a constant
factor, see Fig.~1.

\midinsert
\vskip 2.5truein
\noindent
\includegraphics{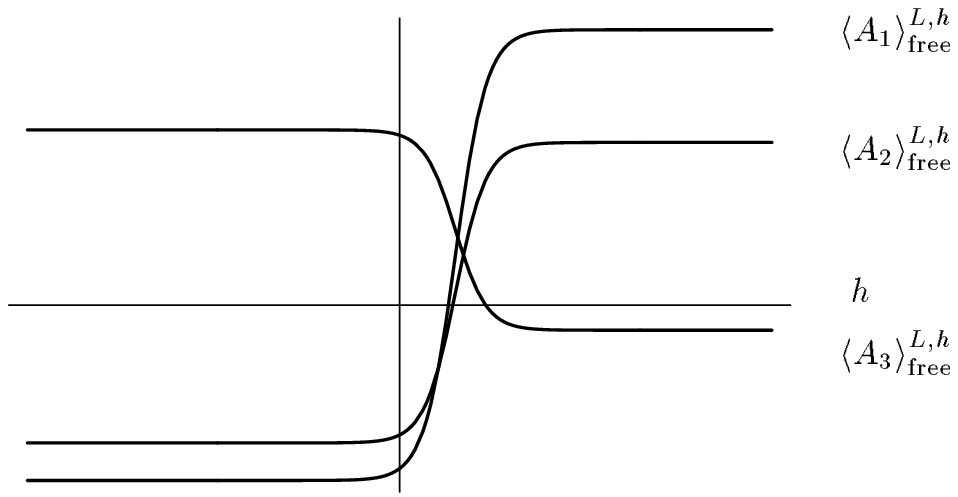}
\botcaption {Fig\. 1}
Finite size scaling of three different observables.
\endcaption
\endinsert
\bigskip

The organization of the paper
is as follows: in the next
section we present, in Theorem A, our main results
for the finite size scaling of the
magnetization and susceptibility in the
context of a field driven transition with
two coexisting phases.
Section 3 is
devoted to the contour representation
of the models considered
in Section 2, together with
our main assumptions and
results for a more abstract
class of models describing the coexistence of $N$ phases.
We state two main theorems concerning the
finite-size scaling:
Theorem 3.1 on partition functions
and other thermodynamical quantities,
and Theorem 3.2 on the finite-size
scaling of local observables.
In Section 4 we will construct
suitable meta-stable free energies
and prove Theorem 3.1, deferring the
technical details to the
appendices. In Section
5 we construct
meta-stable states
and prove
the corresponding theorem,
Theorem 3.2.
In Section 6  we
prove
the results stated in
Section 2, using the abstract
results formulated in Section 3.

\vfill\eject
\head{2. Field driven transitions}
\endhead

\bigskip
\subhead{2.1. Definition of the model}
\endsubhead
\bigskip

In order to explain our main ideas,
we consider an {\it asymmetric} version
of the {\it Ising model}. Working on a finite
lattice
$
\Lambda=\{1, \dots,L\}^d
$,
$
d\geq 2
$,
we consider configurations
$
\sigma_{\Lambda}: i\mapsto
\sigma_i\in \{-1, 1\}
$
and the reduced
Hamiltonian
$$
H(\sigma_{\Lambda})=\frac{J}4
\sum_{\langle ij \rangle \subset
{\Lambda}}
|\sigma_i - \sigma_j|^2 - h
\sum_{i\in {\Lambda}}\sigma_i +
\sum_{A\subset {\Lambda}}\kappa_A
\prod_{i \in A} \sigma_i,
\tag2.1
$$
where $J$ is the reduced coupling
(containing a factor $\beta=1/k_B T$),
the first sum goes over
nearest neighbor pairs
$\langle ij \rangle$,
while the third one is a finite range
(i.e. $\kappa_A=0$ for
$\diam A < R$, where $R<\infty$)
perturbation with translation
invariant coupling constants
$\kappa_A\in\Bbb R$.
While the first two terms in (2.1)
describe the standard
Ising model, the
third term is a perturbation that
may break the $+/-$
symmetry of the Ising
model. 
We will assume that it is small
in the sense that
$$
||\kappa|| =
\sum\Sb A:0 \in A\endSb  
\frac{|\kappa_A|}{|A|} \leq b_0 J
$$
where $b_0>0$ is a constant to be specified in Theorem A below.

The partition function with {\it free
boundary conditions} is
$$
Z_{\free}(L, h) =
\sum_{\sigma_{\Lambda}}
e^{-H(\sigma_{\Lambda})}.
\tag2.2
$$
The derivatives of
its logarithm define the corresponding
{\it magnetization}
$$
m_{\free}(L, h)=
L^{-d}\frac{d}{dh}
\log Z_{\free}(L,h)
\tag2.3
$$
and the {\it susceptibility}
$$
\chi_{\free}(L,
h)=\frac{d}{dh}m_{\free}(L, h).
\tag2.4
$$
The {\it Binder cummulant},
$U_\free(L,h)$,
is given as
$$
U_\free(L,h)=
-{\langle M^4\rangle_c
\over
3\langle M^2\rangle^2}
=
{3
\langle(M-\langle M\rangle)^2\rangle^2
-\langle(M-\langle M\rangle)^4\rangle
\over
3
\langle(M-\langle M\rangle)^2\rangle^2
}
\tag 2.5
$$
where $\langle\cdot\rangle$
denotes expectations with respect to
the Gibbs measure corresponding to (2.1),
$\langle\cdot\rangle_c$
denotes the corresponding truncated expectation
values and $M=\sum_{i\in\Lambda}\sigma_i$.
Note that $U_\free(L,h)\leq 2/3$ by the inequality
$\langle F^2\rangle
\geq
\langle F\rangle^2
$
(applied to $F=(M-\langle M\rangle)^2$).

\bigskip
\subhead{2.2. Heuristic background,
main ideas}
\endsubhead
\bigskip

For low temperatures
(i.e. large $J$), the leading
contributions to the partition function
come from the constant ground state
configurations
$\sigma_{\Lambda}\equiv -1$
and  $\sigma_{\Lambda}\equiv +1$.
In this approximation,
$$
Z_{\free}(L, h)
\cong e^{-E_+(L,h)}+e^{-E_-(L,h)},
\tag2.6
$$
where
$$
E_\pm(L,h) = \sum _{i \in {\Lambda}}
e_\pm(i)
\tag 2.7
$$
with the position dependent
``ground state energies''
$$
e_\alpha(i) =
\sum\Sb A\subset {\Lambda}:
\\ i \in A\endSb \kappa_A \frac
{\alpha^{|A|}}{{|A|}} - h\alpha,\qquad
\alpha = \pm1.
\tag 2.8
$$
In the same approximation,
the magnetization
$m_{\free}(L, h)$
and susceptibility
$\chi_{\free}(L, h)$ are given by
$$
m_{\free}(L, h)\cong
\tanh(\frac{E_-(L,h)-E_+(L,h)}{2})
\tag 2.9
$$
and
$$
\chi_{\free}(L, h)\cong
L^d\cosh^{-2}(\frac{E_-(L,h)-
E_+(L,h)}{2}).
\tag 2.10
$$

Observing that $e_\alpha(i)$
differs from the bulk value
$e_\alpha$
if $i$  is in the vicinity of
$\partial {\Lambda}$, we expand
$E_\pm(L,h)$
into a bulk term $e_{\pm} L^d$
plus boundary terms,
$$
E_\pm(L,h)
= e_\pm(h) L^d + e_\pm^{(d-1)}(h)
2dL^{d-1} + O(L^{d-2}).
\tag 2.11
$$
While, still within the approximation
by ground states,
 the bulk transition point
$h_0$ is the value of $h$ of which
$e_+(h) = e_-(h)$, the finite volume
transition point $h_0(L)$ corresponds
to the equality of $E_+(L,h)$ and
$E_-(L,h)$.
By (2.11), this leads to a shift
$$
h_0(L) - h_0 = O(1/L).
\tag2.12
$$
Notice that for periodic boundary
conditions we get $h_0(L) = h_0$
 for zero temperature and, for
 nonvanishing temperatures,
  a shift $h_0(L) - h_0$
  proportional to
$1/L^{2d}$ for periodic b.c.
\cite{BK90, BK92}.

In order
to make the above considerations
rigorous, one has to take
into account the excitations around
the two ground states
$\sigma_{\Lambda}\equiv\pm 1$.
This is done
in Section 3 and 4 and leads
to a representation
$$
Z_{\free}(L, h)
= \left(
e^{-F_+(L,h)}+e^{-F_-(L,h)}
\right)(1+O(L^de^{-L/L_0})),
\tag2.13
$$
where $L_0$ is a constant
of the order of the
infinite volume
 correlation length and
$F_\pm(L,h)$ have an asymptotic
 expansion similar to
(2.11), namely
$$
F_\pm(L,h) = f_\pm(h) L^d +
f_\pm^{(d-1)}(h)
2dL^{d-1} + O(L^{d-2}),
\tag 2.14
$$
where $f_\pm(h)$ are meta-stable
free energies and
$f_\pm^{(d-1)}(h)$ are
(meta-stable) surface free energies.
Once these results (see
Theorem 3.1 in Section 3 for the precise
statements) are proven,
we obtain the desired finite-size
scaling results by a rigorous version
 of the method presented
in the introduction.

\bigskip
\subhead{2.3. Statements of results}
\endsubhead
\bigskip

In order to state our results in the
form of a theorem,
we introduce,
for $h\neq h_t$, the {\it free energy}
$$
f(h)\equiv f^{(d)}(h)=
-\lim_{L\to\infty}L^{-d}
\log Z_\free(L,h),
\tag 2.15a
$$
the {\it surface free energy}
$$
f^{(d-1)}(h)=-\lim_{L\to\infty}
\frac{1}{2dL^{d-1}}\Bigl[\log
 Z_\free(L,h) + L^{d}f(h)\Bigr],
\tag 2.15b
$$
..., the {\it corner free energy}
$$
f^{(0)}(h)=-\lim_{L\to\infty}
\frac{1}{2^d}\Bigl[\log  Z_\free(L,h)
 + L^{d}f(h)
+\cdots+2^{d-1}dL f^{(1)}(h)\Bigr],
\tag 2.15c
$$
as well as single phase magnetizations
 $m_\pm$ and
surface free energies $\tau_\pm$
 at the transitions point $h_t$,
$$
m_\pm=-\left.\frac{d}{dh}f(h)
\right|_{h_t\pm 0}
\tag 2.16
$$
$$
\tau_\pm=f^{(d-1)}(h_t\pm 0).
\tag 2.17
$$
We also recall that $||\kappa||$ was
defined as
$$
||\kappa|| =
\sum\Sb A:0 \in A\endSb  
\frac{|\kappa_A|}{|A|}
$$

\medskip
\proclaim{Theorem A: Finite size scaling
of $\bold m$ and  $\bold\chi$}

Consider a perturbed Ising model
with a perturbation of
the form (2.1), with translation
invariant coupling
constants  $\kappa_A$ with range
$R<\infty$.
Then there are constants
$J_0<\infty$ and  $b_0>0$
such that, for
$||\kappa||<b_0 J$
and $J>J_0$ ,
the following statements are true.
Let
$$
\align
\Delta F(L)
&= f^{(d-1)}(h_t+0)2d L^{d-1} +
\cdots + f^{(0)}(h_t+0)2^d\cr
&- f^{(d-1)}(h_t-0)2d L^{d-1} -
\cdots - f^{(0)}(h_t-0)2^d
\tag 2.18
\endalign
$$
and define $h_\chi(L)$ and
$h_U(L)$ as
the points where the susceptibility
$\chi_\free(L,h)$ and the Binder cummulant
$U_\free(L,h)$ are maximal.
Then\footnote{Here, and in the following,
$O(L^\alpha)$ stands for an error term
which can be bounded by $KL^\alpha$,
with a constant $K$ that does not depend
on $h$, $J$ and $\kappa$,
as long as $J>J_0$ and
$\Vert \kappa\Vert<b_0 J$.}
$$
m_{\free}(L,h)=\frac{m_++m_-}2
+ \frac{m_+-m_-}2
\tanh
\biggl( \frac{m_+-m_-}2(h-h_\chi(L))
L^d \biggr)
+O((1+\Vert\kappa\Vert)/L)
\tag 2.19
$$
and
$$
\chi_\free(L,h)=
\biggl(\frac{m_+-m_-}2\biggr)^2
\cosh^{-2}
\biggl( \frac{m_+-m_-}2(h-h_\chi(L))
L^d \biggr)
L^d
+ O((1+\Vert\kappa\Vert)L^{d-1})
\tag 2.20
$$
provided 
$|h-h_\chi(L)|\leq O((1+||\kappa||)L^{-1})$.

In addition, for $\Delta F(L) \neq 0$,
the shift $h_\chi(L)$ obeys the
bound
$$
h_\chi(L)=h_t + \frac{\Delta F(L) }{m_+-m_-}
\frac{1}{L^d}(1+O(1/L)).
\tag 2.21a
$$
In the leading order,
the shift of the point $h_U(L)$ with
respect to $h_t$ is the same,
$$
h_U(L)=h_t + \frac{\Delta F(L) }{m_+-m_-}
\frac{1}{L^d}(1+O(1/L)).
\tag 2.21b
$$

\endproclaim

\remark{Remarks}

i)
If $\tau_+\neq \tau_-$, the equation
(2.21a) (and similarly for (2.21b))
can be simplified to
$$
h_\chi(L)=h_t + \frac{\tau_+ -
\tau_-}{m_+-m_-}
\frac{2d}{L}(1+O(1/L))\,,
$$
yielding a shift $\sim 1/L$
which is much larger then the width of the rounding,
which, according to (2.19) and (2.20), is of the order
$1/L^d$.

ii)
It it is interesting to consider
the mutual shift
$h_\chi(L)-h_U(L)$.
While both
$h_\chi(L)-h_t$
and
$h_U(L)-h_t$
are of the order $1/L$,
their mutual shift
is actually much smaller,
namely
$$
h_\chi(L)-h_U(L)=2\frac{\chi_+-\chi_-}{(m_+-m_-)^3}
\frac1{L^{2d}}
+O(\frac1{L^{2d+1}}).
\tag 2.22
$$
It is interesting to notice that,
in the leading order $1/L^{2d}$,
this mutual shift is exactly the same
as the corresponding
shift for periodic boundary conditions.

iii) We stress that the condition
$|h-h_\chi(L)|\leq O((1+||\kappa||)L^{-1})$
is not
a very serious restriction in
our context, because the width
 of the transition in the
volume $L^d$ is only proportional
to $L^{-d}$. In fact,
in Section 6 we will
close the gap left in Theorem A
by showing  that
for
$|h-h_\chi(L)|>\frac{4d}{m_+-m_-}(1+||\kappa||)L^{-1}$, 
one has
$$
|m_{\free}(L,h)-m(h)|\leq O(1/L)
\tag 2.23
$$
and
$$
|\chi_{\free}(L,h)-\chi(h)|\leq O(1/L),
\tag 2.24
$$
where $m(h)$ and $\chi(h)$
are the infinite volume magnetization
and susceptibility of the model (2.1).

iv)  Notice that, for periodic boundary
conditions, it is possible to define
finite size transition points $h_t(L)$
with exponentially small
shift, for example the point
where $m_{\text{per}}(L,h) =
m_{\text{per}}(2L,h)$.
Here, all these definitions  lead to a
shift $\sim 1/L$ yielding no qualitative
improvement with respect to the
point $h_\chi(L)$ or $h_U(L)$.

v) In principle, the coefficients
$m_\pm$, $\tau_\pm$, ..., can be
calculated up to arbitrary precision
using standard
series expansions, provided the
microscopic Hamiltonian is known.
On the other hand, the scaling
(2.19),
(2.20), and (2.21) would allow,
in principle, to obtain the
coefficients $m_+$, $m_-$
and the difference $\tau_+-\tau_-$
from experimental
measurements.

vi) The general context considered
in Section 3 allows to analyze the finite
size scaling with more general boundary
conditions then the free boundary conditions
considered here, including, in particular,
small applied boundary fields favoring
one of the two phases near the boundary.
In order to apply the techniques developed
in this paper, it is necessary, however,
to exclude boundary conditions
which strongly favor
one of the two phases.
Such a condition is
needed to ensure that the main
contributions to the partition functions
do in fact come from small perturbations
of the two ground states
$\sigma_\Lambda\equiv\pm 1$. For large
boundary fields,
the boundary may strongly favor one of
the two phases. The leading contributions
to the partition function then would
include configurations which are in one
phase near the boundary, and in the other
one for the bulk. In such situations,
wetting and roughening effects of the
contour separating the boundary phase
from the bulk phase would be important
physical effects. We are not attempting
to study
these effects in the present paper.
\endremark

\vfill\eject
\head{3. General Setting and Main Theorem}
\endhead

\bigskip
\subhead{3.1. Contour Representation of
  the Ising Model}
\endsubhead
\bigskip

In this section we review the contour
representation for the model (2.1).
To make this subsection as simple as
possible, and to have a concrete
example at hand, we use for illustration
the simplest symmetry breaking term,
namely a perturbation of the form
$$
\kappa\sum_{\langle ijk\rangle\subset\Lambda}
\sigma_i\sigma_j\sigma_k\,,
\tag 2.1${}^\prime$
$$
where the sum goes over all triangles $<ijk>$
made out of two nearest neighbor bonds $<ij>$
and $<jk>$.
See
\cite{PS75, 76} for the contour representation
for the more general model (2.1). It will
be convenient to introduce, in addition to
the finite lattice
${\Lambda}=\{1,\cdots,L\}^d$, the subset
$V=[\half,L+\half]^d$ of ${\Bbb R}^d$
which is obtained from ${\Lambda}$ as the
union of all closed unit cubes $c_i$ with
centers $i\in {\Lambda}$. For a given
configuration $\sigma_{\Lambda}\subset
\{-1,1\}^{\Lambda}$, we then introduce the
set $\partial$ as the boundary between the
region $V_+\subset V$ where $\sigma_i =
+1$ and the region $V_-\subset V$ where
$\sigma_i = -1$, and the contours
$Y_1,\cdots,Y_n$ corresponding to
$\sigma_{\Lambda}$ as the connected
components of $\partial$.

To be more precise, we define an
{\it elementary cube}
as a
closed unit cube with a center in
 ${\Lambda}$ (we sometimes use the symbol
$c_i$ to denote an elementary cube with
center $i\in\Lambda$), and
introduce
$\overline{V}_\pm$ as the union of all
closed elementary cubes $c_i$
for which $\sigma_i = \pm1$,
respectively. The set
$\partial$ is then defined as
$\overline{V}_+ \cap \overline{V}_-$,
and the ``ground state
regions'' $V_\pm$ are defined as
$\overline{V}_\pm \setminus\partial$.
With these definitions, the
partition function
with Hamiltonian (2.1${}^\prime$)
can  be rewritten in the form
$$
Z_{\free}(L, h) = \sum_\partial
{\sum_{\sigma_{\Lambda}}}^{\prime}
e^{- H(\sigma_{\Lambda})},
$$
where the second sum is over all
configurations consistent with $\partial$.

In order to specify the configuration
$\sigma_{\Lambda}$, one has to decide
which
component of $V\setminus\partial$
corresponds to $\sigma_i = +1$ and
which one
to $\sigma_i = -1$. To this end, we
introduce contours with labels. Given a
configuration $\sigma_{\Lambda}$,
the contours
corresponding to $\sigma_{\Lambda}$ are
defined as pairs
$Y=(\operatorname{supp} Y,
\alpha(\cdot))$, where
$\operatorname{supp} Y$ is a connected
component of $\partial$ while $\alpha$
is an assignment of a label
$\alpha (c)\in \{-1, +1\}$ to each
elementary
cube that touches $\operatorname{supp}Y$%
\footnote{In the language of
\cite{HKZ88}, $\operatorname{supp} Y$ is called a
(geometric) contour, while $Y$ is
called a labeled contour.}.
It is chosen in such a way that
$\alpha(c_i)=\sigma_i$. Note that the
labels
of contours corresponding to a
configuration $\sigma_{\Lambda}$ are
matching in the sense that the
labels $\alpha(c)$ are constants on every
component of $V\setminus \partial$.

In fact, a set of contours
$\{Y_1,\dots,Y_n\}$ corresponds to
a configuration $\sigma_{\Lambda}$,
if and only if
\item{i)} $\operatorname{supp}Y_i
\cap \operatorname{supp}Y_j = \emptyset$
for $i\neq j$ and
\item{ii)} the labels of $Y_1,\dots,Y_n$
are matching.

\noindent We call a set of contours
 obeying i) a
{\it set of non-overlapping contours}
and a set of contours obeying i) and ii)
a  {\it set of non-overlapping contours
 with matching labels},
or sometimes just a {\it set of
 matching contours}.

In order to rewrite $Z_{\free}(L, h)$ in
terms of contours, we assign a weight
$\rho (Y)$ to each contour.
This is done in such a way that
$$
c^{-H(\sigma_{\Lambda})}
=e^{-E_+(V_+)}e^{-E_-(V_-)}
\prod_{k=1}^n\rho(Y_k).
\tag 3.1
$$
Here $H(\sigma_{\Lambda}) $ is the
Hamiltonian (2.1${}^\prime$),
$Y_1,\dots Y_n$ are the
contours corresponding to
$\sigma_{\Lambda}$ and
$$
E_\pm(V_\pm)=\sum_{i\in
\Lambda\cap V_\pm} e_\pm(i).
\tag 3.2
$$
For the standard Ising model,
$\rho(Y) = e^{-J |Y|}$, where $|Y|$
is the number of elementary
$(d-1)$-dimensional
faces in
$\operatorname{supp}Y$. The
third term in (2.1${}^\prime$), however,
introduces corrections yielding
a weight of the
form $\rho (Y)=e^{-J|Y| + O(\kappa |Y|)}$.
As a consequence,
$$
|\rho(Y)|\leq e^{-\tau |Y|} \text{ with }
\tau = J-O(\kappa).
\tag 3.3
$$
Similar bounds hold for the
derivatives $| d^k\rho(Y)/dh^k|$.

With the help of (3.1), we rewrite
the partition function $Z_{\free}(L,h)$ as
$$
Z_{\free}(L,h) = \sum_{\{ Y_1,
\dots, Y_n\}}
e^{-E_+(V_+)}e^{-E_-(V_-)}
\prod _{k=1}^h\rho (Y_k),
\tag 3.4
$$
where the sum goes over all
sets of matching contours in $V$.

\bigskip
\subhead{3.2. Assumptions for the
General Model}
\endsubhead
\bigskip

In Section 3.3 below, we
will state our main theorem,
Theorem 3.1, from which we
infer Theorem A of the
preceding section. The setting
of Theorem 3.1 is actually
more general then what is needing
for Theorem A and will include
more general models.
On one hand, we introduce contours
in such a way
that the notion of contours covers
the Ising contours introduced
above as
well as thick Pirogov-Sinai
contours
\cite{PS75, 76, Si82}
constructed as unions of
elementary cubes%
\footnote{The contours are introduced
in such a way that the more general
cases considered in \cite{BW89, 90, HKZ88} are
covered as well.}.
On the other hand, we also consider
the situation of general $N$ phase
coexistence.

As before, we consider the finite lattice
$\Lambda \subset \Bbb Z^d$, $d\geq 2$,
and the
corresponding volume $V \subset \Bbb R^d$.
We introduce the set
$\Cal C$ of {\it elementary cells} as the
set of all elementary cubes in $V$, all
closed $d-1$ dimensional faces of these
cubes, ..., and all closed edges of these
cubes. As usual, we define the boundary $\partial W$
of a
set $W\subset V$ as the set of all points $x$
which have distance zero from both $W$ and $W^c$
and $\overline{W}$ as $W\cup\partial W$.

A contour in $V$ is then a pair
$Y=(\operatorname{supp}Y, \alpha (\cdot))$
where $\operatorname{supp}Y$
is a connected
union of elementary cells and $\alpha
(\cdot)$ is an assignment of a label
$\alpha(c)$ from a finite set
$\{1,\dots,N\}$ to each elementary cube
$c$ in
$\overline{V\setminus \supp Y}$ which touches $Y$ (by touching we
mean that
$c\cap \operatorname{supp}Y\neq\emptyset$
while
$(c\setminus\partial c)\cap\operatorname{supp}Y=
\emptyset$).
As before, we require that $\alpha$ is
constant on each component $C$ of
$V\setminus \operatorname{supp}Y$, and
say that a set $\{Y_1,\dots Y_n\}$ of
contours is a {\it set of matching
contours} (or, more explicitly,
{\it a set of non-overlapping contours
with matching labels}) iff

\flushpar
i) $\operatorname{supp}Y_i \cap
\operatorname{supp}Y_j\neq\emptyset$ for
$i\neq j$ and

\flushpar
ii) the labels of $Y_1,\dots
Y_n$ are matching in the sense
that they are constant on
components of $V\setminus
(\operatorname{supp}Y_1\cup
\dots\cup\operatorname{supp}Y_n)$.

In  this way, each component $C$
of $V\setminus
(\operatorname{supp}Y_1\cup\dots
\cup\operatorname{supp}Y_n)$ has constant
boundary conditions on
$\partial C \setminus\partial V$.
The partition
function of a statistical model with
``weak'' boundary conditions is then
rewritten in terms of contours as
$$
Z(V,h)=\sum_{\{Y_1,\dots, Y_n\}}
\prod^n_{k=1}\rho(Y_k)
\prod_{m=1}^N e^{-E_m(V_m)},
\tag3.5
$$
where the sum goes over
sets of matching contours in $V$
(including the empty set of contours),
 and $V_m$ is the
union of all components of $V\setminus
(\operatorname{supp}Y_1\cup\dots
\cup\operatorname{supp}Y_n)$ that have
boundary condition $m$,
and
$$
E_m(V_m)=\sum_{c\subset\overline{V}_m} e_m(c).
\tag 3.6
$$
We point out that the sum in (3.6)
goes over all elementary cubes in the closure
$\overline{V}_m$
of
$V_m$,
a convention which was chosen to ensure that
all elementary cubes $c$
with center in $V_m$ are taken into account
\footnote{A sum over elementary cubes
$c\subset V_m$ would exclude those elementary
cubes $c\subset\overline{V}_m$ which touch
one of the contours $Y_1,\ldots,Y_n$.}.
Note that
by our definition of $V$ as a closed
 subset
of $\Bbb R^d$, the sum (3.5) contains
contours that touch $\partial V$
(in the sequel, we call these contours
{\it boundary contours},
as well as  contours that do not touch
$\partial V$, {\it ordinary
contours}).
The contribution of the collection
of empty contours to (3.5)
is actually a sum of $N$ terms,
$\sum_m e^{-E_m(V)}$.

In
the equalities
(3.5) and (3.6) we have introduced
``contour
weights'' $\rho(Y)\in \Bbb R$ and
``ground state energies''
$e_m(c)\in\Bbb R$
that depend on a vector parameter
$h\in
\Cal U$,
where $\Cal  U$ is an open
subset of $\Bbb  R^\nu$. We assume that
$\rho(Y)$ and $e_m(c)$ are translation
invariant as long as $Y$ and $c$
do not touch the boundary of $V$. More
generally, we assume translation
invariance along a $(d-k)$
dimensional face
in $\partial V$ as long as $Y$ (or
$c$) does not touch the $(d-k)-1$
dimensional boundary of this face.

As usually, we have to assume the
{\it Peierls condition}, together with
several assumptions on the
ground state energies $e_m(c)$.
Here, we
assume that $e_m(c)$ and $\rho (Y)$
 are $C^6$ functions of $h$
obeying the following bounds:
$$
|\rho(Y)|\leq e^{-\tau |Y| - E_0(Y)},
\tag 3.7
$$
$$
\biggl|\frac{d^k\rho (Y)}
{dh^k}\biggr|\leq
|k|!(C_0|Y|)^{|k|}
e^{-\tau|Y|-E_0(Y)},
\tag 3.8
$$
and
$$
\biggl|\frac{d^ke_m(c)}
{dh^k}\biggr|\leq
C_0^{|k|}.
\tag 3.9
$$
Here $\tau > 0$ is a sufficiently
large constant,
$|Y|$ denotes the number of
 elementary cells
in\footnote{Here, a $k$-dimensional
 cell $c$ in $\supp Y$ is
only counted if there
is no $(k+1)$-dimensional cell
$c^\prime$ in $\supp Y$
with $c\subset c^\prime$.}
$\operatorname{supp} Y$,
$$
E_0(Y)=\sum_{c\subset
\operatorname{supp}Y}e_0(c)
\qquad\text{with}\qquad
e_0(c) = \min_m e_m(c),
\tag 3.10
$$
$k$ is a multi-index
$k=(k_\alpha)_{\alpha = 1,
\dots, \nu}$ with $1\leq |k|\leq 6$,
$|k|=\sum k_{\alpha }$, and $C_0$
is a  constant
independent of $h$ and $\tau$. In
addition, we assume that the
difference between $e_m(c)$ and
the bulk term
$e_m$ is bounded,
$$
|e_m(c) - e_m|\leq\gamma\tau,
\tag 3.11
$$
with a constant
$0<\gamma <1$
to be specified later. This
condition is introduced to avoid
a situation where free b.c. strongly
favor certain phases
$n\in \{1, \dots, N\}$.
Note that
$$
|e_m(c) - e_m|\leq ||\kappa||
$$
for the asymmetric Ising
model (2.1).
For this model, the
condition (3.11) is
therefore satisfied once $||\kappa ||\leq b_0 J$
for a suitable constant $0<b_0<\infty$.

\bigskip
\subhead{3.3. Main Theorem}
\endsubhead
\bigskip

In this section we state our main
result for the general model introduced
in the last section. It actually
generalizes
Theorem A presented in
Sections 2
to a large class of models describing
the coexistence of N phases. As in
Section 2, the leading contribution
to the partition function $Z(V,h)$ is
the sum
$$
\sum_{m=1}^{N} \exp\biggl\{-
\sum_{c\subset V} e_m(c)\biggr\}.
\tag 3.12
$$
Introducing $|\partial_k V|$ as the
joint $k$-dimensional area of all
$k$-dimensional faces of $V$ and
$e_m^{(k)}$ as solutions of equations
$$
\sum_{n=k}^{d}
\binom{d-k}{n-k}  e_m^{(n)} =
e_m(c),\, k=d-1,\dots,0,
\tag 3.13
$$
whenever  $c$ is touching a
$k$-dimensional
face of $V$ and not touching its
$(k-1)$-dimensional boundary%
\footnote{Note that due to translation
invariance properties of $e_m(c)$,
the right hand side of this equation
is constant for all
such elementary cubes $c$.},
we rewrite
$$ \sum_{c\subset V}e_m(c)=e_m|V| +
e_m^{(d-1)}|\partial_{d-1}V| +
\dots + e_m^{(0)}|\partial_0 V|.
\tag 3.14
$$
To see that (3.13) implies (3.14), just notice that
a hypercube $c$ touching a $k$-dimensional face of $V$ and
not touching its
$(k-1)$-dimensional boundary is touching
$\binom{d-k}{n-k}$
different $n$-dimensional faces of $\partial V$.
Each of these faces is specified
 by choosing $n-k$ directions among $d-k$
directions orthogonal to the concerned $k$-dimensional face.

As usually we define the
 bulk free energy $f(h)$ by
$$
f(h) = -\lim_{V\to\Bbb R^d}
\frac1{|V|} \log Z(V, h).
\tag 3.15
$$
and the magnetization
$m(V, h)=(m_\alpha(V, h))_{\alpha=1,
 \dots, \nu}$ by
$$
m(V, h) = \frac1{|V|}\frac d{dh}
\log Z (V, h).
\tag 3.16
$$

\proclaim {Theorem 3.1}
There exist  constants $b>0$,
$\gamma_0>0$ and $\tau_0<\infty$
(where $b$ and $\gamma_0$ depend on $d$
and $\tau_0$ depends on $d$, $N$
and the constant $C_0$ introduced in {\rm (3.8)}
and {\rm (3.9)}), as well as
meta-stable free energies $f_m(h)$,
 surface
free energies $f^{(d - 1)}_m(h),
\dots,$ edge free energies
$f^{(1)}_m(h)$ and corner free
energies $f_m^{(0)}(h)$, such
that the following
statements are true provided the effective
decay constant $\tilde\tau$,
$$
\tilde\tau:=\tau(1- \gamma/\gamma_0) - \tau_0>0
\tag 3.17
$$
(for the definition of $\tau$ and
$\gamma$ see {\rm (3.7), (3.8)} and {\rm (3.11)}).

\item{i)} $\displaystyle{f(h)=
\min_m f_m(h)}$.
\item{ii)} $f_m$ and $f_m^{(l)}$,
$l=d-1,\dots, 0,$ are 6 times
differentiable functions of $h$.
\item{iii)}
If $ |k| \leq 6$,
then
$\displaystyle{\biggl\vert
\frac{d^k}{dh^k}(f_m-e_m)\biggr\vert
\leq e^{-b\tilde\tau}}$
and
$\displaystyle{\biggl\vert
\frac{d^k}{dh^k}(f_m^{(l)}-e_m^{(l)})
\biggr\vert \leq
e^{-b\tilde\tau}}$,
\hfill\newline
where $l=d-1$, $\dots$, $0$.
\item{iv)}
Let
$$
 F_m(V, h) =
f_m(h) |V| + f_m^{(d-1)}(h)
|\partial_{d-1}V| +\dots+ f_m^{(0)}(h)
|\partial_0 V|.
\tag 3.18
$$
Then
$$
\left\vert
\frac{d^k}{dh^k}
\biggl[
Z(V,h) - \sum_{m=1}^N e^{-F_m(V,h)}
\biggr]
\right\vert
\leq
|V|^{|k|+1}
O(e^{- b \tilde\tau L})
\max_m e^{-F_m(V,h)}
\tag 3.19
$$
provided
$\displaystyle{0\leq |k|\leq 6}$.
\item{v)} Let
$\displaystyle{0\leq |k|\leq 5}$
and define
$P_q$ as
$$
P_q=
\biggl[
\sum_{m=1}^N e^{-F_m(V,h)}
\biggr]^{-1}
e^{-F_q(V,h)}.
\tag 3.20
$$
 Then
$$
\left\vert
\frac{d^k}{dh^k}
\biggl[
m_\alpha(V,h)
- \sum_{q=1}^N
{1\over |V|}
\left(
      -{dF_q(V,h)\over dh_\alpha}
\right)
P_q
\biggr]
\right\vert
\leq
|V|^{|k|}
O(e^{- b \tilde\tau L}).
\tag 3.21
$$
\endproclaim
\nobreak
\noindent
Here, as in the rest of this paper,
$O(x)$ stands for a bound $\text{const}\,x$
where the constant depends only on $d$, $N$ and the constant
$C_0$ introduced in (3.8) and (3.9).

Theorem 3.1 is the main theorem of this
paper.
Its proof has three major parts: the
geometric analysis of contours touching
the boundary, a decomposition of $Z(V,h)$
into pure phase partition functions,
and the construction of
meta-stable contour models allowing to
prove the bounds (3.19) and (3.21).
Deferring the technical details to
the appendices, the main steps of this proof are
presented in Section 4.

\bigskip
\subhead{3.4. FSS for Local Observables}
\endsubhead
\bigskip

In addition to the FSS of thermodynamic quantities like
the magnetization or susceptibility, we want to study
the FSS of local observables.
In order to state our results in the context of
the general models considered in Section 3.2,
we introduce the following notation.
An observable
$A$ is a function which associates to each configuration
contributing to (3.5) a real number
$A(Y_1,\cdots,Y_n)$. Its expectation value in the volume
$V$ is defined as
$$
\langle A \rangle^{h}_V
=
{1\over Z(V,h)}Z(A\mid V,h),
\tag 3.22
$$
where
$$
Z(A\mid V,h)=
\sum_{\{Y_1,\dots, Y_n\}}
A(Y_1,\cdots,Y_n)
\prod^n_{k=1}\rho(Y_k)
\prod_{m=1}^N e^{-E_m(V_m)}\,.
\tag 3.23
$$
As in (3.5), the sum in (3.23)
goes over
sets of matching contours in $V$,
 and $V_m$ is the
union of all components of $V\setminus
(\operatorname{supp}Y_1\cup\dots
\cup\operatorname{supp}Y_n)$ that have
the boundary condition $m$.

An observable $A$ is called a local observable,
if there is a finite set of elementary
cubes, denoted  $\supp A$ in the sequel, such that
$A(Y_1,\cdots,Y_n)$ does not depend on
those contours $Y_i$ for which
$\supp A\cap (\supp Y_i\cup \Int Y_i)=\emptyset$,
where $\Int Y_i$ is the interior of $Y_i$
(for the precise definition of $\Int Y_i$ see
Section 4.1 below).

In most applications, local observables
will be bounded, in the sense that the norm
$$
\parallel A \parallel
=\sup_{\{Y_1,\cdots,Y_n\}}
|A(Y_1,\cdots,Y_n)|
\tag 3.24
$$
is finite. In addition,
the observable will either not depend on
the vector parameter $h$ at all, or obey bounds
of the form
$$
\biggl|
\frac{d^k}{dh^k}
A(Y_1,\cdots,Y_n)
\biggr|\leq
|k|!
C_A
(C_0|\supp A|)^{|k|},
\tag 3.25a
$$
where $C_0$ is the constant introduced in
(3.8), $C_A$ is a  constant
and $k$ is a multi-index of order
$0\leq |k|\leq 6$.

Here, we will allow for a slightly
more general situation,
requiring only that
$$
\biggl|
\frac{d^k}{dh^k}
\biggl[
A(Y_1,\cdots,Y_n)
\prod^n_{j=1}\rho(Y_j)
\biggr]
\biggr|
\leq
|k|!C_A (C_0|\overline{\supp Y_A}|)^{|k|}
\prod^n_{j=1}
e^{-\tau|Y_j|-E_0(Y_j)}\,,
\tag 3.25b
$$
where $\overline{\supp Y_A}$ stands for the
set $\supp A\cup\supp Y_1\cup\cdots\cup\supp Y_n$,
$k$ is a multi-index of the order
$0\leq |k|\leq 6$,
$C_0$ is the constant introduced in (3.8)
and $C_A$ is a constant that is
finite\footnote{While we assumed that the
constant $C_0$ is independent of $h$ and $\tau$,
we do not require that $C_A$ is independent
of $h$ and $\tau$.}
for all $h$ and $\tau$.
Assuming this condition\footnote{Note
that (3.7), (3.8) and  (3.25a) imply the bound
(3.25b).} and the
conditions introduced in Section 3.2,
we will be able to prove the following
theorem.

\proclaim {Theorem 3.2}
There are
``meta-stable
expectation functionals''
$\langle\cdot\rangle^{h}_{V,q}$,
$q=1,\cdots,N$, such
that the following
statements are true provided
the effective decay constant
$
\tilde\tau:=\tau(1- \gamma/\gamma_0) - \tau_0
$
defined in Theorem 3.1. is positive
and $0\leq|k|\leq 6$.

\item{i)} For each local observable obeying the bounds
(3.25a) or (3.25b),
one has
$$
\left\vert
\frac{d^k}{dh^k}
\biggl[
\langle A \rangle^{h}_V
- \sum_{q=1}^N
\langle A \rangle^{h}_{V,q}
P_q
\biggr]
\right\vert
\leq
C_A e^{O(\epsilon)|\supp A|}
O(e^{- b \tilde\tau L}),
\tag 3.26
$$
where the probabilities $P_q$
and the constant $b$ are as
in Theorem 3.1 and $\epsilon=e^{-\tilde\tau/2}$.

\item{ii)} For each local observable obeying the bounds
(3.25a) or (3.25b), the limits
$$
\langle A \rangle^{h}_q=
\lim_{V\to{\Bbb R}^d}
\langle A \rangle^{h}_{V,q}
\tag 3.27
$$
exist as $C^6$ functions of $h$,
and obey the bounds
$$
\left\vert
\frac{d^k}{dh^k}
\langle A \rangle^{h}_q
\right\vert
\leq
O(1)
C_A |\supp A|^{|k|}
e^{O(\epsilon)|\supp A|}
\,,
\tag 3.28
$$
where $\epsilon=e^{-\tilde\tau/2}$.
\item{iii)}
For each local observable obeying the bounds
(3.25a) or (3.25b),
one has
$$
\left\vert
\frac{d^k}{dh^k}
\bigl[
\langle A \rangle^{h}_q-
\langle A \rangle^{h}_{V,q}
\bigr]
\right\vert
\leq
C_A |\supp A|^{|k|}
e^{O(\epsilon)|\supp A|}
O(e^{- b \tilde\tau \dist(\supp A,\partial V)})
\,,
\tag 3.29
$$
where $\epsilon=e^{-\tilde\tau/2}$.

\endproclaim

\bigskip
 \demo{Proof}
The proof of Theorem 3.2 is given in Section 5.
\enddemo

\vfill\eject
\head{4. Proof of Theorem 3.1}
\endhead

The proof of Theorem 3.1 has three major parts:
the geometric analysis of contours --- in particular
a bound of the form
$$
N_{\partial V}(\Int Y)
\leq \text{const} |Y|,
$$
where $N_{\partial V}(\Int Y)$ denotes the number
of elementary cubes in\footnote{We
recall that we
use the symbol $\overline{W}$ to
denote the closure of a set $W$.}
$\overline{\Int Y}$
that touch the boundary $\partial V$ of $V$,
the decomposition of $Z(V,h)$
into pure phase
partition functions $Z_1(V,h)$, $\cdots$, $Z_N(V,h)$,
and the construction of suitable meta-stable free
energies $f_1,\cdots,$ $f_n$. Deferring the technical details
to the appendices, we present the main steps in
the following subsections.

\bigskip
\subhead{4.1. The geometry of contours}
\endsubhead
\bigskip

An important notion in the Pirogov-Sinai
theory of contour models is the notion
of the interior and exterior of a contour.
 For ordinary contours
$Y=(\operatorname{supp} Y,
\alpha (\cdot))$,
 one defines $\operatorname{Int} Y$
 as the union of all finite
components of $\Bbb R^d \setminus
\operatorname{supp} Y$
 and $\operatorname{Int}_m Y$ as
the union of all
components of
$\operatorname{Int} Y$ which have
the boundary condition $m$.
 Since ordinary contours do not touch
the boundary $\partial V$ of $V$, the set
 $\operatorname{Ext}Y =V\setminus
(\operatorname{supp}
Y \cup \operatorname{Int} Y)$
is a connected set and $\alpha(c)$
is constant for all cubes $c$
in $\overline{\operatorname{Ext} Y}$
which touch $\operatorname{supp} Y$.
We say that $Y$ is
an $m$-contour,
 if $\alpha (c) = m$ for these
cubes.

We now generalize these
notions to boundary contours.
To this end, we first
introduce, for each corner $k$ of the box
$V$, an ``octant'' $K(k)$. Namely, if $k$
has components $k_1,
\dots k_d,$ with $k_i = 1/2$
for $i\in I_-$ and
$k_i = L+1/2$ for $i\in I_+$, then
$$
K(k):=\{x\in \Bbb R^d \mid x_i\geq 1/2
\text{ for }  i\in I_-,
x_i\leq L+1/2 \text{ for }  i\in I_+\}.
$$
We then say: a contour $Y$ is {\it short}
iff there is a corner $k$ such that
$\operatorname{supp} Y
\cap \partial V\subset\partial K(k)$.
Otherwise $Y$ is called {\it long}.
Note that short contours may be ordinary
contours or boundary contours, while long
contours are always boundary contours.

For a short contour $Y$, we then define
$\operatorname{Int} Y$
as the union of all finite
components of $K(k)\setminus
\operatorname{supp} Y$,
 $\operatorname{Int}_m Y$
as the union of all components
of $\operatorname{Int} Y$ which have
the boundary condition $m$,
$\operatorname{Ext}Y$ as
$V\setminus (\operatorname{supp} Y\cup
\operatorname{Int} Y)$
and $V(Y)$ as
$\operatorname{supp} Y\cup \operatorname{Int}Y$.
As before
$\operatorname{Ext}Y$ is a connected set,
and the notion of an $m$-contour is
defined by the condition that
$\alpha(c)=m$ for all cubes $c$ in $\overline{\Ext Y}$
 that touch $\operatorname{supp} Y$. Note that these
definitions
 are equivalent to the previous ones if
the short contour $Y$ is
in fact an ordinary contour.
Note also that the above
 definitions do not depend on
the choice of the corner
$k$ if there are several
 corners $k$ for which
$\operatorname{supp} Y\cap \partial V
\subset K(k)$.

For long contours, there is a priori
 no natural notion of an exterior or
interior. We chose a
convention that ensures that
that the volume of
a component
 $C_i$ of $\operatorname{Int} Y$
cannot exceed the value $L^d/2$
if $Y$ is a long contour.
Namely,
if $Y$ is a long boundary
contour, and $C_1, \dots C_n$ are
the components
of $V\setminus \operatorname{supp} Y$,
then the component $C_i$ with the
largest volume is called the exterior
$\operatorname{Ext}Y$.
If there are several such components
$C_{i_1}, \dots, C_{i_l}$,
we chose the first
one
 in some arbitrary
fixed
order
(for example the lexicographic order)
 as $\operatorname{Ext}Y$.
 We then define
$\operatorname{Int} Y=
V\setminus(\operatorname{supp} Y\cup
\operatorname{Ext}Y)$,
$V(Y)=\operatorname{supp} Y\cup \operatorname{Int}Y$,
$\operatorname{Int}_m Y$ as
the union of all components of
$\operatorname{Int} Y$ which have the
 boundary condition $m$, and an
 $m$-contour
$Y$ as a contour for which
$\alpha(c)=m$ on all cubes $c$ in $\overline{\Ext Y}$
that touch
 $\operatorname{supp} Y$.

The following
three Lemmas state that
the sets $\operatorname{Ext}Y$ and
$\operatorname{Int} Y$ are defined
in such a way that they have the
main properties of an exterior and
interior of the set
$\operatorname{supp} Y$.
They will be proven in Appendix B.

The first of them expresses the fact that
for two contours $Y_1$ and $Y_2$,
which do not touch each other, $Y_1$ together
with it's interior is necessarily contained
in one of the components of
$\operatorname{Ext}Y_2\cup\operatorname{Int}Y_2$.

\proclaim{Lemma 4.1}
Let $Y_1, Y_2$ be
non-overlapping contours.
Then the following statements are true.
\item{i)} If $\operatorname{supp}
Y_2\subset \operatorname{Ext}Y_1$ and
$\operatorname{supp}
Y_1\subset \operatorname{Ext}Y_2$,
then
$V(Y_2)\subset
\operatorname{Ext}Y_1$
and
$V(Y_1)
\subset \operatorname{Ext}Y_2$.
\item{ii)} If $\operatorname{supp}
Y_1\subset C_2$,
 where $C_2$ is a component of
$\operatorname{Int} Y_2$,
then
$V(Y_1)\subset C_2$.
\item{iii)} If $\operatorname{supp}
 Y_1\subset \operatorname{Int} Y_2$,
then
$V(Y_1)
\subset \operatorname{Int} Y_2$.
\endproclaim
The next lemma expresses the fact
that it is not possible that two
contours which do not touch
are both included in the
interior of each other.

\proclaim{Lemma 4.2}
Let $Y_1$ and $Y_2$ be
 non-overlapping contours.
Then one and only one of the
following three cases is true:
\item{i)} $\operatorname{supp}
 Y_2\subset
\operatorname{Ext}Y_1$ and
$\operatorname{supp} Y_1\subset
\operatorname{Ext}Y_2$,
\item{ii)} $\operatorname{supp}
 Y_2\subset
\operatorname{Ext}Y_1$ and
$\operatorname{supp} Y_1\subset
\operatorname{Int} Y_2$,
\item{iii)} $\operatorname{supp}
 Y_2\subset
\operatorname{Int} Y_1$ and
$\operatorname{supp} Y_1\subset
 \operatorname{Ext}Y_2$.
\endproclaim

\definition{Definition 4.3}
Let $\{Y_1,\dots,Y_n\}$ be a set
of non-overlapping contours.
Then $Y_k\in \{Y_1,\dots,Y_n\}$
is called an
{\it internal contour}
iff there exists a contour
$Y_i\in \{Y_1,\dots,Y_n\}$  with
$\operatorname{supp}Y_k\subset
\operatorname{Int}Y_i$.
Otherwise $Y_k$ is called
an
{\it external contour}.
Finally, $\{Y_1,\dots,Y_n\}$
is called a set of
{\it mutually
external contours}, if all
contours in $\{Y_1,\dots,Y_n\}$
are external.
\enddefinition

The next Lemma will be used in
Section 4.2 to conclude that all
external contours of a given configuration
contributing to (3.5) have the same
external label. This observation
will be an important ingredient in the
decomposition of $Z(V,h)$ into
single phase partition functions
$Z_m(V,h)$, and therefore in the proof of
Theorem 3.1.

\proclaim{Lemma 4.4}
Let $\{Y_1,\dots,Y_n\}$ be a set
of non-overlapping contours in $V$,
and let
$$
\operatorname{Ext}
=V\setminus\bigcup_{i=1}^n
(\operatorname{Int}Y_i\cup
\operatorname{supp}Y_i).
\tag 4.1
$$
Then $\operatorname{Ext}$
is a connected component of
$V\setminus\bigcup_{i=1}^n
\operatorname{supp}Y_i$.
\endproclaim

\remark{Remark}
Let $Y_0$ be a contour, and let $W_0$
 be one of the components
of $\operatorname{Int}Y_0$.
Then Lemma 4.4 remains valid if
$V$ is
replaced by $W_0$, as can be seen
 immediately from the proof in Appendix B.
\endremark

While the preceding three Lemmas,
even though tedious to prove,
just express our intuitive notions about
exteriors and interiors (in fact,
our definitions were chosen in such a way
that they do), the next Lemma is less
obvious.
In order to explain the need for it,
we recall that the ground state energies
$e_m(c)$ may be different from the
corresponding bulk term $e_m$.
As a consequence, the boundary may
favor an otherwise unstable phase.
In the expansion about the leading
contribution $e^{-E_m(V)}$ to the single
phase partition functions $Z_m(V,h)$,
this will have the tendency to
increase the weight of boundary contours
which describe transitions into one of
these ``boundary favored'' phases.
In order to control the contributions coming
from such contours (using the exponential
decay $e^{-\tau\mid Y\mid})$, we need a bound of the
form
$$
N_{\partial V}(\Int Y)
\leq
\text{ const } |Y|,
$$
where
$
N_{\partial V}(\Int Y)
$ denotes the number of elementary cubes in
$\overline{\supp Y}$
that touch the boundary $\partial V$
of $V$.
This is the main statement of the next Lemma.

\proclaim{Lemma 4.5}
Let $Y$ be a contour in $V$, and
let $W_1$, ..., $W_n$ be the components
of $\operatorname{Int}Y$. Then
$$
N_{\partial V}(\Int Y)
\leq C_1 |Y|\,,
\tag 4.2
$$
$$
\sum_{i=1}^n|\partial W_i|
\leq C_2 |Y|\,,
\tag 4.3
$$
and
$$
|\partial V(Y)|\leq C_3 |Y|,
\tag 4.4
$$
where $C_1=2d({2^{1/d}+1})/(
{2^{1/d}-1})$, $C_2=C_1+2d$ and $C_3=C_2+2d$.
\endproclaim

The proof of this lemma relies on a
lattice version of the isoperimetric
inequality and is given in Appendix B.
The proof of the required isoperimetric
inequality is given in Appendix A.

\bigskip
\subhead{4.2.  Decomposition of $Z(V,h)$ into pure phase
partition functions}
\endsubhead
\bigskip

The first step in the proof of Theorem
3.1 is the decomposition of
$Z(V,h)$ into  $N$ terms $Z_q(V,h)$,
$q=1,\ldots,N$,
which are obtained as perturbations
of the leading terms
$e^{-E_q(V)}$. We start with the
observation that all
external contours contributing to
(3.5) touch the set
$\Ext$ introduced in (4.1).
 Given that these contours are
matching, we conclude that all
 external contours of a
given configuration contributing to
(3.5) have the same
label. Therefore
$$
Z(V,h)=\sum_{q=1}^N Z_q(V,h),
\tag 4.5
$$
with
$$
Z_q(V,h)=\sum_{\{Y_1,\dots, Y_n\}}
\prod^n_{k=1}\rho(Y_k)
\prod_{m=1}^N e^{-E_m(V_m)},
\tag 4.6
$$
where the sum goes over
sets of matching contours in $V$
 for which all
external contours are $q$-contours.
As before, $V_m$ is the
union of all components of $V\setminus
(\operatorname{supp}Y_1\cup\dots
\cup\operatorname{supp}Y_n)$ that have
boundary condition $m$, and $E_m(V_m)$
is defined in (3.6).

More generally, let $W$ be a component
of the interior
$\Int Y_0$ of some contour $Y_0$
in $V$, a set of the
form (4.1), or a set obtained
from a component $W_0$
of an interior $\Int Y_0$ by
a similar construction,
$$
W=W_0\setminus\bigcup_{i=1}^n
(\operatorname{Int}Y_i\cup
\operatorname{supp}Y_i)
\tag 4.7a
$$
where $\{Y_1,\ldots,Y_n\}$ is a
set of non-overlapping contours
in $W_0$. We then define
$Z_q(W,h)$ as
$$
Z_q(W,h)=\sum_{\{Y_1,\dots, Y_n\}}
\prod^n_{k=1}\rho(Y_k)
\prod_{m=1}^N e^{-E_m(V_m)},
\tag 4.7b
$$
where the sum goes over
sets of matching contours in
 $V$ for which all
external contours are $q$-contours
 with $V(Y)=(\supp Y\cup\Int Y)\subset W$.
Here, $V_m$ is now defined as the
union of all components of $W\setminus
(\operatorname{supp}Y_1\cup\dots
\cup\operatorname{supp}Y_n)$ that have
boundary condition $m$.
Note that the sum in (4.7b) contains no
contours which surround the holes in $W$.
Finally, given a volume $W$
 which is a disjoint union of
volumes $W_1,\ldots,W_n$
of the form (4.7a), we
define $Z_q(W,h)$ as
 the product of the partition
functions $Z_q(W_i,h)$, $i=1,\ldots,n$.

Returning to (4.6), we derive
a second expression
for $Z_q(V,h)$, which eliminates
the matching condition
for the labels of $Y_1,\ldots,Y_n$.
To this end
we first sum  over all sets
$\{Y_1,\dots, Y_n\}$ with a fixed
collection of external contours.
For each external contour $Y$
this resummation
produces a factor \
$\prod_{m=1}^N Z_m(\Int_m Y,h)$. This
yields the expression
$$
Z_q(V,h)=\sum_{\{Y_1,\dots, Y_n\}_{\text
{ext}}}
e^{-E_q(\Ext)}
\prod^n_{k=1}
\left[\rho(Y_k)\prod_{m=1}^N
 Z_m(\Int_m Y_k,h)\right]
\tag 4.8
$$
where the sum runs over sets
${\{Y_1,\dots, Y_n\}_{\text{ext}}}$
of mutually external $q$-contours in
$V$ and $\Ext$ is the set
defined in (4.1).
Assuming that $Z_q(\Int_mY_k,h)\ne 0$,
 we divide
each \ $Z_m$ \ by the corresponding \
$Z_q$ \ and multiply it back
again in the form (4.7b). Iterating
the same procedure on the terms \
$Z_q(\Int_mY_k,h)$,
we eventually get
$$
Z_q(V,h)=e^{-E_q(V)}
\sum_{\{Y_1,\dots, Y_n\}}
\prod^n_{k=1}K_q(Y_k),
\tag 4.9
$$
where the sum goes over set of
non-overlapping contours
which are all $q$-contours,
while
$$
K_q(Y):=\rho(Y)e^{E_q(Y)}
\prod_{m=1}^N \frac{Z_m(\Int_m Y,h)}
{Z_q(\Int_m Y,h)}.
\tag 4.10
$$
The equality
(4.9) is the desired alternative
expression for $Z_q(V,h)$
which contains no matching condition
 on contours. Assuming
that the new contour activities
$K_q(Y)$ are sufficiently
small (for $h$ in the transition region,
this is actually the case, see Section
4.4), it also expresses the fact that
$e^{-E_q(V)}$ is the leading contribution
to $Z_q(V,h)$.

Obviously, (4.9) can be generalized to volumes $W$ of the form
considered in (4.7). One obtains
$$
Z_q(W,h)=e^{-E_q(W)}
\sum_{\{Y_1,\dots, Y_n\}}
\prod^n_{k=1}K_q(Y_k),
\tag 4.11
$$
where the sum goes over sets of
non-overlapping $q$-contours
$Y_1$, $\dots$, $Y_n$ with $V(Y_i)\subset W$.

\bigskip
\subhead{4.3. Truncated contour models}
\endsubhead
\bigskip

Given the decomposition (4.5) of
$Z(V,h)$
and the representation
(4.9) for
$Z_q(V,h)$,
one might try to to obtain the FSS of
$Z(V,h)$
by a cluster expansion analysis of the partition
functions $Z_q(V,h)$. For such an analysis, one would need
a bound of the form
$
|K_q(Y)|\leq \epsilon^{|Y|}\,\,
$
with a sufficiently small constant $\epsilon>0$. While it
turns out that such a bound can be proven for stable
phases $q$, it is false for unstable phases.

In order to overcome this problem, we will
construct
truncated contour activities
$K_q^\prime(Y)$
and the corresponding
partition functions
$$
Z_q^\prime(W,h)=e^{-E_q(W)}
\sum_{\{Y_1,\dots, Y_n\}}
\prod^n_{k=1}K_q^\prime(Y_k)
\tag 4.12
$$
in such a way that:

\item{i)} The truncated contour activities
$K_q^\prime(Y)$ obey a bound
$$
|K_q^\prime(Y)|\leq \epsilon^{|Y|}\,\,
\tag 4.13
$$
for some small $\epsilon>0$.

\item{ii)} \ $Z_q^\prime(W,h)=Z_q(W,h) $
 \ if \
the corresponding (infinite volume)
free energy, $f_q=f_q(h)$ is equal to
$f\equiv \displaystyle\min_{m\in Q} f_m$, \
so that the truncated model
is identical to the original model
if $f_q = f$ (following
\cite{Si82} and \cite{Z84},
 we call these $q$
 ``stable'').
\item{iii)} The truncated contour
 activities, and the corresponding
free energies, are smooth
functions of the external fields $h$.

\noindent
Heuristically, the truncated model will be
a model where contours corresponding to
supercritical droplets in the corresponding
droplet model are suppressed with the help
of a smoothed characteristic function. In
the case of a two phase model, this idea
could be implemented by defining
$$
\align
K_+^\prime(Y)
&=K_+(Y)\chi(\alpha|Y|-(f_+-f_-)|V(Y)|)
\\
K_-^\prime(Y)
&=K_-(Y)\chi(\alpha|Y|-(f_--f_+)|V(Y)|),
\endalign
$$
where
$\chi$ is a smoothed characteristic function and
$\alpha$ is a constant of the order of $\tau$,
for example $\alpha=\tau/2$.
While the presence of the characteristic
function would not affect the stable phases
since $f_+-f_-\geq 0$ if $+$ is stable
(and $f_--f_+\geq 0$ if $-$ is stable), it
would suppress contours immersed into an
unstable phase $+$ as soon as the volume
term $(f_+-f_-)|V(Y)|$ is bigger then the
decay term proportional to $|Y|$. As a consequence,
all contours contributing to the ``meta-stable''
partition function $Z_q^\prime$ obey a bound
of the form (4.13) as desired.

Unfortunately, the above definition
of
$K_q^\prime(Y)$ is
circular because it uses
free energies $f_q$
that are defined
as free energies of a model with activities
$K_q^\prime(Y)$. To overcome
this problem, we will use
the following
inductive procedure.

Assume
that $K_q^\prime(Y)$ has already been
defined for all $q$ and all
contours $Y$ with $|V(Y)| < n$,
$n\in\Bbb N$, and that
it obeys a bound
of the form (4.13). Introduce $f_q^{(n-1)}$
as the free energy of a contour model
with activities
$$
K^{(n-1)} (Y^q) = \cases K'(Y^q) \ \ & \text{\rm if} \ |V(Y^q)| \leq n-1 \\
0 & \text{\rm otherwise}.\endcases \tag"{(4.14)}"
$$
Consider then a
contour $Y$ with $|V(Y)|=n$.
Since $|V(\tilde Y)| < n$ for all
contours $\tilde Y$ in $\Int Y$,
the truncated partition functions
$Z_q^\prime(\Int_m Y,h)$
are well defined for all $q$ and $m$.
Their logarithm can be controlled by a convergent cluster
expansion, and
$Z_q^\prime(\Int_m Y,h)\neq 0$ for all $q$ and $m$.
We therefore may define
$K_q^\prime(Y)$ for a $q$-contours $Y$ with $|V(Y)|=n$ by
$$
K_q^\prime(Y)=\chi_q^\prime(Y)
\rho(Y)e^{E_q(Y)}
\prod_m{Z_m(\Int_m Y,h)\over
Z_q^\prime(\Int_m Y,h)}
\,\,,
\tag 4.15a
$$
with
$$
\chi_q^\prime(Y)=\prod_{m\neq q}\chi
\left(
\alpha |Y|
- (f^{(n-1)}_q-f^{(n-1)}_m) |V(Y)|
\right).
\tag 4.15b
$$
Here $\alpha$ is a constant that
 will be chosen later
and
$\chi$ is a smoothed characteristic
function.
We assume that $\chi$ has been
defined in such a way
that $\chi$ is a $C^6$ function
that obeys the conditions
$$
0\leq\chi (x)\leq 1\,\,,
\tag 4.16a
$$
$$
\chi(x)=0
\quad\text{\it if}\quad x
\leq -1
\quad\text{\it and}\quad\chi(x)=
1\quad\text{\it if}\quad x\geq 1\,\,,
\tag 4.16b
$$
$$
0\leq\frac{d}{dx}\chi(x)\leq
1\,\,,\quad\text{and}
\tag 4.16c
$$
$$
\left|\frac{d^k}{dx^k}\chi(x)\right|
\leq {\tilde C}_0\quad
\text { for all } k\leq 6,
\tag 4.16d
$$
for some constant $\tilde C_0$.

As the final element of
the construction of $K_q^\prime$,
we have to establish the bound
(4.13) for contours $Y$ with $|V(Y)|=n$.
We defer the proof,
together with the proof of the following
Lemma 4.6, to Appendix C.

 We use $f_q=f_q(h)$ to denote the free
energy corresponding to the partition
function $Z_q^\prime(V,h)$,
$$
f_q= - \lim_{V\to\Bbb R^d}\frac {1}{|V|}
 \text{log}\,
Z_q^\prime(V,h)\,\,,
\tag 4.17
$$
and
introduce
$f=f(h)$ and $a_q=a_q(h)$ as
$$
\align
f &= \min_m f_m\,\,,
\tag 4.18a\\
a_q&=f_q - f\,\,.
\tag 4.18b
\endalign
$$
Finally, we recall that for a volume $W$ of the form
(4.7a), $|W|$ denotes the euclidean volume of $W$, while
for a contour $Y$ or for the boundary $\partial W$ of a
volume $W$, $|Y|$ and $|\partial W|$ are used
to denote the number of elementary cells, i.e. the
number of elementary cubes, plaquettes, ..., and
bonds in $Y$ and $\partial W$, respectively.

\proclaim{Lemma 4.6} Assume that
$\rho(\cdot)$ and $e_q(\cdot)$
obey the conditions (3.7) and (3.11),
and let
$$
\align
\epsilon&=
e^{2+\alpha}
e^{-\tau(1-(1+2C_1))\gamma},
\tag 4.19a
\\
\overline{\alpha}&=
{2d\over C_3}(\alpha-2).
\tag 4.19b
\endalign
$$
Then there
exists a
constant $\epsilon_0>0$
(depending only on $d$ and $N$)
such that the following statements
hold provided $\epsilon<\epsilon_0$ and
$\overline\alpha\geq 1$.

\medskip\noindent
\item{i)} The contour activities
$K_q^\prime(Y)$ are well defined
for all  $Y$ and obey (4.13).
\item{ii)} {If}
$a_q\vert V(Y)\vert^{1/d}\leq\overline\alpha$,
then $\chi_q(Y)=1$ and $K_q(Y)=K_q^\prime(Y)$.
\item{iii)} If
$a_q\vert W\vert^{1/d}\leq \overline\alpha$,
then $Z_q(W,h)=Z_q^\prime(W,h)$.
\item{iv)} For all volumes $W$ of the form (4.7a)
one has
$$
|Z_q(W,h)|
\leq
e^{-f|W|)+O(\epsilon)|\partial W|
   +\gamma\tau N_{\partial V} (W)}
,
\tag 4.20
$$
where $N_{\partial V}(W)$ is the number of elementary cubes in
$\overline W$ which touch $\partial V$.
\item{v)} For $W=V$ the bound (4.20) can be sharpened to
$$
|Z_q(V,h)|
\leq
e^{-f|V|}
e^{(1+\gamma\tau)|\partial V|}
\max
\left\{
e^{-\frac{a_q|V|}{4}}
,
e^{-(4 C_3)^{-1}\tau|\partial V|}
\right\},
\tag 4.21
$$
where $C_3=C_3(d)$ is the constant defined in
Lemma 4.5.
\endproclaim

\remark{Remarks}

i) Due to the bound (4.13),
the partition function
$Z_q^\prime(V,h)$ can be analyzed
by a convergent
cluster expansion. As a consequence,
one can prove the usual volume,
surface, ..., corner asymptotics
for
its logarithm. Namely,
using $f_q^{(d)}$,
$f_q^{(d-1)}$, ..., $f_q^{(0)}$
 to denote the bulk,
surface, ..., corner free energies
corresponding to
$Z_q^\prime(V,h)$,
and introducing
 $F_q(W)$ as
$$
F_q(W)=\sum_{c\in\bar W} f_q(c),
\tag 4.22
$$
where $f_q(c)=f_q$ if $c$
does not touch the boundary $\partial V$
of $V$, and --- in analogy to (3.13) --- we have
$$
f_q(c) = \sum_{n=k}^{d}\binom{d-k}{n-k}  f_m^{(n)},
\quad k=d-1,\dots,0,
\tag 4.23
$$
whenever  $c$ is touching a
$k$-dimensional
face of $V$ and not touching its
$(k-1)$-dimensional boundary,
and as a result we get
$$
\left|\,\text{log}\,Z_q^\prime(V,h)
 + F_q(V)\,\right|
\leq
|V|
O((K\epsilon )^L)
\tag 4.24
$$
for some $K<\infty$ depending
only on $N$ and $d$.

ii) It is interesting to
present
a heuristic derivation of the
bound (4.21) in the approximation of the droplet model.
To this end, we recall that the diameter of a critical
droplet is proportional to $\tau/a_q$.
Assume now that $a_q L/\tau$ is small.
Then
the size of a critical droplet is larger
then the system
size, and $Z_q(V,h)$ is a partition function
describing small perturbations around a meta-stable
ground state, with
the weight
$$
Z_q(V,h)
\sim e^{-f_q|V|+O(|\partial V|)}
=e^{-f|V|+O(|\partial V|)}
e^{-a_q|V|}.
\tag 4.25a
$$
For large values of
$a_q L/\tau$, on the other hand, supercritical droplets
do fit into the volume $V$. As a consequence,
the leading configuration contributing to
$Z_q(V,h)$ contains a big contour
(with an interior that is
essential all of $V$) describing a transition
from the unstable boundary condition $q$ to a stable
phase
$\overline q$  with $f_{\overline q}=f$.
We conclude that
$$
Z_q(V,h)
\sim e^{-f|V|+O(|\partial V|)}
e^{-O(\tau)|\partial V|}
\tag 4.25b
$$
if $a_q L/\tau$ is large. Except for
the numerical value of the involved constants,
the bound (4.21) exactly describes this behavior.

\item{iii)} The fact that $\chi_q(Y)$ suppresses
supercritical droplets manifests itself in the fact that
$$
\chi_q(Y)=0
\quad\text{unless}\quad
a_q|V(Y)|\leq (\alpha+1+O(\epsilon))|Y|
\,,
\tag 4.26
$$
see Appendix C for the proof of (4.26).

\endremark

\bigskip

\bigskip
\subhead{4.4. Bounds on Derivatives}
\endsubhead
\bigskip

We finally turn to the continuity
 properties of $Z_q$ and $Z_q^\prime$.
As a finite sum of $C^6$ functions,
 $Z_q(V,h)$
is a $C^6$ function of $h$.
The following lemma
yields a
bound on the derivatives of
$Z_q(V,h)$.

\proclaim{Lemma 4.7} There is a constant $K$
(depending on $d$, $N$ and the constants introduced
in (3.8), (3.9) and (4.16)) such that the following
statements are true
provided $\epsilon<\epsilon_0$ and
$\overline\alpha\geq 1$.

\item{i)}  $Z_q(W,h)$
is a $C^6$ function of $h$ and
$$
\left\vert
{d^k \over dh^k}
Z_q(W,h)
\right\vert
\leq
|k|!
\left(
(C_0+O(\epsilon)) \vert W\vert
\right)^{|k|}
e^{-f|V|}
e^{O(\epsilon)\vert\partial W\vert}
e^{\gamma\tau N_{\partial V}(W)}
\tag 4.27
$$
for all multi-indices $k$ of order
$1\leq |k| \leq 6$.

\item{ii)} $K_q^\prime(Y)$
is a $C^6$ functions of $h$,
and
$$
\left|\frac{d^k}{dh^k}
K_q^\prime(Y)\right|
\leq (K\epsilon)^{|Y|}
\tag 4.28
$$
for all multi-indices $k$
of order $1\leq |k|\leq 6$.

\item{iii)} {\rm log}$\,Z_q^\prime(W,h)$
is a $C^6$ functions of $h$,
and
$$
\left|
\frac{d^k}{dh^k}
\log Z_q^\prime(W,h)
\right|
\leq (C_0^{|k|}+O(\epsilon))|W|
\tag 4.29
$$
for all multi-indices $k$
of order $1\leq |k|\leq 6$.

\item{iv)}
For $W=V$ (and $1\leq |k|\leq 6$),
the bound (4.27) can be sharpened
to
$$
\align
\left|
\frac{d^k}{dh^k}
Z_q(V,h)
\right|
\leq
&|k|!
\left(
(C_0+O(\epsilon)) |V|
\right)^{|k|}
e^{-f|V|}
e^{(1+\gamma\tau)|\partial V|}
\\
&\times\max
\left\{
e^{-\frac{a_q|V|}{4}}
,
e^{-(4 C_3)^{-1}\tau|\partial V|}
\right\}\,.
\tag 4.30
\endalign
$$

\endproclaim

\demo{Proof} The proof of
this lemma is given
in Appendix D.
\enddemo

\remark{Remarks}

i) For many models, including the perturbed Ising model
introduced in Section 2, it is possible to
prove a {\it degeneracy removing condition}. In the context
of a model with $N$ ground states and a driving
parameter $h\in\Bbb R^{N-1}$
($N=2$ for the perturbed Ising model), one considers
the matrix
$$
\Bbb E =\,\left({d\over d h_i}\;
(e_q-e_N)\right)_{q,i=1,
\ldots,N-1}
\tag 4.31
$$
and its inverse $\Bbb E^{-1}$.
One then proves that for some value $h_0$ of $h$,
all ground state energies are equal, and that
$\Bbb E^{-1}$ obeys a bound
$$
\|\Bbb E^{-1}\|_\infty
=\max_i\sum_q |(\Bbb E^{-1})_{iq}|
\leq \text{const}
\tag 4.32
$$
in a neighborhood $\Cal U$ of $h_0$, which does not
depend on $\tau$.

On the other hand,
$s_q=f_q-e_q$
is a $C^6$ function of $h$ with
$$
|f_q-e_q|\leq O(\epsilon)
\tag 4.33
$$
and
$$
\left|\frac{d}{dh_i}(f_q-e_q)
\right|\leq O(\epsilon)\,\,.
\tag 4.34
$$
by Lemmas 4.6 and 4.7.
As a consequence, the inverse of the matrix
$$
\Bbb F=\,\left({d\over d h_i}\;
(f_q-f_N)\right)_{q,i=1,
\ldots,N-1}
\tag 4.35
$$
obeys a bound of the same form as
$\Bbb E^{-1}$,
 with a slightly larger constant
on the right hand side; combined
with the inverse function theorem and the fact that
$f_q(h_0)-f_N(h_0)\leq O(\epsilon)$, one
immediately obtains the existence of
a point
$h_t\in \Cal U$,
$|h_t-h_0|\leq O(\epsilon)$,
for which all \ $a_q$ \ are zero,
 {\it i.e.}, all
phases are
stable. More generally,
one may construct
differentiable curves \ $h_q(t)$,
starting at $h_t$, on which
only
the phase
\ $q$ \ is unstable,
surfaces \ $h_{q\bar q}(t,s)$ \ on
 which
phases $q$ and $\bar q$  are unstable,
etc. A possible parametrization of
these curves, surfaces, etc., is given
by
\ $a_m(h_q(t))=\delta_{mq}\,t$, \
$a_m(h_{q\bar q}(t,s))=\delta_{mq}\,
t+\delta_{m\bar q}\,s,\,\cdots$.

ii) Due to Lemma 4.7 ii), the bound (4.24) can be
generalized to the first six derivatives of
$\log Z_q^\prime(V,h)$. Namely,
$$
\left|
\frac{d^k}{dh^k}
\left(
\log Z_q^\prime(V,h) + F_q(V)
\right)
\right|
\leq
|V|
O((K\epsilon )^L)
\tag 4.36
$$
for all multi-indices $k$ of order $1\leq|k|\leq 6$.
\endremark

\bigskip
\subhead{4.5. Proof of Theorem 3.1}
\endsubhead
\bigskip

\noindent
In order to prove Theorem 3.1, we introduce
the sets
$$
Q=\{1,\cdots,N\}\quad and \quad
S=\{q\in Q\mid a_q L < \overline\alpha \}.
\tag 4.37
$$
Using the decomposition (4.5) together with
Lemma 4.6 iii), we bound
$$
\multline
\Biggl\vert
{d^k\over dh^k}
\Bigl[
Z(V,h)-\sum_{q=1}^N e^{-F_q(V)}
\Bigr]
\Biggr\vert
\leq
\sum_{q\in S}
\left\vert
{d^k\over dh^k}
\left[
Z_q^\prime(V,h) - e^{-F_q(V)}
\right]
\right\vert+
\\
+
\sum_{q\notin S}
\left\vert
{d^k\over dh^k}
Z_q(V,h)
\right\vert
+
\sum_{q\notin S}
\left\vert
{d^k\over dh^k}
e^{-F_q(V)}
\right\vert,
\endmultline
\tag 4.38
$$
where $k$ is an arbitrary multi-index of order
$0\leq |k| \leq 6$.

Next, we observe that for
$1\leq |k| \leq 6$,
$$
\left\vert
{d^k\over dh^k}
{F_q(V)}
\right\vert
\leq
O(1) |V|
\tag 4.39
$$
by the assumption (3.9) and the fact that
$F_q(V)-E_q(V)$ can be analyzed by a convergent
expansion using Lemmas 4.6 and 4.7.

For $q\in S$, we then rewrite
$$
\left[
Z_q^\prime(V,h) - e^{-F_q(V)}
\right]
=
-e^{-F_q(V)}
\left[
1-e^{ F_q(V) + \log Z_q^\prime(V,h) }
\right].
$$
Using the bounds (4.24), (4.36) and (4.39),
we obtain the following bound
on the first sum on the r.h.s.
of
(4.38),
$$
\multline
\sum_{q\in S}
\left\vert
{d^k\over dh^k}
\left[
Z_q^\prime(V,h) - e^{-F_q(V)}
\right]
\right\vert
\leq
O((K\epsilon)^L)
|V|^{|k|+1}
\sum_{q\in S}
e^{-F_q(V)}\leq
\\
\leq
O((K\epsilon)^L)
|V|^{|k|+1}
\max_{q\in S}e^{-F_q(V)}
\leq
O((K\epsilon)^L)
|V|^{|k|+1}
\max_{q\in Q}e^{-F_q(V)}.
\endmultline
\tag 4.40
$$

In order to bound the last sum in (4.38),
we observe that for $q\notin S$
one has
$$
\align
\left\vert
{d^k\over dh^k}
e^{-F_q(V)}
\right\vert
&\leq
O(1) |V|^{|k|} e^{-F_q(V)}
\leq
O(1) |V|^{|k|}
e^{(\gamma\tau+O(\epsilon))|\partial V|}
e^{-f_q|V|}
\\
&=
O(1) |V|^{|k|}
e^{(\gamma\tau+O(\epsilon))|\partial V|}
e^{-a_q|V|} e^{-f|V|}
\\
&\leq
O(1) |V|^{|k|}
e^{(2\gamma\tau+O(\epsilon))|\partial V|}
e^{-a_q|V|}
\max_{q\in Q} e^{-F_q(V)}
\\
&\leq
O(1) |V|^{|k|}
e^{-(\overline\alpha/2d-2\gamma\tau-O(\epsilon))|\partial V|}
\max_{q\in Q} e^{-F_q(V)},
\tag 4.41
\endalign
$$
where we used the definition (4.37) of $S$
and, in the last step, the fact
that $L^{-1}|V|=(1/2d)|\partial V|$.

Finally,
again for $q\notin S$, we have
$$
\multline
\left\vert
{d^k\over dh^k}
Z_q(V,h)
\right\vert
\leq
O(1) |V|^{|k|}
e^{(1+\gamma\tau)|\partial V|}
\max
\{
e^{-{\overline\alpha/ 4}|V|}
,
e^{-{\tau/4 C_3}|\partial V|}
\}
e^{-f|V|}\leq
\\
\leq
O(1) |V|^{|k|}
e^{
(1+2\gamma\tau -
\min\{\overline\alpha/8d,\tau/4 C_3\}
)|\partial V|}
\max_{q\in Q} e^{-F_q(V)}
\endmultline
\tag 4.42
$$
by (4.21) and (4.30).

Inserting the bounds (4.40) through (4.42) into
(4.38),
and observing that
$|\partial V|\geq 2dL$ for all $d\geq 2$, we finally
obtain the bound
$$
\Biggl\vert
{d^k\over dh^k}
\Bigl[
Z(V,h)-\sum_{q=1}^N e^{-F_q(V)}
\Bigr]
\Biggr\vert
\leq
O(e^{-L/L_0}) |V|^{|k|+1}
\max_{q\in Q} e^{-F_q(V)},
\tag 4.43
$$
where
$$
\align
{1\over L_0}
&:=
\min
\bigl\{
-\log (K\epsilon),
\overline\alpha -4d\gamma\tau-O(\epsilon),
\overline\alpha/4-2d -4d\gamma\tau,
{2d\over 4C_3}\tau-2d -4d\gamma\tau
\bigr\}
\\
&=
\min
\bigl\{
-\log (K\epsilon),
\overline\alpha/4-2d -4d\gamma\tau,
{2d\over 4C_3}\tau-2d -4d\gamma\tau
\bigr\}.
\tag 4.44
\endalign
$$
Recalling the definitions (4.19) of $\overline\alpha$
and $\epsilon$, together with the fact
that $C_3=C_1+4d$, we now rewrite
$$
-\log (K\epsilon)
=\tau-\alpha-(1+2C_1)\gamma\tau -O(1),
\tag 4.45a
$$
$$
\overline\alpha/4-2d -4d\gamma\tau
={2d\over 4C_3}(\alpha-(32d+8C_1)\gamma\tau)-O(1),
\tag 4.45b
$$
and
$$
\align
{2d\over 4C_3}
\tau-2d -4d\gamma\tau
&=
{2d\over 4C_3}
\left(\tau-(32d+8C_1)\gamma\tau\right)
-O(1)\\
&=
2{2d\over 4C_3}
\left({\tau\over 2}-(16d+4C_1)\gamma\tau\right)
-O(1).
\tag 4.45c
\endalign
$$
Choosing $\alpha={\tau\over 2}+(16d+3C_1-1/2)\gamma\tau$,
we obtain
$$
-\log (K\epsilon)
={\tau\over 2}-(1/2+5C_1+16d)\gamma\tau -O(1),
\tag 4.46a
$$
$$
\overline\alpha/4-2d -4d\gamma\tau
={2d\over 4C_3}
\left({\tau\over 2}-(1/2+16d+5C_1)\gamma\tau\right)
-O(1),
\tag 4.46b
$$
and
$$
{1\over L_0}
={2d\over 4C_3}
\left({\tau\over 2}-(1/2+16d+5C_1)\gamma\tau\right)
-O(1))
={d\over 4C_3}
\left(\tau-(1+32d+10C_1)\gamma\tau\right)-\tau_0),
\tag 4.46c
$$
where $\tau_0$ is a constant that depends
on $N$, $d$, and the constants
introduced in (3.8), (3.9) and (4.16).

Defining
$$
b=b(d)={d\over 4C_3},
\tag 4.47a
$$
$$
\gamma_0=\gamma_0(d)={1\over 1+32d+ 10 C_1},
\tag 4.47b
$$
and
$$
\tilde\tau=\tau(1-\gamma/\gamma_0)-\tau_0,
\tag 4.47c
$$
we obtain
${1/ L_0}=b\tilde\tau$
and hence the bound (3.19) of Theorem 3.1.

Observing that
$$
\epsilon=e^{2-{\tau\over 2}(1-(1+16d+10C_1)\gamma)}
=e^2 e^{-{\tau\over 2}(1-\gamma/\gamma_0)},
\tag 4.48
$$
we note that the condition $\tilde\tau>0$
implies
the inequality $\epsilon<\epsilon_0$
provided $\tau_0$ is chosen large enough.
The condition $\bar\alpha\geq 1$, on the
other hand, is trivial, since
$
\bar\alpha={2d\over C_3}(\alpha -2)
\geq
{d\over C_3}
\tau -O(1).
$

It remains to prove statements iii) and v).
While v) is a direct consequence of iv),
iii) follows from the fact that
$(f_m-e_m)$ and $(f_m^{(l)}-e_m^{(l)})$
can be analyzed by a convergent cluster
expansion involving the decay
constant $\epsilon$.
Observing that
$O(\epsilon)\leq O(e^{-\tilde\tau})$
can be bounded by $e^{-b\tilde\tau}$,
this proves iii).
\qed

\vfill\eject
\head{5. Proof of Theorem 3.2}
\endhead

\bigskip
\subhead{5.1. Decomposition of $Z(A|V,h)$ into pure phase
partition functions}
\endsubhead
\bigskip

The first step in the proof of Theorem 3.2
is the same as
the first step in the proof of Theorem 3.1.
Namely, we decompose $Z(A|V,h)$ as
$$
Z(A|V,h)
=\sum_{q=1}^N Z_q(A|V,h),
\tag 5.1
$$
with
$$
Z_q(A|V,h)=\sum_{\{Y_1,\dots, Y_n\}}
A(Y_1,\cdots,Y_n)
\prod^n_{k=1}\rho(Y_k)
\prod_{m=1}^N e^{-E_m(V_m)}
.
\tag 5.2
$$
Here the sum goes over
sets of matching contours in $V$
 for which all
external contours are $q$-contours.

Next, we group all contours $Y_i$ for which
$V(Y_i)\cap\supp A\neq\emptyset$
into a new contour $Y_A$, and introduce
the sets
$$
\align
&\supp Y_A=\bigcup_{Y\in Y_A}\supp Y,
\quad
   V(Y_A)=\bigcup_{Y\in Y_A} V(Y)\,,\\
&\Int Y_A=V(Y_A)\setminus \supp Y_A\,,
\quad
\text{and}\quad
\Ext Y_A=V\setminus V(Y_A)\,,
\endalign
$$
as well as
$$
\align
&\overline{\supp Y_A}=\supp Y_A\cup\supp A,
\quad
\overline{V(Y_A)}=V(Y_A)\cup\supp A\,,\\
&\Int^{(0)} Y_A=\Int Y_A\setminus\supp A\,,
\quad
\text{and}\quad
\Ext^{(0)} Y_A=\Ext Y_A\setminus\supp A\,.
\endalign
$$
As usual,
$\Int_mY_A$
is the union of all components
of $\Int Y_A$ which have boundary condition $m$,
$\Int_mY_A=\Int Y_A\cap V_m$,
while
$\Int_m^{(0)} Y_A=\Int^{(0)} Y_A\cap V_m$.

Recalling that $A$ does only depend on
those contours for which $V(Y)\cap\supp A\neq\emptyset$,
we then define
$$
\rho(Y_A)=
A(Y_1^\prime,\cdots,Y_{n'}^\prime)
\prod^{n'}_{k=1}\rho(Y_k^\prime)
\prod_{m=1}^N e^{-E_m(V_m\cap (\supp A\setminus\supp Y_A))},
\tag 5.3
$$
where $Y_1^\prime,\cdots,Y_{n'}^\prime$ are the
contours in $Y_A$.
Fixing now, for a moment, all contours $Y_i$
in (5.2) for which $V(Y_i)\cap\supp A\neq\emptyset$,
and resumming the rest, we obtain
$$
Z_q(A|V,h)=\sum_{Y_A}\rho(Y_A)
Z_q(\Ext^{(0)} Y_A,h)
\prod_{m=1}^N
Z_m(\Int_m^{(0)} Y_A,h)
\,.
\tag 5.4
$$
Introducing
$$
K_q(Y_A)
=
\rho(Y_A)
e^{E_q(\overline{\supp Y_A})}
\prod_{m=1}^N
{Z_m(\Int_m^{(0)} Y_A,h)
\over
Z_q(\Int_m^{(0)} Y_A,h)},
\tag 5.5
$$
we further rewrite (5.4) as
$$
Z_q(A|V,h)
=\sum_{Y_A}
K(Y_A) e^{-E_q(\overline{\supp Y_A)}}
Z_q(\Ext^{(0)} Y_A,h) Z_q(\Int^{(0)} Y_A,h)\,.
\tag 5.6
$$
Using finally the representation (4.11)
for
$Z_q(\Ext^{(0)} Y_A,h)$ and
$Z_q(\Int^{(0)} Y_A,h)$,
we get
$$
Z_q(A|V,h)
=
e^{-E_q(V)}
\sum_{Y_A}
K_q(Y_A)
\sum_{\{Y_1,\dots, Y_n\}}
\prod^n_{k=1}K_q(Y_k).
\tag 5.7
$$
Here the second sum goes over set of
non-overlapping $q$ contours
$Y_1,\dots, Y_n$, such that for all contours $Y_i$,
the set $V(Y_i)$ does not intersect the set
$\overline{\supp Y_A}$.

In order to make the connection to the standard
Mayer expansion for polymer systems, we then
introduce
$G(Y_A,Y_1,\cdots,Y_n)$
as the
graph
on the vertex set $\{0,1,\cdots,n\}$
which has
an edge
between two vertices $i\geq 1$
and $j\geq 1$, $i\neq j$, whenever
$\supp Y_i\cap\supp Y_j\neq\emptyset$,
and
an edge
 between the vertex $0$ and a
vertex $i\neq 0$ whenever
$V(Y_i)\cap\overline{\supp Y_A}\neq \emptyset$.
Implementing the non-overlap
constraint in (5.7)
by a characteristic function
$\phi(Y_A,Y_1,\cdots,Y_n)$
which is zero whenever the
graph $G$ has less then $n+1$
components, the standard
Mayer expansion for polymer
systems (see,
for example \cite{Sei82})
then
yields
$$
{Z_q(A\mid V,h)\over Z_q(V,h)}
=
\sum_{Y_A}
K_q(Y_A)
\sum_{n=0}^\infty
{1\over n!}
\sum_{\{Y_1,\dots, Y_n\}}
\left[
\prod^n_{k=1}K_q(Y_k)
\right]
\phi_c(Y_A,Y_1,\cdots,Y_n)
\,.
\tag 5.8
$$
Here
$\phi_c(Y_A,Y_1,\cdots,Y_n)$
is a combinatoric factor
defined in term of the
connectivity properties of the
graph $G(Y_A,Y_1,\cdots,Y_n)$,
see
\cite{Sei82}. It
vanishes if
$G(Y_A,Y_1,\cdots,Y_n)$
has more then one component.

\bigskip
\subhead{5.2. Truncated Expectation Values}
\endsubhead
\bigskip

In the context of Section 5.1, the expansion
(5.8) is a formal power series in the
activities $K(Y_i)$. In order to use this
expansion, one has to prove its convergence.
As in Section 4, it is useful to introduce
truncated models.

For a contour $Y$ with
$V(Y)\cap\supp A=\emptyset$,
we define $K_q^\prime(Y)$ as before, see (4.15a),
while for $Y_A=\{Y_1,\cdots,Y_n\}$,
where
$\{Y_1,\cdots,Y_n\}$ is a set of contours
with
$V(Y)\cap\supp A\neq\emptyset$ for all $Y\in Y_A$,
we define
$$
K_q^\prime(Y_A)
=
\rho(Y_A)
e^{E_q(\overline{\supp Y_A})}
\prod_{m=1}^N
{Z_m(\Int_m^{(0)} Y_A,h)
\over
Z_q^\prime(\Int_m^{(0)} Y_A,h)}
\prod_{Y\in Y_A}\chi_q(Y)\,,
\tag 5.9
$$
with $\chi_q(Y)$ as in (4.15b).
Given this definition,
we introduce
$$
Z_q^\prime(A|V,h)
=
e^{-E_q(V)}
\sum_{Y_A}
K_q^\prime(Y_A)
\sum_{\{Y_1,\dots, Y_n\}}
\prod^n_{k=1}K_q^\prime(Y_k)
\,
\tag 5.10
$$
and
$$
\langle A \rangle_{V,q}^h=
{Z_q^\prime(A\mid V,h)\over Z_q^\prime(V,h)}
\,,
\tag 5.11
$$
which can again be expanded as
$$
\langle A \rangle_{V,q}^h
=
\sum_{Y_A}
K_q^\prime(Y_A)
\sum_{n=0}^\infty
{1\over n!}
\sum_{\{Y_1,\dots, Y_n\}}
\left[
\prod^n_{k=1}K_q^\prime(Y_k)
\right]
\phi_c(Y_A,Y_1,\cdots,Y_n)
\,.
\tag 5.12
$$
The following Lemma will allow us
to prove absolute convergence
of the expansion (5.12),
which immediately
yields the
statements ii) through iv)
of Theorem 3.2.

\medskip
\proclaim{Lemma 5.1} Let $\epsilon$,
$\epsilon_0$ and $\overline\alpha$
be as defined in Lemma 4.6, and assume
that $\epsilon<\epsilon_0$ and
$\overline\alpha\geq 1$.
Then the following statements are true:

\medskip\noindent
\item{i)} {If}
$$
a_q \max_{Y\in Y_A}
\vert V(Y)\vert^{1/d}
\leq\overline\alpha\,,
\tag 5.13
$$
then $K_q(Y_A)=K_q^\prime(Y_A)$.

\item{ii)}  Let
$$
|Y_A|=\sum_{Y\in Y_A} |\supp Y|\,.
\tag 5.14
$$
Then
$$
\left|
K_q^\prime(Y_A)
\right|
\leq
C_A e^{O(\epsilon)|\supp A|}
\epsilon^{|Y_A|}
\,.
\tag 5.15
$$

\item{iii)} Let $k$ be a multi-index
of order $1\leq |k| \leq 6$. Then
$$
\left|
{d^k\over dh^k}
K_q^\prime(Y_A)
\right|
\leq
|\supp A|^{|k|} C_A
e^{O(\epsilon)|\supp A|}
(K\epsilon)^{|Y_A|}
\,,
\tag 5.16
$$
where $K$ is a constant that depends
only on $d$, $N$, and the constants
introduced in (3.8), (3.9) and (4.16).

\endproclaim

\demo{Proof} The proof of
Lemma 5.1 is given in Appendix E.
\enddemo

Using standard estimates for polymer
expansions, see
for example \cite{Sei82},
the bounds of Lemma 4.6 and Lemma 5.1
immediately imply the
absolute convergence of the
expansion (5.12),
$$
\align
\left|
\langle A \rangle_{V,q}^h
\right|
&\leq
\sum_{Y_A}
\left|
K_q^\prime(Y_A)
\right|
\sum_{n=0}^\infty
{1\over n!}
\sum_{\{Y_1,\dots, Y_n\}}
\prod^n_{k=1}
\left|
K_q^\prime(Y_k)
\right|
\left|
\phi_c(Y_A,Y_1,\cdots,Y_n)
\right|
\\
&\leq O(1)
C_A e^{O(\epsilon)|\supp A|}
\,,
\tag 5.16a
\endalign
$$
and similar bounds for the derivatives,
in particular,
$$
\left|
{d^k\over dh^k}
\langle A \rangle_{V,q}^h
\right|
\leq O(1) |\supp A|^{|k|}
C_A e^{O(\epsilon)|\supp A|}
\,.
\tag 5.16b
$$
Theorem 3.2 ii) through iv) then
follows using standard arguments.

\bigskip
\subhead{5.3. Bounds on $Z_q(A|V,h)$}
\endsubhead
\bigskip

In conjunction with Lemma 4.6, Lemma 5.1 allows to
analyze $Z_q(A|V,h)/Z_q(V,h)$
provided $a_q L\leq \overline\alpha$.
In order to prove Theorem 3.2 in the case
where $a_q L>\overline\alpha$ for some
of the phases $q$, we need an analogue of the
bounds (3.21) and (4.30) for
$Z_q(A|V,h)$.

\medskip
\proclaim{Lemma 5.2} Let $\epsilon$,
$\epsilon_0$ and $\overline\alpha$
be as defined in Lemma 4.6, let
$\tilde\epsilon=\max\{\epsilon,e^{-3\tau/4}\}$,
and assume
that $\tilde\epsilon<\epsilon_0$ and
$\overline\alpha\geq 1$.
Then the following statements are true:

\medskip\noindent
\item{i)}
$\displaystyle
\left|
Z_q(A|V,h)
\right|
\leq
C_A e^{O(\tilde\epsilon)|\operatorname{supp} A|}
e^{(\gamma\tau+O(\tilde\epsilon))|\partial V|}
e^{-f|V|}
\max
\left\{
e^{-\frac{a_q}{4}|V|}\,,\,
e^{-(\tau/4C_3)|\partial V|}
\right\}
$.

\item{ii)} Let $k$ be a multi-index
of order $1\leq |k| \leq 6$. Then
$$
\multline
\left|
\frac{d^k}{dh^k}
Z_q(A|V,h)
\right|
\leq
|k|!
\left(
(C_0+O(\epsilon)) |V|
\right)^{|k|}
C_A e^{O(\tilde\epsilon)|\operatorname{supp} A|}\, \times
\\
\,\,\times
e^{(\gamma\tau+O(\tilde\epsilon))|\partial V|}
e^{-f|V|}
\max
\left\{
e^{-\frac{a_q}{4}|V|}\,,\,
e^{-(\tau/4C_3)|\partial V|}
\right\}.
\endmultline
$$
\endproclaim

\demo{Proof} The proof of
Lemma 5.2 is given in Appendix E.
\enddemo

\bigskip
\subhead{5.4. Proof of Theorem 3.2}
\endsubhead
\bigskip

As pointed out before, the absolute convergence
of the cluster expansion (5.12) immediately
implies the statements ii) through iv).
In order to prove Theorem 3.2 i), we
proceed as in the proof of Theorem 3.1,
using the decomposition (5.1),
Lemma 5.1 and Lemma 5.2 instead of
the decomposition
(4.5), Theorem 4.6 and
Theorem 4.7.
Defining $S$
as in Section 4.5, and observing that
$Z_q(A|V,h)=Z_q^\prime(A|V,h)$ if $q\in S$,
we bound
$$
\multline
\left\vert
{d^k\over dh^k}
\Bigl[
Z(A|V,h)-\sum_{q=1}^N e^{-F_q(V)}\langle A\rangle_{V,q}^h
\Bigr]
\right\vert
\leq
\sum_{q\in S}
\left\vert
{d^k\over dh^k}
\left[
\langle A\rangle_{V,q}^h
\left(
Z_q^\prime(V,h) - e^{-F_q(V)}
\right)
\right]
\right\vert
\\
+
\sum_{q\notin S}
\left\vert
{d^k\over dh^k}
Z_q(A|V,h)
\right\vert
+
\sum_{q\notin S}
\left\vert
{d^k\over dh^k}
\left[
\langle A\rangle_{V,q}^h
e^{-F_q(V)}
\right]
\right\vert\,,
\endmultline
\tag 5.17
$$
where $k$ is an arbitrary multi-index of order
$0\leq |k| \leq 6$.

Combining the bounds
(4.40) and (5.16), and bounding
terms of the form $|\supp A|^{|k|}$ and
$|V|^{|k|+1}$ by $e^{O(1)L}$,
we get an estimate for the first sum on the right
hand side
of (5.17)
by
$$
\sum_{q\in S}
\left\vert
{d^k\over dh^k}
\left[
Z_q^\prime(A|V,h)
-
\langle A\rangle_{V,q}^h
e^{-F_q(V)}
\right]
\right\vert
\leq
C_A
e^{O(\epsilon)|\supp A|}
(K\epsilon)^L
\max_{q\in Q} e^{-F_q(V)}
\,.
\tag 5.18
$$
Here
$K$ is a constant that depends only on $N$,
$d$ and the constants introduced in (3.8),
(3.9) and (4.16).
The terms for $q\notin S$ are bound in a similar way,
leading to
$$
\left\vert
{d^k\over dh^k}
\left[
\langle A\rangle_{V,q}^h e^{-F_q(V)}
\right]
\right\vert
\leq
C_A
e^{O(\epsilon)|\supp A|}
e^{-(\overline\alpha/2d-2\gamma\tau-O(1))|\partial V|}
\max_{q\in Q} e^{-F_q(V)}
\tag 5.19
$$
and
$$
\left\vert
{d^k\over dh^k}
Z_q(A|V,h)
\right\vert
\leq
C_A
e^{O(\tilde\epsilon)|\supp A|}
e^{(2\gamma\tau +O(1)-
   \min\{\overline\alpha/8d,\tau/4 C_3\})|\partial V|}
\max_{q\in Q} e^{-F_q(V)}
\,.
\tag 5.20
$$
Inserting the bounds (5.18) through (5.20)
into (5.17),
and choosing $\alpha$ as in Section 4.5,
we get
$\tilde\epsilon=\epsilon$ and
$$
\left\vert
{d^k\over dh^k}
\Bigl[
Z(A|V,h)-\sum_{q=1}^N e^{-F_q(V)}\langle A\rangle_{V,q}^h
\Bigr]
\right\vert
\leq
C_A
e^{O(\epsilon)|\supp A|}
O(e^{-b\tilde\tau L})
\max_{q\in Q} e^{-F_q(V)}
\,,
\tag 5.21
$$
where $b$ and $\tilde\tau$ are the constants
introduced in (4.47).
Together with Theorem 3.1 and the observation
that a prefactor $|V|^{|k|+1}$ can be absorbed into
the exponential decay term
$e^{-b\tilde\tau L}$, the bound (5.21) implies
Theorem 3.2 i).
\hfill\qed

%
%

\vfill\eject
\head{6. Proof of Theorem A}
\endhead

Even though the statements of this
Section have a generalization (sometimes
a very straightforward one) to the case
of several phases, we will restrict
ourselves to the situation
where only 2 phases, plus and minus, come
to play and the driving parameter is an
external field $h$. However, we do not restrict
ourselves
to the model (2.1) (for which Theorem A
is stated), but consider the two phase case in the general
setting of Section 3. In particular, we have two ground
state energies $e_\pm$ satisfying, for $h$ in an interval
$\Cal U$ (containing the point $h_0$ for which $e_+(h)=
e_-(h)$), the (nondegeneracy) bounds
$$
0<\bar a \le \frac{d}{dh}(e_-(h)-e_+(h)) \le \bar A
\tag 6.1
$$
that imply the bounds
$$
0<  a \le \frac{d}{dh}(f_-(h)-f_+(h)) \le   A
\tag 6.2
$$
on the free energies $f_\pm(h)$ from Theorem 3.1 (cf\.
also (4.17)).
Actually, $\bar A=2C_0$ according to the assumption (3.9).
In the situation of Theorem A  we  have
$\frac{d}{dh}(e_-(h)-e_+(h))=2$.
Considering now the free
energies\footnote{We take here
$\partial_d V\equiv V$.}
$$
F_\pm (L,h)=\sum_{k=0}^d f_\pm^{(k)}(h)|\partial_k V|,
\tag 6.3
$$
cf\. (3.18), and their derivative
$$
M_\pm (L,h)=\sum_{k=0}^d m_\pm^{(k)}(h)|\partial_k V|,
\tag 6.4
$$
where $m_\pm^{(k)}(h)=-df_\pm^{(k)}(h)/dh$,
and introducing
$$
\Delta F(L,h)=F_+(L,h)-F_-(L,h)
\,,
\tag 6.5a
$$
$$
\Delta M(L,h)=M_+(L,h)-M_-(L,h)
\tag 6.5b
$$
$$
F_0(L,h)={F_+(L,h)+F_-(L,h)\over 2}
\,,
\tag 6.5c
$$
and
$$
M_0(L,h)={M_+(L,h)+M_-(L,h)\over 2}
\,,
\tag 6.5d
$$
the bounds (3.21) of Theorem 3.1 can
be, for the two phase case,
reformulated as
$$
\Biggl|\frac{d^k}{dh^k}
  \Bigl[
    m(L,h) -
    \Bigl(
    \frac 1{L^d}  M_0(L,h)
     +
    \frac 1{L^d}  \frac {\Delta M(L,h)}{2}
    \tanh(-\frac{\Delta F(L,h)}2)
    \Bigr)
  \Bigr]
\Biggr|
\leq
e^{-b\tilde\tau L}
\tag 6.6
$$
with $0\leq k\leq 5$.
For the magnetization $m(h,L)$
and its derivative, the susceptibility  $\chi(h,L)=dm(h,L)/dh$,
these bounds yield
$$
m(L,h) =
\frac 1{L^d}  M_0(L,h)
 +
\frac 1{L^d}  \frac {\Delta M(L,h)}{2}
\tanh(-\frac{\Delta F(L,h)}2)
+    O(e^{-b\tilde\tau L})
\tag 6.7
$$
and
$$
\align
\chi(L,h) &=
\frac 1{L^d}  \chi_0(L,h)
 +
\frac 1{L^d}  \frac {\Delta \chi(L,h)}{2}
\tanh(-\frac{\Delta F(L,h)}2)
\\
&+
\frac 1{L^d}
\Bigl(\frac {\Delta M(L,h)}{2}\Bigr)^2
\cosh^{-2}(-\frac{\Delta F(L,h)}2)
+    O(e^{-b\tilde\tau L}).
\tag 6.8
\endalign
$$
Here
$\chi_0(L,h)=dM_0(L,h)/dh$
and
$\Delta\chi(L,h)=d\Delta M(L,h)/dh$ .

In order to obtain Theorem A,
and more generally the corrections
to it in terms of an asymptotic power
series in $1/L$, we proceed in several steps:

\item{i)} We expand the functions $\Delta F(L,h)$, $M_0(L,h)$,
$\Delta M(L,h)$, $\chi_0(L,h)$ and $\Delta \chi(L,h)$
around the point $h_t(L)$ where $\Delta F(L,h)=0$,
obtaining a power series in
$(h-h_t(L))$ with coefficients that are
derivatives of $\Delta F(L,h)$ and $F_0(L,h)$
at the point $h_t(L)$.

\item{ii)} We Taylor expand the coefficients
in i) into a power series in $(h_t(L)-h_t)$.
Combined with the volume, surface, ..., corner
expansion for the derivatives of
$F_\pm(L,h)$ and the fact that
$h_t(L)-h_t$ can be represented as an
asymptotic expansion in powers of $1/L$, we obtain
the coefficients of i) as power series in $1/L$, with
coefficients that are
derivatives of the infinite volume
free energies $f_\pm(h)$,
surface free energies $f_\pm^{(d-1)}(h)$,
...,
and corner free energies $f_\pm^{(0)}(h)$
at the infinite volume transition
point $h_t$.

\item{iii)} At $h_t$, the derivatives of $f_{\pm}^{(k)}(h)$
are identified with the one-sided derivatives of the free energies
$f^{(k)}(h)$ defined by (2.15).

\item{iv)} We use Lemma 6.1 below to
replace  the argument of
the hyperbolic functions in (6.7)
and (6.8) by few expansion terms with an additive error.

\item{v)} In a final step, we use Lemma 6.3
 to replace $h-h_t(L)$ by $h-h_\chi(L)$,
where $h_\chi(L)$ is the position of
the susceptibility maximum.

\proclaim{Lemma 6.1}
Let $x$ and $y$ two nonzero real numbers which
have the same sign.
Then
$$
\left|
{\tanh x - \tanh y}
\right|
\leq
\min(\frac{\tanh x}{x},\frac{\tanh y}{y}) |x-y|
\,
\tag 6.9a
$$
and
$$
\left|
{\cosh^{-2} x} - {\cosh^{-2} y}
\right|
\leq
2\min(\frac{\tanh x}{x},\frac{\tanh y}{y}) |x-y|
\,.
\tag 6.9b
$$
\endproclaim

\proclaim{Lemma 6.2}
For large $L$ there exists a unique point $h_t(L)\in\Cal U$
for which \hfill\newline
$F_+(L,h)=F_-(L,h)$. This point
satisfies the bound
$$
h_t(L)=h_t+\frac{\Delta F(L,h_t)}{m_+-m_-}\frac1{L^d}
\bigl(1+O(\frac1L)\bigr)\,,
\tag6.10
$$

\endproclaim

\proclaim{Lemma 6.3}
For large $L$ there exists
a unique point $h_{\chi}(L)\in\Cal U$
as well as a unique point $h_U(L)\in\Cal U$
for which the susceptibility
$\chi(L,h)$ and the Binder cummulant
$U(L,h)$, respectively, attain its maximum.
To the leading order in $1/L$, their shift with
respect of the point $h_t(L)$ is given by
$$
h_{\chi}(L)=h_t(L) + 6\frac{\chi_+-\chi_-}{(m_+-m_-)^3}
\frac1{L^{2d}}
+O(\frac1{L^{3d}})
\tag6.11
$$
 and
$$
h_U(L)=h_t(L) + 4\frac{\chi_+-\chi_-}{(m_+-m_-)^3}
\frac1{L^{2d}}
+O(\frac1{L^{3d}}).
\tag6.12
$$
\endproclaim

\demo{Proof of Theorem A}
Let us begin with the identification iii).
Introducing
$$
m_\pm^{(k)}(h)=-\frac{df_\pm^{(k)}(h)}{dh},\,
\chi_\pm^{(k)}(h)=-\frac
{d^2f_\pm^{(k)}(h)}{dh^2},\,\, k=d,\dots, 0,
\tag 6.13
$$
we get
$$
f^{(k)}_{\pm}(h_t)= \lim_{h\to h_t\pm 0} f^{(k)}(h)
\tag6.14
$$
and
$$
m_\pm^{(k)}(h_t)=-\frac{df^{(k)}(h)}{dh}\Big|_{h_t\pm0},\,
\chi_\pm^{(k)}(h_t)=
-\frac{d^2f^{(k)}(h)}{dh^2}\Big|_{h_t\pm0}
\tag 6.15
$$
for  $k=d,\dots,0$.
In particular,  the one-sided derivatives (2.16)
as well as the limits (2.17) and (2.18) are expressed
in terms of derivatives and limits
of the corresponding differentiable functions $f^{(k)}_\pm$:
$$
m_\pm=-\frac{d}{dh}f_\pm(h)\Bigr|_{h=h_t}
\tag 6.16
$$
and
$$
\tau_\pm=f^{(d-1)}_\pm(h_t).
\tag 6.17
$$
Also
$$
\Delta F(L)=\Delta F(L,h_t)
\tag 6.18
$$
with $\Delta F(L,h) =(F_+(L,h)
- F_+(L,h))$.
To show all this, we first notice that,
 for the two phase case, the bound (3.19) from
Theorem 3.1 reads
$$
\Biggl|\frac{d^k}{dh^k}\bigl[Z(L,h)- e^{-F_+(L,h)}
-e^{-F_-(L,h)}\bigr]
\Biggr|\leq |V|^{k+1}\max
(e^{-F_+(L,h)},e^{-F_-(L,h)})
O(e^{-b\tilde\tau L}).
\tag 6.19
$$
Taking into account that $F_{\pm}(L,h)$ are
asymptotically dominated by $f_\pm (h)  L^d$, the bound
(6.19) implies that for
$h>h_t$ the free energies $f^{(k)}(h)$, $ k=d,\dots,0$,
 defined by
 (2.15)
actually equal the corresponding free energies
$f_+^{(k)}(h)$, $k=d, \dots, 0$,
from Theorem 3.1 (we have chosen the notation for which
$\text{min}
(f_+(h), f_-(h))=f_+(h)$ for $h>h_t$). Similarly,
$f^{(k)}(h)=f_-^{(k)}(h)$,
$k=d, \dots, 0$, for every $h<h_t$.%
\footnote{
For $h=h_t$, the asymptotic behavior will
be determined by the first $k=d-1, \dots, 0$,
for which $f_+^{(k)}(h_t)\neq
f_-^{(k)}(h_t)$. For example, if $f_+^{(d-1)}(h_t)>
f_-^{(d-1)}(h_t)$,
then $f^{(k)}(h_t)=f^{(k)}_-(h_t)$ {\it for all}
 $ k=d, \dots, 0$
(of course, $f^{(d)}_+(h_t)=f_-^{(d)}(h_t)$).}
This identification immediately implies the equalities
(6.14)--(6.18).

Notice also that by (3.9) and Theorem 3.1 iii) one has
$$
\Biggl| \frac{d^kf_{\pm}^{(\ell)}}{dh^k}\Biggr|
\le C_0^k +e^{-b\tilde \tau}
\tag6.20
$$
and thus also
$$
m_{\pm}(h_t(L))=m_{\pm}+ O(|h_t(L)-h_t|)=m_{\pm}+
O(\frac{1+\Vert\kappa\Vert}L)
\tag6.21
$$
according to Lemma 6.2, where we evaluate
$\Delta F(L)$ with the help of (2.8) and Theorem 3.1 iii).

Expanding now $M_{\pm}(L,h)$ and $F_{\pm}(L,h)$,
 we have
$$
\align
M_\pm(L,h)
&=
M_\pm(L,h_t(L)) + O((h-h_t(L))L^d)
\\
&=m_\pm(h_t(L))L^d+ O((h-h_t(L))L^d)+ O(L^{d-1})
\\
&=m_\pm L^d +O((h-h_t(L))L^d)
+O((1+\Vert\kappa\Vert) L^{d-1})
\,,
\tag 6.22
\endalign
$$
and
$$
\align
-\Delta &F(L,h)=
\biggl(M_+(L,h_t(L))-M_-(L,h_t(L)\biggr)
(h-h_t(L))
+O((h-h_t(L))^2L^d)
\\
&=
(m_+-m_-)(h-h_t(L))L^d
+O((h-h_t(L))^2L^d)
+ O((1+\Vert\kappa\Vert) (h-h_t(L))L^{d-1})
\\
&=2x(
     1+O((h-h_t(L)))
     + O((1+\Vert\kappa\Vert) L^{-1}
    )
\,.
\tag 6.23
\endalign
$$
Here
$$
x=\frac{m_+-m_-}2
(h-h_t(L))L^d
\,.
\tag 6.24
$$
Using Lemma 6.1
to replace  the  argument of the hyperbolic functions in
(6.7) and (6.8)
by $x$, we  get
$$
m(L,h)=\frac{m_++m_-}2
+ \frac{m_+-m_-}2
\tanh (x)
+O((1+\Vert\kappa\Vert) L^{-1})
+ O((h-h_t(L)))
\tag 6.25
$$
and
$$
\chi(L,h)=
\left(\frac{m_+-m_-}2\right)^2
\cosh^{-2}(x) L^d
+ O((1+\Vert\kappa\Vert) L^{d-1})
+ O((h-h_t(L))L^d)
\,.
\tag 6.26
$$
In order to replace further the argument $x$ of the
hyperbolic functions by the argument
$$
\tilde x=\frac{m_+-m_-}2 (h-h_\chi(L))L^d
\tag 6.27
$$
used in (2.19) and (2.20),
we finally use Lemma 6.3 to bound
$$
|\tanh x- \tanh \tilde x|
\leq
|x-\tilde x|\leq O(L^{-d})
\tag 6.28a
$$
and
$$
|\cosh^{-2} x - \cosh^{-2}\tilde x|
\leq
|x-\tilde x|\leq O(L^{-d})
\,.
\tag 6.28b
$$
Combining the bounds (6.24) and (6.28) with
the assumption
    $|h-h_\chi(L)|\leq O((1+||\kappa||)L^{-1})$,
we get the bounds (2.19) and (2.20).

The shifts (2.21a) and (2.21b) as well as the bound
(2.22) on the mutual shift $h_U(L) - h_\chi(L)$ follow
from Lemma 6.2 and 6.3.
\qed
\enddemo

\demo{Proof of Lemma 6.1}
Without loss of generality, we may assume that
$x>y>0$.  Since
$\frac {\tanh t}t$ is a decreasing function of $t$,
we have
$\frac {\tanh y}y > \frac {\tanh x}x$ and thus
$$
\left|
\frac{\tanh x - \tanh y}{\tanh x}
\right|
=
1-\frac{\tanh y}{\tanh x}
\leq
1-\frac yx  = \frac{|x-y|}{|x|}
\,.
\tag6.29
$$
This concludes the proof of (6.9a).
In order to prove (6.9b), we bound
$$
\left|
{\cosh^{-2} x} - {\cosh^{-2} y}
\right|
=
2\left|
\int_x^y \frac{\sinh t}{\cosh^{3}t} \,dt
\right|
\leq
2\left|
\int_x^y \cosh^{-2}t \,dt
\right|
=2
\left|
{\tanh x - \tanh y}
\right|
\tag6.30
$$
and use (6.9a).
\qed
\enddemo

\demo{Proof of Lemma 6.2}
Using the bounds (6.2) we get, for sufficiently large $L$,
the bound
$$
\frac{a}2 L^d
\le
-\frac{d\Delta F(L,h)}{dh}
\le
2A L^d.
\tag6.31
$$
Since $\Delta F(L)=\Delta F(L,h_t)=O(L^{d-1})$,
we get the existence of a unique $h_t(L)$ for which
$\Delta F(L,h)=0$.
Moreover, $h_t(L)\in (h_t-\frac{\bar B}{L},h_t+\frac{\bar
B}{L})$ for some $\bar B$.
For $h$ in this interval, the Taylor expansion of $\Delta
F(L,h)$ around $h_t$  yields
$$
\Delta F(L,h)=\Delta F(L) - (m_+-m_-)L^d (h-h_t) +
(h-h_t) O(L^{d-1})
\tag6.32
$$
whenever $L\ge \bar B$.
This implies (6.7) (valid also for $\Delta F(L)=0$
when $ h_t(L)=h_t$).
\qed
\enddemo

\demo{Proof of Lemma 6.3}
To get (6.11), we  may actually follow the proof of Theorem
(3.3) in \cite{BK90}, replacing only $h_t$ by $h_t(L)$.
Thus we use first (6.8) combined with the bound
$$
\Biggl|\frac{d^kF_\pm(L,h)}{dh^k}\Biggr|\leq \bar C_k L^d
\tag6.33
$$
(here $\bar C_k$ may be chosen
$\bar C_k=(C_0)^{k}+O(e^{-b\tilde\tau})$
by (3.9) and Theorem 3.1 iii))
to get
$$
\bigl|
\chi(L,h)-\frac{1}{4L^d}\bigl(\Delta
M(L,h)\bigr)^2 \cosh^{-2} (\frac{\Delta
F(L,h)}{2})\bigr|\le 1
+2\bar C_2.
\tag6.34
$$
Hence
$$
\multline
\chi(L,h_t(L))-\chi(L,h)\ge \frac{1}{4L^d}
\bigl(\Delta
M(L,h_t(L))\bigr)^2-\\-\frac{1}{4L^d}
\bigl(\Delta
M(L,h)\bigr)^2\cosh^{-2}(\frac{\Delta
F(L,h)}{2})-2
-4\bar C_2.
\endmultline
\tag6.35
$$
Next, we use (6.31) and (6.33)
to bound
$
\left|
\frac{d}{dh}
(\Delta M(L,h))^2
\right|
$
by
$
8A \bar C_2L^{2d}
$.
As a consequence,
\smallskip
$$
\left|
(\Delta M(L,h))^2-
(\Delta M(L,h_t(L)))^2
\right|
\leq
8A \bar C_2L^{2d}
|h-h_t(L)|
\,
\tag6.36
$$
\noindent and
$$
\multline
\chi(L,h_t(L))-\chi(L,h)\ge \frac{1}{4L^d}
\bigl(\Delta
M(L,h_t(L))\bigr)^2\bigl(1-\cosh^{-2}
\bigl(\frac{\Delta
F(L,h)}{2}\bigr)\bigr)
-\\
-2A\bar C_2
|h-h_t(L)|L^d
\cosh^{-2}\bigl(\frac{\Delta
F(L,h)}{2}\bigr)-2-4\bar C_2.
\endmultline
\tag6.37
$$
On the other hand,
$$
\frac a4 |h-h_t(L)| L^d
\leq
\left|\frac{\Delta F(L,h)}2\right|
\leq
A|h-h_t(L)|L^d
\tag6.38
$$
by (6.31) and the fact that
$\Delta F(L,h_t(L))=0$.
Using the lower bound
$$
\cosh^2\alpha\geq\bigl(1+\frac{\alpha^2}{2}\bigr)^2\geq
1+\alpha^2
\geq 2|\alpha|
\tag 6.39
$$
valid for any $\alpha$,
we imply that
$$
\cosh^{-2}\bigl(\frac{\Delta
F(L,h)}{2}\bigr)\le \frac{2}{a|h-h_t(L)|L^d}.
\tag6.40
$$
Thus,
using once more (6.31) and (6.38), we get
$$
\chi(L,h_t(L))-\chi(L,h)
\ge \frac{1}{4L^d}
\bigl(\frac{a}{2}L^d\bigr)^2
(1-\cosh^{-2}(\frac a4 B))
-\frac{4A\bar C_2}{a}-2-4\bar C_2
\tag6.41
$$
whenever we suppose that $|h-h_t(L)|\ge B L^{-d}$,
where $B>0$ will be chosen later.
Observing that
$$
\cosh^{-2}(\frac a4 B)
<1,
\tag6.42
$$
it is clear that
the right hand side of (6.41) is positive, once $L$ is
sufficiently large
(how large depends on the choice of $B$).

Thus, it remains to consider the case $|h-h_t(L)|< B
L^{-d}$.
Taking into account that $\Delta F(L,h_t(L))=0$ we get
$$
\multline
\frac{d\chi (L,h)}{dh}\biggr|_{h_t(L)}=
-\frac1{L^d}\frac{d^2M_0(L,h)}{dh^2}\biggr|_{h_t(L)}+\\
+\frac3{4L^d}\frac{d\Delta M(L,h)}{dh}\biggr|_{h_t(L)}
\Delta M(L,h_t(L)) +O(e^{-b\tilde
\tau L}).
\endmultline
\tag 6.43
$$
Using the bound (6.33) we get
$$
\frac{d\chi (L,h)}{dh}\bigr|_{h_t(L)}=
\frac3{4}(\chi_+-\chi_-)
(m_+-m_-)L^d +O(L^{d-1}).
\tag 6.44
$$
Applying once more the bound (6.6), this time for $k=3$,
and using (6.31) we get, for
$|h-h_t(L)|< B L^{-d}$,
the bound
$$
\frac{d^2\chi (L,h)}{dh^2}=
-2\bigl(\frac12 \Delta M(L,h)\bigr)^4 L^{-d}
\frac{1-3\tanh^2(\frac{\Delta
F(L,h)}{2})}{\cosh^2(\frac{\Delta F(L,h)}{2})}+O(L^{2d}).
\tag 6.45
$$
Taking into account that according to
(6.38) one has
$|\Delta F(L,h)|\leq 2A|h-h_t(L)|L^d$
and  choosing
$B>0$ so that
$$
\varepsilon
:=
\frac{1-3\tanh^2(AB)}{\cosh^2({AB})}
>0
\,,
\tag 6.46
$$
we get
$$
\frac{d^2\chi (L,h)}{dh^2}\le
-\frac1{16}(m_+-m_-)^4 L^{3d}\varepsilon
\tag 6.47
$$
for $|h-h_t(L)|< B L^{-d}$ and
$L$ large enough.
The bound
(6.47)
together with the bound (6.44) implies
that there exists a unique zero $h_\chi(L)$ of
$\frac{d\chi(L,h)}{dh}$ in the interval
$(h_t(L)-\frac{B}{L^d},h_t(L)+\frac{B}{L^d})$ and that
$h_\chi(L)-h_t(L)=O(L^{-2d})$.

For $h-h_t(L)=O(\frac1{L^{2d}})$ we have $\Delta F(L,h)=
O(\frac1{L^{d}})$ by (6.38) and thus, using (6.45) and
(6.36) we get
$$
\frac{d^2\chi (L,h)}{dh^2}=
-2(\frac12(m_+-m_-))^4 L^{3d}+
O({L^{3d-1}}).
\tag 6.48
$$
Taking into account (6.44) we conclude the bound (6.11).

To prove the bound (6.12), we first notice,
by a straightforward computation, that
$$
U(h_t(L)) =\frac23  \bigl(
1-4\frac{\chi_++\chi_-}{(m_+-m_-)^2}
\frac1{L^{d}}
+O(\frac1{L^{d+1}})\bigr).
\tag6.49
$$
Using the fact that, for $L$ large,
$$
\multline
\frac{d^2\chi}{dh^2}=-\frac1{L^d}\frac{d^3M_0(L,h)}{dh^3}+
\frac1{2L^d}\frac{d^3\Delta M(L,h)}{dh^3}
\tanh(\frac{\Delta F(L,h)}{2})+\\+
\frac1{L^d}\frac{d^2\Delta M(L,h)}{dh^2}
\frac{\Delta
M(L,h)}{\cosh^2(\frac{\Delta F(L,h)}{2})}
+\frac3{4L^d}\bigl(\frac{d\Delta M(L,h)}{dh}\bigr)^2
\frac1{\cosh^2(\frac{\Delta F(L,h)}{2})}-\\-
\frac3{2L^d}\frac{d\Delta M(L,h)}{dh}
(\Delta M(L,h))^2
\frac{\tanh(\frac{\Delta F(L,h)}{2})}{\cosh^2(\frac{\Delta
F(L,h)}{2})}-
\frac1{8L^d}
(\Delta M(L,h))^4
\frac{1-3\tanh^2(\frac{\Delta
F(L,h)}{2})}{\cosh^2(\frac{\Delta F(L,h)}{2})}
\le\\  \le O(L^{2d}) - L^{3d}\frac18\bigl(\frac{a}2\bigr)^4
\frac{1-3\tanh^2(\frac{\Delta
F(L,h)}{2})}{\cosh^2(\frac{\Delta F(L,h)}{2})},
\endmultline
\tag6.50
$$
we find that $U(L,h)$ is negative (and thus smaller than
$U(h_t(L))$ whenever
$$
\frac{1-3\tanh^2(\frac{\Delta
F(L,h)}{2})}{\cosh^2(\frac{\Delta F(L,h)}{2})}\le
-\varepsilon
\tag6.51
$$
for some positive
$\varepsilon$.
To meet this condition, it suffices to take
$|h-h_t(L)|>\frac{B}{L^d}$ with $B$ such that
$$
\cosh^2(\frac{aB}4)
\geq \frac32 (1+\tilde\epsilon)
\tag6.52
$$
for some $\tilde\epsilon>0$.
Indeed, using (6.38) we get $|\Delta F(L,h)|\ge \frac{a}2  L^d
|h-h_t(L)|$
and thus
$\cosh^2(\frac{\Delta F(L,h)}{2})>\frac32(1+\tilde\epsilon)$,
which
implies
(6.51) with $\epsilon=2\tilde\epsilon/(1+\epsilon)$.

In the interval $|h-h_t(L)|\le\frac{B}{L^d}$ we consider
the leading terms to $\frac{dU(L,h)}{dh}$ and
$\frac{d^2U(L,h)}{dh^2}$.
Namely,
$$
\multline
\frac{dU(L,h)}{dh}\sim -\frac83 L^d
\cosh(\frac{\Delta
F(L,h)}{2})
\bigl[
-\frac{\Delta M(L,h)}{2}\sinh(\frac{\Delta
F(L,h)}{2})-\\-
\frac{\frac{d\Delta M(L,h)}{dh}}{\Delta
M(L,h)}\cosh^3(\frac{\Delta
F(L,h)}{2})-3\frac{\frac{d\Delta M_0(L,h)}{dh}}{\Delta
M(L,h)} \sinh(\frac{\Delta
F(L,h)}{2}))
\bigr]
\endmultline
\tag6.53
$$
yielding
$$
\frac{dU(L,h)}{dh}\biggr|_{h_t(L)}\sim \frac83L^d
\frac{\chi_+-\chi_-}{m_+-m_-}
\tag6.54
$$
and
$$
\multline
\frac{d^2U(L,h)}{dh^2}\sim -\frac23 L^{3d}
(m_+-m_-)^2(1+ 2\sinh^2(\frac{\Delta
F(L,h)}{2}) )+O(L^{3d-1})\le \\
\le-\frac13 L^{3d}(m_+-m_-)^2.
\endmultline
\tag6.55
$$
Thus, there exists a unique root $h_U(L)$ of the equation
$dU(L,h)/{dh}=0$ in the interval
$|h-h_t(L)|\le\frac{B}{L^d}$
and $h_U(L)-h_t(L)=O(L^{-2d})$.
Moreover, for $h-h_t(L)=O(L^{-2d})$
 we have
$\Delta F(L,h)=O(L^{-d})$
and thus
${d^2U(L,h)}/{dh^2}=-\frac23 L^{3d}
(m_+-m_-)^2+O(L^{3d-1})$.
Hence,
$$
\frac{dU(L,h)}{dh}= \frac{dU(L,h)}{dh}\biggr|_{h_t(L)}
+\frac{d^2U(L,h)}{dh^2}\biggr|_{h_t(L)+\xi (h- h_t(L) )}
|h-h_t(L)|=0
\tag6.56
$$
yields
the shift (6.12) claimed in the lemma.
\qed
\enddemo

We are left with the proof of (2.23) and (2.24)
for
$$
|h-h_\chi(L)|>\frac{4d(1+||\kappa||)}{(m_+-m_-)L}
\,.
\tag 6.57
$$
Using (6.2), (6.11) and the bound (6.38), we first
note that the condition (6.57) implies
$$
\frac{\Delta F(L,h)}2>\frac {a4d}{4AL} L^{d}+O(L^{-d})
\geq \frac {aL}A +O(L^{-d})
\,.
\tag 6.58
$$
Combined with (6.7) and (6.8) we obtain
$$
\align
m(L,h)&=L^{-d}M_+(L,h)+O(e^{-aL/A})\\
\chi(L,h)&=L^{-d}\frac{dM_+(L,h)}{dh}+O(e^{-aL/A})
\endalign
$$
if $h> h_\chi(L)+\frac{4d(1+||\kappa||)}{(m_+-m_-)L}$
and
$$
\align
m(L,h)&=L^{-d}M_-(L,h)+O(e^{-aL/A})\\
\chi(L,h)&=L^{-d}\frac{dM_-(L,h)}{dh}+O(e^{-aL/A})
\endalign
$$
if $h<h_\chi(L)-\frac{4d(1+||\kappa||)}{(m_+-m_-)L}$.
Expanding $M_\pm(L,h)$ and its derivative into volume,
surface, ..., corner terms,
this leads to
$$
\align
m(L,h)&=m_+(h)+O(1/L)\\
\chi(L,h)&=\chi_+(h)+O(1/L)
\endalign
$$
if $h> h_\chi(L)+\frac{4d(1+||\kappa||)}{(m_+-m_-)L}$
and
$$
\align
m(L,h)&=m_-(h)+O(1/L)\\
\chi(L,h)&=\chi_-(h)+O(1/L)
\endalign
$$
if $h< h_\chi(L)+\frac{4d(1+||\kappa||)}{(m_+-m_-)L}$

Next, we recall
$|e_\pm(c)-e_\pm|\leq ||\kappa||$
for the asymmetric Ising
model (2.1).
As a consequence,
$|\Delta F(L)|
\leq
2dL^{d-1}(2||\kappa||+O(e^{-b\tilde\tau}))$.
Combined with Lemmas 6.2 and 6.3 we conclude
that
$$
|h_\chi(L)-h_t|\leq
\frac{4d(||\kappa||+1)}{(|m_+-m_-|)L}
\,.
\tag 6.58
$$
As a consequence,
$h< h_\chi(L)+\frac{4d(1+||\kappa||)}{(m_+-m_-)L}$
implies $h<h_t$ and hence $m(h)=m_-(h)$ and
$\chi(h)=\chi_-(h)$,
while
$h> h_\chi(L)+\frac{4d(1+||\kappa||)}{(m_+-m_-)L}$
implies $h>h_t$ and hence $m(h)=m_+(h)$ and
$\chi(h)=\chi_+(h)$. The condition (6.57)
therefore implies the bounds (2.23) and (2.24).

%
%

\vfill\eject
\head{Appendix A: Strong isoperimetric inequality}
\endhead

Using the standard isoperimetric inequality,
$$
|\partial W|\ge \frac{\sqrt \pi}
{\Gamma(\frac{d}2 +1)^{\frac1d}} d |W|^{\frac{d-1}{d}},
\tag A.1
$$
in the proof of Lemma B.3 below, we would get, for $d\ge 4$,
a negative factor on the right hand side of the bound (B.2).
We strengthen (A.1) with the help of an additional
information --- the fact that considered sets $W$
 are finite unions of closed elementary cubes.
\proclaim{Lemma A.1} Let $W$ be a union of closed
elementary cubes. Then
$$
|\partial W|\ge 2 d |W|^{\frac{d-1}{d}}.
\tag A.2
$$
\endproclaim
\demo{Proof}
The proof is just a particularly simple case of the proof
of optimality of the Wulff shape \cite{T87}. Namely,
$$
|\partial W|=\lim_{\varepsilon\to 0}
\frac{|W+\varepsilon C|-|W|}{\varepsilon},
\tag A.3
$$
where $\varepsilon C$ is the rescaling, by the factor
$\varepsilon$, of the (hyper)cube $C$ of side 2 with
the center at the origin, and
$$
W+\varepsilon C= \{ x+y : x\in W, y\in \varepsilon C\}
\tag A.4
$$
is the $\varepsilon$-neighborhood of $W$ in the maximum
metric. The Brunn-Minkowski inequality (valid also for
nonconvex $W$, see for example \cite{Fe69}) yields
$$
|W+\varepsilon C|^{\frac1d}\ge |W|^{\frac1d}
+| \varepsilon C |^{\frac1d}.
\tag A.5
$$
Thus
$$
\lim_{\varepsilon\to 0}
\frac{|W+\varepsilon C|-|W|}{\varepsilon}\ge
\lim_{\varepsilon\to 0}
\frac{(|W|^{\frac1d}+|\varepsilon
C|^{\frac1d})^d-|W|}{\varepsilon}=
d|C|^{\frac1d}|W|^{\frac{d-1}{d}}=2d|W|^{\frac{d-1}{d}}.
\tag A.6
$$
\hfil\rightline{\raise 6pt
\hbox{\qed}}
\enddemo

%
%

\vfill\eject
\head{Appendix B: Proof of Lemmas 4.1 -- 4.5}
\endhead

We start with two Lemmas, B.1 and B.2,
that
 are an important technical
ingredient to prove
Lemmas 4.1, 4.2 and 4.4.

\proclaim{Lemma B.1}
 Let $Y$ be a short contour with
$\operatorname{supp}
 Y \cap\partial V\subset
\partial K(k).$ Then
\item{i)} $\operatorname{supp} Y \cap
\partial V =\operatorname{supp} Y\cap
 \partial K(k)=
\operatorname{supp} Y\cap(\partial
V\cap\partial K(k))$
\item{ii)}$\partial \operatorname{Int} Y
\subset\partial K(k)\cup\partial
\operatorname{supp} Y$
\item{iii)}$\operatorname{Int}
 Y\cap \partial V=\operatorname{Int}
 Y\cap\partial
 K(k)=\operatorname{Int} Y
\cap(
\partial V
\cap
\partial K(k)
)$
\endproclaim

\demo{Proof}
{i)} $\operatorname{supp} Y
\cap\partial K(k)=\operatorname{supp} Y
\cap\partial K(k)\cap V=
\operatorname{supp} Y\cap\partial
K(k)\cap\partial V$
since $V\cap\partial K(k)=
\partial V\cap\partial K(k)$ and
$\operatorname{supp} Y\subset V$.
 On the other hand
$\operatorname{supp} Y\cap\partial
V\subset\partial K(k)$
 implies
$\operatorname{supp} Y\cap\partial
V\subset
\partial K(k)
\cap\partial V$ and hence
$\operatorname{supp} Y
\cap\partial V \subset
\operatorname{supp} Y\cap\partial K(k)\cap
\partial V$. Combining this with
$\operatorname{supp} Y\cap\partial
K(k)\cap\partial V\subset
\operatorname{supp} Y\cap\partial V$
we obtain i).

{ii)} Follows from the fact that all
components of $\operatorname{Int} Y$ are
components of $K(k)\setminus
 \operatorname{supp} Y$.

{iii)} In order to prove iii), we first prove
$\operatorname{Int}
Y\cap \partial V
 \subset\partial \operatorname{Int}
 Y\cap\partial V$. This can be proven as
follows:
the inclusion $\operatorname{Int} Y\subset
V$ implies $ V^c\subset(\operatorname{Int}
Y)^c$ implies $ \operatorname{dist} (x,
V^c)\geq  \operatorname{dist}
 (x, (\operatorname{Int} Y)^c)$. Therefore
$\operatorname{dist} (x,
(\operatorname{Int} Y)^c)=0$
for all $x\in \partial V$
and hence for all
$x\in \operatorname{Int} Y\cap
 \partial V$.
Since $x\in \operatorname{Int} Y$ and
$\operatorname{dist} (x,
(\operatorname{Int} Y)^c)=0$ implies
 $x\in\partial\operatorname{Int} Y$,
this
proves $\operatorname{Int}
Y\cap \partial V
 \subset\partial \operatorname{Int}
 Y\cap\partial V$.

Using ii), the (just proven) fact
that $\operatorname{Int} Y\cap\partial
V\subset \partial \operatorname{Int} Y
\cap\partial V$ and the fact that
$\partial \operatorname{supp} Y
\cap\partial V \subset
\partial K(k)$, one proves that
$
\operatorname{Int} Y\cap\partial V
\subset
\partial K(k)
$.
Intersecting both sides with
$\operatorname{Int} Y$ and
observing that $\operatorname{Int}
Y\subset
V$ while $V\cap\partial K(k)=
\partial K(k)\cap\partial V$,
one concludes that
$$
\operatorname{Int} Y\cap\partial V\subset
 \operatorname{Int} Y \cap
\partial K(k)
= \operatorname{Int}
Y\cap \partial K(k)\cap\partial V.
$$
This combined with the fact that
$$
\operatorname{Int} Y\cap\partial K(k)\cap
\partial V\subset
\operatorname{Int} Y\cap\partial V,
$$
proves iii).
\qed\enddemo

\proclaim{Lemma B.2}
Let $Y_1$ and $Y_2$ be two
 non-overlapping
 contours with $\operatorname{supp}
 Y_1\subset \operatorname{Int} Y_2$.
Assume that $Y_2$ is a short contour
 with
$\partial V\cap \operatorname{supp}
Y_2\subset \partial K(k)$ some corner $k$.
Then $Y_1$ is a short contour with
$\partial V\cap \operatorname{supp}
Y_1\subset\partial K(k)$ as well.
 In addition $\operatorname{supp} Y_2\cap
 \operatorname{Int} Y_1
=\emptyset$.
\endproclaim

\demo{Proof}
By Lemma B.1 iii) and the fact that
$\operatorname{supp} Y_1\subset
 \operatorname{Int} Y_2$,
 $\operatorname{supp} Y_1
\cap\partial V\subset \operatorname{Int}
Y_2\cap \partial K(k)
\subset\partial K(k)$.
Let now $x_0\in \operatorname{supp}
Y_1\subset \operatorname{Int} Y_2$.
By the definition of
$\operatorname{Int} Y_2$
and $\operatorname{Ext}Y_2$, and
by the fact that
$\operatorname{supp} Y_2$
is connected, we may construct
a path $\omega$ in $K(k)$ which connects
$x_0=\omega(0)$ to infinity such that
$$
\align
\omega(t)\in \operatorname{Int} Y_2
\qquad&\text{for}\ t\in [0, 1)\cr
\omega(t)\in \operatorname{supp} Y_2
\qquad&\text{for}\ t\in [1, 2)\cr
\omega(t)\in \operatorname{Ext} Y_2
\qquad&\text{for}\ t\in [2, 3)\cr
\omega(t)\in K(k)\setminus V
\qquad&\text{for}\ t\in [3, \infty)\cr
\endalign
$$
Assume now that
$\operatorname{supp} Y_2\cap
 \operatorname{Int} Y_1\not=
 \emptyset$. Since
$\operatorname{supp} Y_2$
is a connected set
 which does not intersect
 $\operatorname{supp} Y_1$, this
implies that $\operatorname{supp}
Y_2\subset \operatorname{Int} Y_1$.
 As  a consequence $x_2=
\omega(2)\in \operatorname{Int} Y_1$
and $\omega|_{[2, \infty)}$
is a path in $K(k)$ which
connects $x_2\in
\operatorname{Int} Y_1$ to infinity.
But this implies that
$\omega|_{[2, \infty)}$
must intersect the set
$\operatorname{supp} Y_1$, and hence,
 by the assumption that
$\operatorname{supp}
Y_1\subset \operatorname{Int} Y_2$,
the set $\operatorname{Int} Y_2$.
This is a contradiction, because $\omega$
 was constructed
in such a way that $\omega(t)\not\in
 \operatorname{Int} Y_2$ for $t\geq 1$.
\qed
\enddemo

\subhead{Proof of Lemma 4.1}
\endsubhead

{i)} Since
 $\operatorname{supp} Y_2\subset
\operatorname{Ext}Y_1$,
$\operatorname{supp} Y_2\cap
\operatorname{Int} Y_1=\emptyset$. It
follows that each point in
$\operatorname{Int} Y_1$ can be connected
to $\partial\operatorname{Int} Y_1$ (and
therefore to $\operatorname{supp} Y_1$) by
a path $\omega$ which does not intersect
$\operatorname{supp} Y_2$. As a
consequence, all points in
$\operatorname{Int} Y_1$ lay in the same
connectivity component of $V\setminus
\operatorname{supp}Y_2$ as
$\operatorname{supp} Y_1$. Since
$\operatorname{supp} Y_1\subset
\operatorname{Ext}Y_2$, we concluded that
$\operatorname{Int} Y_1\subset
\operatorname{Ext}Y_2$, and hence
$\operatorname{Int} Y_1\cup
\operatorname{supp} Y_1 \subset
\operatorname{Ext}Y_2$. In a similar way,
$\operatorname{Int} Y_2\cup
\operatorname{supp} Y_2\subset
\operatorname{Ext}Y_1$.

{ii)} Let us first assume that $Y_2$
is a short contour. Then $Y_1$
is a small contour as well and
$\operatorname{supp} Y_2 \cap
\operatorname{Int} Y_1=\emptyset$
 by Lemma B.2.
Continuing as in the proof of i)
we obtain that $\operatorname{Int}
Y_1\cup \operatorname{supp} Y_1
\subset C_2$.

Next, consider the case where $Y_1$
is short while $Y_2$ is long,
and
assume
that $\operatorname{supp} Y_2\cap
\operatorname{Int} Y_1 \not=\emptyset$.
 Since $\operatorname{supp} Y_2$
is connected, this would
imply that  $\operatorname{supp}
Y_2 \subset \operatorname{Int} Y_1$.
 By Lemma B.2, this would
imply that $Y_2$
is short as well. Therefore
$\operatorname{supp} Y_2 \cap
\operatorname{Int} Y_1$ must be empty.
Again, this implies $\operatorname{Int}
 Y_1\cup
 \operatorname{supp} Y_1\subset C_2$.

As the last case, assume now that
 both $Y_1$ and $Y_2$ are long.
Since $C_2\subset
 \operatorname{Int} Y_2$,
$ |C_2|\leq L^d/ 2$
by the definition
of the exterior for long contours.
 Since $\operatorname{supp} Y_2$
is a connected set, while $C_2$
is a connected
 component of $V\setminus
\operatorname{supp} Y_2$,
both $C_2$ and $V\setminus C_2$
 are connected sets.
Observing that
$\operatorname{supp} Y_1\subset C_2$
 implies $V\setminus
 \operatorname{supp} Y_1\supset
V\setminus C_2$, we then
introduce the component
$C_1$ of $V\setminus
\operatorname{supp} Y_1$ which contains
$V\setminus C_2$. A minute
 of reflection shows that
$|C_1|>|V\setminus C_2|$, which,
 by the fact that $|V\setminus C_2|\geq
L^d/2$ implies that $C_1=
\operatorname{Ext}Y_1$.
 As a consequence
$$
\operatorname{Int} Y_1\subset V
\setminus \operatorname{Ext}Y_1
\subset C_2,
$$
which concludes the proof of ii).

iii) follows from ii).
\qed

\subhead{Proof of Lemma 4.2}
\endsubhead

We only have to show that
$\operatorname{supp} Y_1\subset
\operatorname{Int} Y_2$
and $\operatorname{supp} Y_2\subset
 \operatorname{Int} Y_1$
leads to a contradiction. In fact,
$\operatorname{supp} Y_1\subset
\operatorname{Int} Y_2$ implies
that $\operatorname{supp} Y_1\cup
\operatorname{Int} Y_1\subset
\operatorname{Int} Y_2$
by Lemma 4.1. As a consequence
$\operatorname{Ext} Y_1 \supset
 \operatorname{supp} Y_2 \cup
\operatorname{Int} Y_2$
which implies that $\operatorname{supp}
 Y_2\subset \operatorname{Ext}Y_1$.
But this is incompatible
with $\operatorname{supp} Y_2\subset
\operatorname{Int} Y_1$.
\qed

\subhead{Proof of Lemma 4.4}
\endsubhead

Let  $Y_k$ be an internal contour.
Due to Lemma 4.2, this implies that
$(\operatorname{Int}Y_k\cup
\operatorname{supp}Y_k)
\subset\operatorname{Int}Y_j$
for some $j\neq k$ and hence
$\operatorname{Ext}
=V\setminus\bigcup_{i\neq k}
(\operatorname{Int}Y_i\cup
\operatorname{supp}Y_i)$.
Iterating this argument, we get that
the set $\operatorname{Ext}$
is given by
$$
\operatorname{Ext}
=V\setminus\bigcup_{i=1}^{\tilde n}
(\operatorname{Int}Y_i^e\cup
\operatorname{supp}Y_i^e),
\tag B.1
$$
where
$\{Y_1^e,\dots,Y_{\tilde n}^e\}$
are the external contours
in $\{Y_1,\dots,Y_n\}$. Obviously,
 $\operatorname{Ext}$ is
separated from the rest of $V$
by the support of the contours
$Y_1^e,\dots,Y_{\tilde n}^e$.
We therefore only have to show
that $\operatorname{Ext}$ is connected.

Let $E_1=V\setminus
(\operatorname{Int}Y_1^e\cup
\operatorname{supp}Y_1^e)
=\operatorname{Ext}Y_1^e$ and
$E_k=E_{k-1}\setminus(
\operatorname{Int}Y_k^e\cup
\operatorname{supp}Y_k^e)$.
Assume by induction that $E_{k-1}$ is
connected. Let $x,y\in E_k$. We have to
show that $x$ and $y$ can be connected by
a path $\omega_{k}$ in $E_k$. Using the
inductive assumption, $x$ and $y$ can be
connected by a path $\omega_{k-1}$ in
$E_{k-1}$. Assume without loss of
generality that $\omega_{k-1}$ intersects
the set
$W=\operatorname{Int}Y_k^e\cup
\operatorname{supp}Y_k^e$,
and let $x_1$ be the first and $y_1$ the
last intersection point of $\omega_{k-1}$
with $W$. Since both $W$ and $V\setminus
W$ are connected, the boundary
$$
\partial_V W=\{x\in V \mid
\operatorname{dist}(x,W)=
\operatorname{dist}(x,V\setminus W)=0\}
$$
is connected, and $x_1$ and $y_1$ can be
connected by a path $\omega$ in
$\partial_V W$. Using the path
$\omega_{k-1}$ from $x$ to $x_1$,
the path
$\omega$ from $x_1$ to $y_1$, and again
the path $\omega_{k-1}$ from $y_1$ to $y$,
we obtain a path $\tilde\omega_{k}$ in
$E_k\cup \partial_V W$ which connects $x$
to $y$. The desired path $\omega_{k}$ is
obtained by a small deformation of
 $\tilde\omega_{k}$ which ensures that
 $\omega_{k}$ is a path in $E_k$.
\qed

In order to prove Lemma 4.5, we need the following Lemma,
which is based on the strong isoperimetric inequality
proven in Appendix A.

\proclaim{Lemma B.3}
Let $W$ be a union of
elementary cubes in $V$, with $|W|\leq
L^d/2$. Then
$$
|\partial W \cap \partial V|
\leq\frac{2^{1/d}+1}{2^{1/d}-1}
|\partial W \setminus \partial V|
\tag B.2
$$
$$
|\partial W|
\leq\left(1+\frac{2^{1/d}+1}{2^{1/d}-1}
\right)
|\partial W \setminus \partial V|
\tag B.3
$$
\endproclaim

\demo{Proof}
We introduce the $(d-1)$ dimensional
faces
$$
F_i=\{x\in\Bbb R^d \mid x_i=1/2,
 \; 1/2 \leq x_i \leq L+1/2\}
\qquad i=1,\ldots,d
$$
$$
F_{d+i}=\{x\in\Bbb R^d \mid x_i=L+1/2,
 \; 1/2 \leq x_i \leq L+1/2\}
\qquad i=1,\ldots,d,
$$
together with the projections
$\pi_i:V\to F_i$,
$\pi_{d+i}:V\to F_{d+i}$,
where $x^\prime=\pi_i(x)$ has
 coordinates $x_k^\prime=x_k$
for $k\neq i$ and $x_i^\prime=1/2$,
 while
$x_i^\prime=L+1/2$ for
$x^\prime=\pi_{d+i}(x)$.
Finally, for each elementary
 $(d-1)$-cell
$p\in\Cal  C$,
we define $\pi_-(p)$ as
 the projection $\pi_i(p)$
onto the face $F_i$ which
 is parallel to $p$, and
$\pi_+(p)$ as $\pi_{d+i}(p)$.

Let $G_i=F_i\cap\partial W$,
 $i=1,\ldots,2d$,
and consider an elementary
$(d-1)$-cell $p\in G_i$,
together with the line $\ell$
that links the center
of $\pi_-(p)$ to the center of $\pi_+(p)$. Then
$\ell$ must intersect
$\partial W$ an even number of times.
Define
$$
H_i=\{p\in G_i \mid
\text{ there does not exist }
p^\prime\in\partial W\setminus
\partial V\quad
\text{with}\quad
\pi_-(p^\prime)=\pi_-(p)\}
$$
and consider an elementary
$(d-1)$-cell $p\in G_i\setminus H_i$,
$i=1,\ldots,2d$.
Then either both $\pi_-(p)$
and $\pi_+(p)$ lay in
$\cup_{j=1}^{2d}G_j\setminus H_j$,
 in which case there are at least two
elementary $(d-1)$-cell
$p^\prime\in\partial W\setminus\partial V$
with $\pi_-(p^\prime)=\pi_-(p)$,
or only one of
$\pi_-(p)$ and $\pi_+(p)$ lies in
$\cup_{j=1}^{2d}G_j\setminus H_j$,
 in which case there is at least one
elementary $(d-1)$-cell
$p^\prime\in\partial W\setminus\partial V$
with $\pi_-(p^\prime)=\pi_-(p)$.
As a consequence,
$$
\sum_{i=1}^{2d}|G_i\setminus H_i|
 \leq |\partial W\setminus\partial V|.
\tag B.4
$$
On the other hand,
$$
|H_i|\leq|W|\,L^{-1}
\leq\left(\frac{1}{2}\right)^{1/d}
|W|^{1-1/d}
$$
by the fact that $|W|\leq L^d/2$.
Using the strong isoperimetric inequality
 (see Appendix A),
we obtain that
$$
|H_i|\leq\frac{1}{2d}2^{-1/d}|\partial W|.
\tag B.5
$$
Combining (B.4) and (B.5), we get
$$
\align
|\partial W\cap\partial V|
&=\sum_{i=1}^{2d}|G_i|
=\sum_{i=1}^{2d}|G_i\setminus H_i| +
\sum_{i=1}^{2d}|H_i|\cr
&\leq |\partial W\setminus\partial V| +
2^{-1/d}|\partial W|\cr
&= (1+2^{-1/d}) |\partial
 W\setminus\partial V|
   + 2^{-1/d} |\partial
W\cap\partial V|\cr
\endalign
$$
and hence
$$
|\partial W\cap\partial V|
\leq
\frac{1+2^{-1/d}}{1-2^{-1/d}}
|\partial W\setminus\partial V|
$$
which implies (B.2).
The bound (B.3)
follows from
(B.2).
\qed
\enddemo

\demo{Proof of Lemma 4.5}
We start with the observation that
$$
|Y|=|Y|_d +|Y|_{d-1} +\cdots+|Y|_1,
\tag B.6a
$$
where $|Y|_k$ denotes the number of $k$-dimensional
elementary cells in
\footnote{As in Section 3, a $k$-dimensional
 cell $c$ in $\supp Y$
is
only counted if there
is no $(k+1)$-dimensional cell
$c^\prime$ in $\supp Y$
with $c\subset c^\prime$.}
 $Y$, and similarly for the
boundary $\partial W_i$ of a
component  $W_i$ of
$\operatorname{Int}Y$,
$$
|\partial W_i|=|\partial W_i|_{d-1}
+  \cdots+|\partial W_i|_1.
\tag B.6b
$$
Using the fact that
$
|\partial W_i\cap\partial V|_{d-1}
=
|\partial \overline{W}_i\cap\partial V|_{d-1}
$
we then decompose $|\partial W_i|$ as
$$
|\partial W_i|=
|\partial W_i\setminus \partial V|_{d-1}
+
|\partial \overline{W}_i\cap\partial V|_{d-1}
+\sum_{k=1}^{d-2}|\partial W_i|_k\,.
\tag B.7
$$
For a long contour $Y$, we  use
Lemma B.3 applied to the set
$\overline{W}_i$, together with the fact that
$
\partial \overline{W}_i
\subset \partial {W}_i
$
to bound
$$
|\partial \overline{W}_i\cap \partial V|_{d-1}
\leq
\left(
\frac{2^{1/d}+1}{2^{1/d}-1}
\right)
|\partial \overline{W}_i\setminus \partial V|_{d-1}.
\leq
\left(
\frac{2^{1/d}+1}{2^{1/d}-1}
\right)
|\partial {W}_i\setminus \partial V|_{d-1}.
\tag B.8
$$
For short contours $Y$, on the other hand,
$
|\partial \overline{W}_i \cap \partial V|
\leq
|\partial \overline{W}_i\setminus\partial V|
$
which implies (B.8) with a better constant.
Therefore (B.8) is valid for both long and short
contours.

As a last step we observe that each cube $c$
in $\supp Y$ can be shared by at most $2d$
elementary faces in
$
(\partial W_1\setminus V)
\,\cup\,
\cdots
\,\cup\,
(\partial W_n\setminus V)
$,
while each
$d-1$ dimensional elementary face in $Y$
may be shared by the boundary of at most
two different components
of $\Int Y\cup\Ext Y$. Since
all lower dimensional elementary
cells in $Y$ belong to a unique component
of $\Int Y\cup\Ext Y$,
we get that
$$
\sum_{i=1}^n
|\partial {W}_i\setminus\partial V|_{d-1}
+\sum_{i=1}^n
\sum_{k=1}^{d-2}|\partial W_i|_k
\leq
2d |Y|.
\tag B.9
$$
Combining (B.7) with (B.8) and (B.9) and
the fact that
$$
N_{\partial V}(\Int Y)
\leq
\sum_{i=1}^n|\partial\overline{W}_i\cap\partial V|
\,,
\tag B.10
$$
we obtain the first two bounds
of the lemma.

In order to prove (4.4) we observe that
$
V(Y)=\supp Y \cup W_1
\cup\cdots\cup
W_n
$
which in turn implies
$
\partial V(Y)
\subset
\partial\supp Y
\cup \partial W_1
\cup\cdots\cup
\partial W_n
$
and hence
$$
|\partial V(Y)|
\leq
|\partial\supp Y|
+
|\partial W_1|
+\cdots+
|\partial W_n|
$$
Combined with the bound
$
|\partial\supp Y|\leq 2d |Y|
$
we obtain the remaining bound of
Lemma 4.5.
\qed
\enddemo
%
%

\vfill\eject
\head{Appendix C:  Inductive Proof of Lemma 4.6}
\endhead
\bigskip

In this appendix, we prove Lemma 4.6.
Actually, we will first prove the
following Lemma C.1.  In order to state the lemma, we
recall the definition of $f_q^{(n)}$ as the free
energy of an
auxiliary contour model with activities
$$
K^{(n)} (Y^q) =
\cases K^\prime (Y^q) \ \ & \text{\rm if} \ |V(Y^q)| \leq n, \\
0 & \text{\rm otherwise},\endcases \tag"{(C.1)}"
$$
and define
$$
\align
f^{(n)} & = \min_q f_q^{(n)}, \tag"{(C.2)}" \\
a_q^{(n)} & = f_q^{(n)} - f. \tag"{(C.3)}"
\endalign
$$
We also assign a number $v(W)$ to each volume of the form (4.7a),
$$
v(W) = \max_{Y \text{\rm in} W} |V(Y)|, \tag"{(C.4)}"
$$
where the maximum goes over all contours $Y$ with
$\supp Y \subset W$.
Obviously, $v(\Int  Y) \leq |V(Y)|$ for all contours $Y$.
 In fact,
$$
v(\Int Y) < |V(Y)| \tag"{(C.5)}"
$$
due to the fact that  $\dist(\tilde Y, Y) \geq 1$
if $\tilde Y$ is a
contour in $\Int Y$.

Finally, we recall that for a volume $W$
of the form (4.7a), $|W|$ is used to denote
 the euclidean volume of $W$,
while for a contour $Y$ and the boundary
$\partial W$ of a volume $W, \
|Y|$ and $|\partial W|$ are used
to denote the number of elementary
cells in $Y$ and $\partial W$,
respectively (see Equation (B.6a) and (B.6b)).
\medskip
\proclaim{Lemma C.1}
{Let
$$
\epsilon = e^{-\tau(1-(2C_1 + 1)\gamma)} e^{\alpha+2}
\qquad\text{and}\qquad
\bar \alpha = \frac{(\alpha - 2) 2d}{C_3}.
\tag"{(C.6)}"
$$
Then there is a constant $\epsilon_0$,
depending only on $d$ and $N$,
such that the following statements
are true for all $\epsilon <
\epsilon_0$ and all $n \geq 0$,
provided $|V(Y)| \leq n, \ v(W) \leq n$,
and $\bar \alpha \geq 1$}.
\medskip
i) \ \ $|K_q^\prime (Y)| \leq \epsilon^{|Y|}$.

ii) \ \ If $a_q^{(n)}|V(Y)|^{1/d}\leq\bar\alpha$
\quad then \quad $\chi_q^\prime (Y) = 1$.

iii) \ \ If $a_q^{(n)}|V(Y)|^{1/d}\leq\bar\alpha$
\quad then \quad $K_q^\prime (Y) = K_q(Y)$.

iv) \ \ If $a_q^{(n)}|W|^{1/d}\leq\bar\alpha$
\quad then \quad $Z_q(W, h) = Z^\prime_q(W, h)$.

v) \ \
    $
    |Z_q(W,h)|
    \leq
    e^{-f^{(n)}|W|} e^{O(\epsilon)|\partial W|}
    e^{\gamma \tau N_{\partial V}(W)}.
    $
\endproclaim
\bigskip

\demo{Proof} We proceed by induction on $n$,
first proving the lemma for $n=0$ and then
for any given
$n\in\Bbb N$, assuming that it has
been
already proven for all integers smaller
than $n$.
\enddemo

\bigskip
\subhead{Proof of Lemma C.1 for n=0}
\endsubhead
\bigskip

For $|V(Y)| = 0$
we have
 $\chi_q^\prime (Y) = 1$
and
thus $K_q^\prime (Y) = K_q(Y) = \rho(Y)$.  This makes i),
ii) and iii)
trivial statements.  Using iii)
for $|V(Y)| = 0$, we then
conclude that $Z_q(W, h) = Z_q^\prime (W, h)$
for $v(W) = 0$.  By i)
$Z_q(W, h)
= Z_q^\prime (W,h)$
and thus the partition function can be analyzed
by a convergent expansion
yielding
$$
|Z_q(W, h)|
\leq
e^{-f_q^{(0)} |W|}
e^{O(\epsilon) |\partial W|}
e^{(e_q|W| - E_q(W))}
\leq
e^{-f_q^{(0)}|W|} e^{O(\epsilon)|\partial W|}
e^{\gamma \tau N_{\partial V} (W)}.
$$
Observing that $f_q^{(0)} \geq f^{(0)}$,
this concludes the proof of Lemma
C.1 for $n = 0$.

\bigskip
\subhead{Proof of Lemma C.1 i) for $|V(Y)| = n$}
\endsubhead
\bigskip

Due to (C.5),
$v(\Int Y) < n$, and all contours $\tilde Y$ contributing to
$Z^\prime_q(\Int_m Y, h)$ obey the condition $|V(\tilde Y)| < n$.
This implies that
$|K_q^\prime (\tilde Y)|
\leq \epsilon^{|\tilde Y|}$ by the
inductive assumption i).
As a consequence, the logarithm of
$Z^\prime_q(\text{\rm Int}_m Y, h)$
can be analyzed by a convergent
expansion, and
$$
\left|
\log Z_q^\prime  (\text{\rm Int}_m Y, h)
 + f_q^{(n-1)}|\text{\rm Int}_m Y|
\right|
\leq
O(\epsilon) |\partial \text{\rm Int}_m Y|
+ \gamma\tau N_{\partial V} (\Int_m Y).
\tag"{(C.7)}"
$$
Combining (C.7) with the induction assumption v), we
get
$$
\align
\prod_m
\left|
\frac{Z_m(\text{\rm Int}_m Y, h)}
{Z^\prime_q(\text{\rm Int}_m Y, h)}
\right|
&\leq
e^{a_q^{(n-1)} |\Int Y|}
e^{2\gamma\tau N_{\partial V}(\Int Y)}
e^{O(\epsilon)\sum_m |\partial \text{\rm Int}_m Y|} \\
& \leq e^{a_q^{(n-1)} |\Int Y|} e^{(2 C_1 \gamma\tau +
O(\epsilon))|Y|},\tag"{(C.8)}" \\
\endalign
$$
where we have used Lemma 4.5
in the last step. Observing that
$$
|e_m - f_m^{(n-1)}| \leq O(\epsilon),
\tag"{(C.9a)}"
$$
which implies
the bound
$$
|(e_q - e_0) - a_q^{(n-1)}| \leq O(\epsilon),
\tag"{(C.9b)}"
$$
we
use the assumptions (3.7) and (3.11) to bound
$$
 |\rho(Y) e^{E_q(Y)}|
\leq e^{-\tau|Y|}
e^{\gamma\tau N_{\partial V}(\supp Y)}
e^{(e_q - e_0)|Y|_d}
\leq e^{-(\tau - \gamma\tau - O(\epsilon))|Y|}
e^{a_q^{(n-1)} |Y|_d}.
\tag"{(C.10)}"
$$
Here $|Y|_d$ is defined as the
number of $d$-cells in $Y$ and thus
$|V(Y)| = | \text{\rm Int} \, Y| + |Y|_d$.
Combining now (C.10) with (C.8), we obtain
$$
|K_q^\prime (Y)| \leq \chi_q^\prime (Y) e^{a_q^{(n-1)} |V(Y)|}
 e^{-(\tau - O(\epsilon) - (1 + 2 C_1) \gamma\tau)|Y|}.
\tag C.11
$$
Without loss of generality,
we may now assume that $\chi^\prime_q(Y) > 0$
(otherwise $K^\prime_q(Y) = 0$
and the statement i) is trivial).
By the definition of $\chi^\prime_q(Y)$,
this implies
$$
 (f_q^{(n-1)} - f_m^{(n-1)}) |V(Y)|
\leq 1 + \alpha |Y| \leq (1 + \alpha) |Y|
$$
for all $m \neq q$.  As a consequence,
$$
a_q^{(n-1)} |V(Y)| \leq (1 + \alpha )
|Y|, \tag"{(C.12)}"
$$
provided $\chi_q^\prime  (Y) \neq 0$.
Combined with (C.11) and the fact that
$\chi_q^\prime (Y) \leq 1$, this implies that
$$
|K_q^\prime (Y)| \leq
e^{-[\tau - 1 - O(\epsilon) - \alpha -
   (1 + 2 C_1) \gamma\tau]|Y|},
\tag"{(C.13)}"
$$
which
yields the desired bound i) for
$|V(Y)| = n$.

\bigskip
\subhead{Proof  of Lemma C.1 ii)
for
$k=|V(Y)| \leq n$
and
$a_q^{(n)} |V(Y)|^{1/d} \leq \bar \alpha$
}
\endsubhead
\bigskip

We just have proved that i)
is true for all contours $Y$ with
$|V(Y)| \leq n$.  As a consequence,
both $f_m^{(k)}$ and
$f_m^{(n)}$
may be analyzed by a
convergent cluster expansion.
On the other hand,
$$
|V(Y)|^{\frac{d-1}{d}} \leq \frac{1}{2d} |\partial V(Y)| \leq
\frac{C_3}{2d} |Y|, \tag"{(C.14)}"
$$
by the isoperimetric inequality and
Lemma 4.5. Using this bound and the definition
of $f_m^{(n)}$, one may easily
see that all contours $Y$ contributing to the
cluster expansion of the difference
$f_m^{(k)} - f_m^{(n)}$ obey the bound
$$
|Y|
\geq
\frac{2d}{C_3}(k+1)^{(d-1)/d}
\geq
\frac{2d}{C_3}k^{1/d}=:n_0.
$$
As a consequence,
$$
|f_m^{(k)} - f_m^{(n)}|
\leq (K \epsilon)^{n_0},
$$
where $K$ is a constant depending
only on the dimension $d$ and the number
of phases $N$.
Using the bound (C.14)
for the second time and recalling
that $|V(Y)|=k$, we
get
$$
|f_m^{(k)} - f_m^{(n)}|
|V(Y)|
\leq
(K \epsilon)^{n_0}
|V(Y)|^{1/d} \frac{C_3}{2d}|Y|
=O(1) n_0 (K \epsilon)^{n_0} |Y|\leq O(\epsilon)|Y|.
\tag C.15
$$
Combining (C.15) with the assumption
$a_q^{(n)} |V(Y)|^{1/d} \leq \bar \alpha$
and the bound (C.14),
 we obtain
the lower bound
$$
\align
 \alpha |Y| - [f_q^{(k)}-f_m^{(k)}] |V(Y)|
&\geq
 \alpha |Y| -  a_q^{(n)}|V(Y)|-O(\epsilon)|Y|
\geq\\
& \geq
\left(
\alpha - \bar\alpha \frac{C_3}{2d}
- O(\epsilon)
\right)
|Y|
=(2-O(\epsilon))|Y|
\geq 2-O(\epsilon),
\endalign
$$
where,
in the next to the last step,
we used the definition of $\bar \alpha$, see (C.6).
Combined with (4.16b) we obtain
the equality $\chi_q^\prime (Y) = 1$.

\bigskip
\subhead{Proof of Lemma C.1 iii) and iv)}
\endsubhead
\bigskip

Given ii) and the
definition of
$K_q^\prime (Y)$ and $Z^\prime_q(W, h)$,
the statement is obvious.  See
\cite{BK90} for a formal proof using induction on the subvolumes of $W$ and
$\Int Y$.

\bigskip
\subhead{Proof of Lemma C.1 v)}
\endsubhead
\bigskip

We say a contours $Y$ is
{\it small}
if
$a_q^{(n)} |V(Y)|^{1/d} \leq \bar\alpha$
while it is
{\it large} if
$a_q^{(n)} |V(Y)|^{1/d} > \bar\alpha$. We
then use the relation (4.8) to
rewrite
$Z_q(W, h)$
by splitting the set
$\{ Y_1, \cdots , Y_k\}_{\text{\rm ext}}$
of external contours into
$
 \{ X_1, \cdots , X_{k^\prime } \}
 \cup
 \{ Z_1, \cdots , Z_{k^{\prime\prime}} \}
$,
where
$Z_1, \cdots , Z_{k^{\prime\prime}}$ are
the small contours in
$\{ Y_1, \cdots , Y_k\}_{\text{\rm ext}}$
and
$X_1, \cdots , X_{k^\prime }$
are the large contours in
$\{ Y_1, \cdots , Y_k\}_{\text{\rm ext}}$.
Note
that for a fixed set
$\{ X_1, \cdots X_{k^\prime } \}$,
the sum over
$\{ Z_1, \cdots Z_{k^{\prime\prime}} \}$
runs over sets of mutually external small
$q$-contours in
$\text{\rm Ext} = W \backslash \overset k^\prime  \to{\underset i
= 1 \to \cup} V(X_i)$.  
Resumming the small  contours, we thus obtain
$$
Z_q(W,h) = \sum_{\{ X_1,\cdots ,
X_{k^\prime }\}_{\text{\rm ext}}}
 \
Z_q^{\text{\rm small}} (\text{\rm Ext}, h)
\prod_{i=1}^{k^\prime }
\biggl[
\rho(X_i) \prod_m Z_m (\text{\rm Int}_m X_i, h)
\biggr].
\tag"{(C.16)}"
$$
Here the sum goes over sets of mutually
external large contours in $W$
and $Z_q^{\text{\rm small}}(\text{\rm Ext}, h)$
is obtained from
$Z_q(\text{\rm Ext}, h)$ by dropping all
large external $q$-contours.

Due to the inductive assumption iii),
$K_q(Y) = K^\prime_q(Y)$ if $Y$ is
small.
Since
$|K^\prime_q(Y)| \leq \epsilon^{|Y|}$
by i), $Z_q^{\text{\rm small}}(\text{\rm Ext}, h)$
can be controlled
by a convergent cluster expansion, and
$$
\left| Z_q^{\text{\rm small}} (\text{\rm Ext}, h) \right| \leq
e^{-f_q^{\text{\rm small}}|\text{\rm Ext}|}
e^{O(\epsilon)|\partial \text{\rm Ext}|}
e^{\gamma \tau N_{\partial V} (\text{\rm Ext})},
\tag"{(C.17)}"
$$
where $f_q^{\text{\rm small}}$
is the free energy of the contour model with
activities
$$
K_q^{\text{\rm small}} (Y) =
\cases K^\prime_q(Y) \ \ & \text{\rm if} \ |V(Y)|
\leq n \ \text{\rm and} \ Y \ \text{\rm is small}, \\
0 & \text{\rm otherwise}.
\endcases
\tag"{(C.18)}"
$$
On the other hand,
$$
\prod_m  | Z_m(\text{\rm Int}_m X_i, h) |
\leq
e^{-f^{(n-1}) |\Int X_i|}
e^{O(\epsilon)|\partial \Int X_i|}
e^{\gamma\tau N_{\partial V} (\Int X_i)}
$$
by the induction assumption v).
Observing that
the smallest contours contributing to the difference of
$f_m^{(n)}$ and $f_m^{(n-1)}$ obey
the bound
$$
|Y|
\geq
\frac{2d}{C_3}
n^{(d-1)/d}
\geq
\frac{2d}{C_3}
n^{1/d}
=:n_0\,,
$$
while $|V(X_i)|\leq n$, we may continue as in the proof of
(C.15) to bound
$$
|f^{(n-1)} - f^{(n)}||\Int X_i|
\leq
|f^{(n-1)} - f^{(n)}||V(X_i)|
\leq O(1) n_0 (K \epsilon)^{n_0} \leq O(\epsilon)\,.
$$
Since $|\Int X_i|\leq O(1)|X_i|$ by Lemma 4.5,
we conclude that
$$
\prod_m |Z_m(\Int X_i, h)|
\leq
e^{-f^{(n)}|\Int X_i|} e^{O(\epsilon) |X_i|}
e^{\gamma\tau N_{\partial V} (\Int X_i)}\,.
\tag"{(C.19)}"
$$
Combining (C.17) and (C.19) with the bounds
$$
|\rho(X_i)| \leq
e^{-\tau|X_i|-e_0|X_i|_d}
e^{\gamma\tau N_{\partial V}(\supp X_i)}
\leq e^{-(\tau-O(\epsilon))|X_i|}
e^{-f^{(n)}|X_i|_d}
e^{\gamma\tau N_{\partial V}(\supp X_i)}
 \tag"{(C.20)}"
$$
and
$$
|\partial\text{\rm Ext}|
\leq
|\partial W| + \sum_{i=1}^{k^\prime } |\partial V(X_i) |
\leq
|\partial W| + C_3 \sum_{i=1}^{k^\prime } |X_i| ,
\tag"{(C.21)}"
$$
and the
equality
$
N_{\partial V}(\Ext)
+\sum_{i=1}^{k^\prime }
[
N_{\partial V}(\supp X_i)
+
N_{\partial V}(\Int X_i)
]
=
N_{\partial V}(W)
$,
we conclude that
$$
\multline
|Z_q (W, h)|
\leq e^{O(\epsilon)|\partial W|}
e^{\gamma\tau N_{\partial V}(W)}
e^{-f^{(n)} |W|}
\times 
\\  
\times
\sum_{\{X_1, \cdots , X_{k^\prime } \}_{\text{\rm ext}}}
e^{-[f_q^{\text{\rm small}} - f^{(n)}] |\text{\rm Ext}|}
\prod_{i=1}^{k^\prime } e^{-(\tau-O(\epsilon))|X_i|}.
\endmultline
\tag"{(C.22)}"
\endalign
$$
Next, we bound the difference $f_q^{\text{\rm small}} - f_q^{(n)}$.  In a
first step, we use the isoperimetric inequality together with Lemma 4.5 and
the definition of
large contours to bound
$$
|X|
\geq
\frac{1}{C_3} |\partial V(X)|
\geq
\frac{2d}{C_3} |V(X)|^{\frac{d-1}{d}}
\geq
\frac{2d}{C_3} |V(X)|^{1/d}
\geq
\ell_0 := \frac{2d \bar \alpha}{C_3} \frac{1}{a_q^{(n)}}
\tag"{(C.23)}"
$$
for all large contours $X$.  Next, we observe that
$$
|f_q^{(n)} - f_q^{\text{\rm small}}|
\leq (K\epsilon)^{\ell_0} \leq
\frac{1}{-\ell_0 \log (K\epsilon)},
\tag"{(C.24)}"
$$
where $K$ is a constant
depending only on $d$ and $N$.
Recalling the
condition $\bar \alpha \geq 1$, we
get
$$
|f_q^{(n)} - f_q^{\text{\rm small}}| \leq \frac{1}{2} a_q^{(n)},
\tag"{(C.25)}"
$$
provided $\epsilon$ is chosen small enough.
Combining (C.22) with
(C.25), we finally obtain
$$
|Z_q (W, h)|
\leq
e^{O(\epsilon)|\partial W|}
e^{\gamma\tau N_{\partial V} (W)}
e^{-f^{(n)}|W|}
\sum_{\{X_1, \cdots , X_{k^\prime } \}_{\text{\rm ext}}}
e^{-\frac{a_q^{(n)}}{2} |\text{\rm Ext}|}
\prod_{i=1}^{k^\prime } e^{-\tilde \tau |X_i|}
\tag"{(C.26)}"
$$
with
$$
\tilde \tau = (\tau - 1). \tag"{(C.27)}"
$$
At this point we need the following Lemma C.2,
which is a variant of a
lemma first proven in \cite{Z84}, see also \cite{BI89}.
\bigskip
\proclaim{Lemma C.2}
{Consider an arbitrary contour functional $\tilde
K_q(Y) \geq 0$, and let $\tilde Z_q$ be the partition function}
$$
\tilde Z_q(W) =
\sum_{\{Y_1, \cdots , Y_n \}} \prod_{i=1}^h (\tilde K_q
(Y_i) e^{|Y_i|}). \tag"{(C.28)}"
$$
{\it Let $\tilde s_q$ be the corresponding free energy, and assume
that $\tilde
K_q(Y) \leq \tilde \epsilon^{|Y|}$, where
$\tilde \epsilon$ is small
(depending on $N$ and $d$).  Then for any
$\tilde a \geq -\tilde s_q$
the following bound is true}
$$
\sum_{\{Y_1, \cdots , Y_k \}_{\text{\rm ext}}}
e^{\tilde a|\text{\rm Ext}|}
\prod_i \tilde K_q (Y_i)
\leq
e^{O(\tilde \epsilon)|\partial W|}\,,
\tag"{(C.29)}"
$$
{where the sum goes over sets of
mutually external $q$-contours in $W$}.
\endproclaim
\bigskip

In order to apply the lemma, we define
$\tilde K_q(Y) = e^{-\tilde
\tau|Y|}$ if $Y$ is a large $q$-contour,
 and $\tilde K_q(Y) = 0$
otherwise.  With this choice,
$$
0 \leq -\tilde s_q \leq (K\epsilon)^{\ell_0} \leq \frac{1}{-\ell_0
\log(K\epsilon)},
 \tag"{(C.30)}"
$$
where $\ell_0$ is the constant from (C.23).  As a consequence,
$$
-\tilde s_q \leq \tilde a := \frac{a_q^{(n)}}{2} \tag"{(C.31)}"
$$
provided $\epsilon$ is small enough.  Applying Lemma C.2 to the
right hand side  of (C.26), and observing that $\tilde
\epsilon := e^{-\tilde\tau} \leq \epsilon$, we finally obtain
the desired inequality
$$
|Z_q(W,h)| \leq e^{O(\epsilon)|\partial W|}
e^{\gamma\tau N_{\partial V} (W)}
e^{-f_0^{(n)} |W|}.
$$
This concludes the inductive proof of Lemma C.1. \hfill \qed

\bigskip
\subhead{Proof of Lemma C.2}
\endsubhead
\bigskip

The partition function
$\tilde Z_q$ is
defined in terms  of the polymer model with activities
$
K^*(Y) = \tilde K_q(Y) e^{|Y|}
$.
For $\tilde \epsilon$ small enough,
$\tilde Z_q$ can be controlled by a
convergent cluster expansion and
$$
| \log \tilde Z_q(\Int Y) + \tilde s_q | \Int Y||
\leq O(\tilde\epsilon) |\partial \Int Y| \leq
O(\tilde\epsilon) |Y|.
$$
\comment
Combined with the fact that 
$-\tilde a \leq \tilde s_q = O(\tilde\epsilon)$
and the equality
$|W|=|\text{\rm Ext}|
+
\sum_i\left(|\Int Y_i|+|Y_i|\right)
$
provided
${\{Y_1, \cdots , Y_k \}_{\text{\rm ext}}} $
is a set of mutually external contours in $W$,
\endcomment
On the other hand,
$|W|=|\text{\rm Ext}|
+
\sum_i\left(|\Int Y_i|+|Y_i|\right)
$
if
${\{Y_1, \cdots , Y_k \}_{\text{\rm ext}}} $
is a set of mutually external contours in $W$.
Combined with 
the fact that 
$-\tilde a \leq \tilde s_q = O(\tilde\epsilon)$,
we obtain
$$
\align
\sum_{\{Y_1, \cdots , Y_k \}_{\text{\rm ext}}}
e^{-\tilde a|\text{\rm Ext}|}
\prod_{i=1}^k \tilde K_q(Y_i)
& \leq
e^{\tilde s_q|W|}
\sum_{\{Y_1, \cdots , Y_k \}_{\text{\rm ext}}}
\prod_{i=1}^k \tilde K_q(Y_i)
e^{ - \tilde s_q(|\Int Y_i|+|Y_i|_d} \\
& \leq
e^{\tilde s_q|W|}
\sum_{\{Y_1, \cdots , Y_k \}_{\text{\rm ext}}}
\prod_{i=1}^k \tilde K_q(Y_i)
\tilde Z_q(\Int Y_i)
e^{O(\tilde\epsilon)|Y_i| - \tilde s_q|Y_i|_d} \\
& \leq e^{\tilde s_q|W|} \sum_{\{Y_1, \cdots , Y_k \}_{\text{\rm ext}}}
\prod_{i=1}^k \tilde K_q(Y_i) e^{|Y_i|} \tilde Z_q(\Int Y_i) \\
& = e^{\tilde s_q|W|} \tilde Z_q(W) \leq e^{O(\tilde \epsilon)|\partial
W|}.\tag"{\qed}"
\endalign
$$

\bigskip
\subhead{Proof of Lemma 4.6}
\endsubhead
\bigskip

Lemma 4.6 i) through iv) follows from
Lemma C.1 and the fact that
$f = \underset n \rightarrow \infty \to
\lim f^{(n)}$ and $a_q = \underset n \rightarrow \infty \to \lim
a_q^{(n)}$.

\medskip
In order to prove the statement v), we
extract
the factor
$$
\align
\max_{\{X_1, \cdots , X_{k^\prime } \}}
e^{-\frac{a_q^{(n)}}{4}|\Ext|}
e^{-\frac{\tau}{4} \sum_i |X_i|}
& \leq
\max_{\{X_1, \cdots , X_{k^\prime }\}}
e^{-\frac{a_q^{(n)}}{4}|\Ext|}
e^{-(\tau/4C_3)\sum_i|\partial V(X_i)| }
\\
& \leq
\max_{U \subset W}
e^{-\frac{a_q^{(n)}}{4}|W\setminus U|}
e^{-(\tau/4C_3)|\partial U| }
\endalign
$$
from the right hand side of (C.26),
and bound the remaining sum as before.
Taking the limit $n\to\infty$
in the resulting bound,
this
yields
$$
|Z_q(W, h)|
\leq
e^{\gamma\tau N_{\partial V} (W)}
e^{(O(\epsilon)+O(e^{-3\tau/4}))|\partial W|} e^{-f|W|}
\max_{U \subset W}
e^{-\frac{a_q}{4}|W\setminus U|}
e^{-(\tau/4C_3)|\partial U| }
\,.
\tag C.32
$$
We conclude, with the help of the isoperimetric inequality,
that
$$
\multline
|Z_q(W, h)|
\leq
e^{\gamma\tau N_{\partial V} (W)}
e^{(O(\epsilon)+O(e^{-3\tau/4}))|\partial W|} e^{-f|W|}
\max_{U \subset W}
e^{-\frac{a_q}{4}|W\setminus U|
-(2d\tau/4C_3)|U|^{d/(d-1)}}
\\
=
e^{\gamma\tau N_{\partial V} (W)}
e^{(O(\epsilon)+O(e^{-3\tau/4}))|\partial W|} e^{-f|W|}
\max
\left\{
        e^{-\frac{a_q}{4}|W|}, e^{-(2d\tau/4C_3)|W|^{d/(d-1)}}
\right\}
\,,
\endmultline
\tag"{(C.33)}"
\endalign
$$
where we
used the fact that the maximum over $U$
is obtained for either
$U=W$ or $U=\emptyset$.

Observing that $2d |V|^{d/(d-1)}=|\partial V|$
and that $N_{\partial V}(V)$ can be bounded by
$|\partial V|$, the bound (C.33)
implies Lemma 4.6 v).\hfill\qed

\bigskip
\subhead{Proof of the bound (4.26)}
\endsubhead
\bigskip

Due to the bound (C.12)
we have
$a_q^{(n-1)} |V(Y)|\leq (1+\alpha)|Y|$
if
$\chi_q(Y)\neq 0$. Using the strategy which was used
to prove (C.15), we replace
$a_q^{(n-1)}$ by $a_q$, concluding
that
$
\chi_q(Y)\neq 0
$
implies
$
a_q |V(Y)|\leq (1+O(\epsilon)+\alpha)|Y|
$.
\hfill\qed

%
%

\vfill\eject
\head{Appendix D: Proof of Lemma 4.7}
\endhead

We start with a combinatoric Lemma
that will be
used throughout this appendix.

\medskip

\proclaim{Lemma D.1} Let $k_0$ be a positive
integer and let
$G(h)$ be a function which satisfies
the bounds
$$
\left|
{d^{k}\over dh^{k}}
{G(h)}
\right|
\leq
\lambda^{|k|}
$$
for all multi-indices $k$ with
$1\leq|k|\leq k_0$
and some $\lambda>0$. Then
$$
\left|
{d^k\over dh^k}
e^{G(h)}
\right|
\leq
|k|!\lambda^{|k|}
e^{G(h)}
$$
for all multi-indices $k$ with
$1\leq|k|\leq k_0$.
\endproclaim

\demo{Proof} Observing that
$$
{d^k\over dh^k}
e^{G(h)}
=H_k(h)e^{G(h)}\,,
$$
where $H_k(h)$ is a polynomial of degree
$|k|$ in the derivatives of $G$,
the the Lemma is
immediately obtained by
induction on $|k|$.
\qed
\enddemo

\comment
The next Lemma is an immediate generalization
of Lemma D.1.

\medskip

\proclaim{Lemma D.2} Let $k_0$ be a positive
integer and let
$G(h)$ be a function which satisfies
the bounds
$$
\left|
{d^{k}\over dh^{k}}
{G(h)}
\right|
\leq
\lambda^{|k|}
$$
for all multi-indices $k$ with
$1\leq|k|\leq k_0$
and some $\lambda>0$. Assume that $\chi:\Bbb R\to\Bbb R$
is $k_0$ times differentiable, with derivatives
$\chi^{(l)}$, $l=1,\cdots,k_0$.
If
$1\leq|k|\leq k_0$,
then
$$
{d^k\over dh^k}
\chi({G(h)})
=\sum_{l=1}^|k| H_{k,l}(h) \chi^{(l)}(G(h)),
$$
where
$H_{k,l}(h)$
are polynomials of degree $l$
in the derivatives of $G$ which obey the bounds
$$
\sum_{l=1}^{|k|}|H_{k,l}(h)|\leq |k|! \lambda^{|k|}.
$$
\endproclaim
\endcomment

Keeping the notation of Appendix C, we
now prove the following Lemma, which contains
statements i) through iii) of Lemma 4.7.

\proclaim{Lemma D.2} There is a constant
$K<\infty$,
depending only on $N$, $d$ and the constants introduced in
(3.8),
(3.9), and (4.16),
such that the following statements
are true provided $\epsilon<\epsilon_0$,
$\overline\alpha\geq 1$ and $n\geq 0$.

\item{i)} For $|V(Y)|\leq n$
and $h_0\in\Cal  U$
one has
$$
\left|\frac{d^k}{dh^k}
K_q^\prime(Y)\right|_{h=h_0}
\leq
(K\epsilon)^{|Y|}
\tag D.1
$$
provided $1\leq |k|\leq 6$.

\item{ii)} For $v(W)\leq n$
and $h_0\in\Cal  U$
one has
$$
\left|
\frac{d^k}{dh^k}
\log Z_q^\prime(W,h)
\right|_{h=h_0}
\leq (C_0^{|k|}+O(\epsilon))|W|
\tag D.2
$$
provided $1\leq |k|\leq 6$.

\item{iii)} For $v(W)\leq n$
and $h_0\in\Cal  U$
one has
$$
\left\vert
{d^k \over dh^k}
Z_q(W,h)
\right\vert_{h=h_0}
\leq
|k|!
\left(
C_0
+O(\epsilon))|W|
\right)^{|k|}
e^{-f|W|}
e^{O(\epsilon)\vert\partial W\vert}
e^{\gamma\tau N_{\partial V}(W)}
\tag D.3
$$
provided $1\leq |k|\leq 6$.

\endproclaim

\demo{Proof} As in the proof of Lemma C.1
we proceed by induction on $n$.
\enddemo

\bigskip
\subhead{Proof of Lemma
D.2 for n=0}
\endsubhead
\bigskip

For $|V(Y)| = 0$,
$K_q^\prime(Y) = K_q(Y) = \rho(Y)$,
which makes i) a trivial statement. As a consequence,
the left hand side of (D.2) can be analyzed by
a convergent cluster expansion, leading immediately
to the bound (D.2) for $v(W)=0$.
Bounding finally
$(C_0^{|k|}+O(\epsilon))|W|$ by
$\{(C_0+O(\epsilon))|W|\}^{|k|}$ and observing
that $Z_q(W, h) = Z_q^\prime(W, h)$ if $v(W) = 0$,
we obtain iii) with the help of Lemma D.1.

\bigskip
\subhead{Proof of Lemma D.2 i) for $|$V(Y)$|$ = n}
\endsubhead
\bigskip

Using the assumptions (3.8) and (3.9)
together with Lemma D.1, the bound
(C.10) can be easily generalized to derivatives,
giving
$$
\left|
{d^k\over dh^k}
\left[
\rho(Y) e^{E_q(Y)}
\right]
\right|
\leq
|k|!
\left[
2 C_0 |Y|
\right]^{|k|}
e^{-(\tau - \gamma\tau - O(\epsilon))|Y|}
e^{a_q |Y|_d}\,.
\tag"{(D.4))}"
$$
In a similar way, the bound (C.8) can be generalized to
derivatives, using the inductive assumptions
ii) and iii) together with Lemma D.1 and Lemma 4.5.
This gives
$$
\Bigl|
{d^k\over dh^k}
\Bigl[
\prod_m
\frac{Z_m(\Int_m Y, h)}{Z^\prime_q(\Int_m Y, h)}
\Bigr]
\Bigr|
\leq
|k|!
\left[
(2 C_0+O(\epsilon))|\Int Y|
\right]^{|k|}
e^{a_q |\Int Y|} e^{(2 C_1 \gamma\tau +
O(\epsilon))|Y|}\,.
\tag"{(D.5)}"
$$
Using finally
the possibility to analyze the
derivatives of $f^{(n-1)}_m(h)$   by a convergent
expansion due to the inductive assumption i),
we bound
$$
\Bigl|
{d^k f^{(n-1)}_m(h)\over dh^k}
\Bigr|
\leq
C_0^{|k|}+O(\epsilon)
\leq
(C_0+O(\epsilon))^{|k|}
\,.
$$
As a consequence,
$$
\left|
{d^k\over dh^k}
\chi_q^\prime(Y)
\right|
\leq
(\overline C_1 |V(Y)|)^{|k|}
\,
\tag D.6
$$
for all multi-indices of order $|k|\leq 6$.
Here $\overline C_1$ is a constant that depends on
$N$ and the constants introduced in
(3.8), (3.9) and (4.16).
Combining (D.4) through (D.6) and bounding terms
of the form $O(1)|V(Y)|$ and $O(1)|Y|$
by $e^{O(1)|Y|}$,
we obtain the bound
(D.1).

\bigskip
\subhead{Proof  of Lemma D.2 ii)
for v(W)=n}
\endsubhead
\bigskip

We just have proved that i)
is true for all contours $Y$ with
$|V(Y)| \leq n$.  As a consequence
the derivatives of
$\log Z_q^\prime(W,h)$
can be analyzed by a
convergent cluster expansion. The
bound (D.2) immediately follows.

\bigskip
\subhead{Proof of Lemma D.2 v) for v(W)=n}
\endsubhead
\bigskip

We define: a contours $Y$ is
{\it small} if
$a_q(h_0) |V(Y)|^{1/d} \leq \bar\alpha$,
while a contour $Y$ is called
{\it large} if
$a_q(h_0) |V(Y)|^{1/d} > \bar\alpha$. As in Appendix C
we then
rewrite
$Z_q(W, h)$ as
$$
Z_q(W,h) =
\sum_{\{ X_1,\cdots , X_{n}\}_{\text{\rm ext}}} \
Z_q^{\text{\rm small}} (\text{\rm Ext}, h)
\prod_{i=1}^{n}
\biggl[
\rho(X_i) \prod_m Z_m (\text{\rm Int}_m X_i, h)
\biggr],
\tag D.7
$$
where the sum goes over sets of mutually external
 large contours in $W$
and $Z_q^{\text{\rm small}}(\text{\rm Ext}, h)$
is obtained from
$Z_q(\text{\rm Ext}, h)$ by dropping all
 large external $q$-contours.

Due to Lemma 4.6,  $K_q(Y) = K^\prime_q(Y)$ if $Y$ is
small and $h=h_0$.
Combining this with the bound (4.13) and the
inductive assumption (D.1), we conclude
that
$$
\left|\frac{d^k}{dh^k}
K_q^\prime(Y)\right|
\leq
(2K\epsilon)^{|Y|}
\tag D.8
$$
in a certain neighborhood $\Cal  U_0$ of $h_0$.
As a consequence, the derivatives of
$\log Z_q^{\text{\rm small}}(\text{\rm Ext}, h)$ can be controlled
by a convergent cluster expansion, and
$$
\biggl|\frac{d^k}{dh^k}
\log Z_q^{\text{\rm small}} (\text{\rm Ext}, h)
\biggr|
\leq
\bigl[C_0^{|k|}+O(\epsilon)\bigr]
|\Ext|
\leq
\left[(C_0+O(\epsilon))|\Ext|\right]^{|k|}
\tag D.9
$$
provided $h\in \Cal  U_0$. Combining (D.9) with the bound
$$
\left| Z_q^{\text{\rm small}} (\text{\rm Ext}, h_0) \right| \leq
e^{-f_q^{\text{\rm small}}|\text{\rm Ext}|}
e^{O(\epsilon)|\partial \text{\rm Ext}|}
e^{\gamma \tau N_{\partial V} (\text{\rm Ext})},
\tag"{(D.10)}"
$$
where $f_q^{\text{\rm small}}$ is the
free energy of the contour model with
activities
$$
K_q^{\text{\rm small}} (Y) =
\cases K^\prime_q(Y) \ \ &
\text{\rm if} \ Y \ \text{\rm is small}, \\
       0                 & \text{\rm if} \ Y \
\text{\rm is large},
\endcases
\tag D.11
$$
we obtain
$$
\left|\frac{d^k}{dh^k}
Z_q^{\text{\rm small}} 
(\text{\rm Ext}, h)\right|_{h=h_0}
\leq
|k|!
\left[(C_0+O(\epsilon))|\Ext|\right]^{|k|}
e^{-f_q^{\text{\rm small}}|\text{\rm Ext}|}
e^{O(\epsilon)|\partial \text{\rm Ext}|}
e^{\gamma \tau N_{\partial V} (\text{\rm Ext})}.
\tag"{(D.12)}"
$$
On the other hand
$$
\multline
\left|
{d^k\over dh^k}
\rho(X_i)
\right|
\leq
|k|!(C_0|X_i|)^{|k|}e^{-(\tau-O(\epsilon))|X_i|}
e^{-f|X_i|_d+N_{\partial V}(\supp X_i)}\leq
\\
\leq
|k|!C_0^{|k|}e^{-(\tau-|k|/e-O(\epsilon))|X_i|}
e^{-f|X_i|_d+N_{\partial V}(\supp X_i)}\,.
\endmultline
\tag D.13
$$

Combining (D.7) with the inductive assumption (D.3)
and the bounds (D.12) and (D.13),
we may continue as in Appendix C to get
$$
\multline
\left|
{d^k \over dh^k}
Z_q(W,h)\right|_{h=h_0}
\leq
|k|!
\left(
C_0
+O(\epsilon))|W|
\right)^{|k|}
e^{-f|W|}e^{\gamma\tau N_{\partial V}(W)}
e^{O(\epsilon)\vert\partial W\vert}\times
\\
\times
 \sum_{\{ X_1,\cdots , X_{n}\}_{\text{\rm ext}}} \
e^{-\frac{a_q}{2}|\Ext|} \prod_{i=1}^{n}
e^{-\tilde\tau|X_i|},
\endmultline
\tag D.14
$$
where now
$$
\tilde\tau=\tau-6/e-1\,.
\tag D.15
$$
Note the extra term $6/e$ with respect to $(C.27)$,
which comes from the term $|k|/e$ in (C.13) (recall that
we assumed $|k|\geq 6$).

Given the bound (D.14), the proof of (D.3)
for $v(W)=n$ now follows
using Lemma C.2 from Appendix C.
This concludes the proof of Lemma D.2.
\hfill\qed

\bigskip
\subhead{Proof of Lemma 4.7 iv)}
\endsubhead
\bigskip

Starting from (D.14), statement iv) of Lemma 4.7 is
obtained in the same way as statement v) of Lemma 4.6
was obtained in Appendix C.
\hfill\qed
\medskip

As a corollary of this proof, one obtains
the analogue of (C.32) and (C.33) for
derivatives, namely
$$
\multline
\left|
{d^k \over dh^k}
Z_q(W,h)\right|_{h=h_0}
\leq
|k|!
\left(
C_0
+O(\epsilon))|W|
\right)^{|k|}
e^{\gamma\tau N_{\partial V}(W)}
e^{(O(\epsilon)+O(e^{-3\tau/4})\vert\partial W\vert}
e^{-f|W|}\times
\\
\times
\max_{U \subset W}
e^{-\frac{a_q}{4}|W\setminus U|}
e^{-(\tau/4C_3)|\partial U| }
\endmultline
\tag D.16
$$
and
$$
\multline
\left|
{d^k \over dh^k}
Z_q(W,h)\right|_{h=h_0}
\leq
|k|!
\left(
C_0 + O(\epsilon))|W|
\right)^{|k|}
e^{\gamma\tau N_{\partial V} (W)}
e^{(O(\epsilon)+O(e^{-3\tau/4}))|\partial W|}
e^{-f|W|}\times
\\
\times
\max
\left\{
        e^{-\frac{a_q}{4}|W|}, e^{-(2d\tau/4C_3)|W|^{d/(d-1)}}
\right\}
\,.
\endmultline
\tag D.17
$$

%
%

\vfill\eject
\head{Appendix E: Proof of Lemma 5.1 and 5.2}
\endhead

\bigskip
\subhead{Proof of Lemma 5.1}
\endsubhead
\bigskip

Observing that all components $W$ of $\Int^{(0)} Y_A$
obey the bound
$
\displaystyle
|W|\leq \max_{Y\in Y_A} |V(Y)|
$,
the statement i) of Lemma 5.1 immediately follows from
Lemma 4.6.

In order to prove ii), we first note
that  for
$Y_A=\{Y_1,\cdots,Y_n\}$,
$$
\multline
\rho(Y_A) e^{E_q(\overline{\supp Y_A})}=\\
=
A(Y_1,\cdots,Y_n)
\prod_{i=1}^n
\rho(Y_i)e^{E_q(Y_i)}\prod_{m\neq q}
e^{E_q(\Int_m Y_A\cap\,\supp A)-
E_m(\Int_m Y_A\cap\,\supp A)}
\,,
\endmultline
\tag E.1
$$
which implies that
$$
\multline
\left|
\rho(Y_A) e^{E_q(\overline{\supp Y_A})}
\right|
\leq
C_A
e^{-(\tau-\gamma\tau)|Y_A|}
e^{2\gamma\tau N_\partial(\Int Y_A\cap\,\supp A)}\times
\\
\times
e^{(e_q-e_0)(|\supp Y_A|_d+|\Int Y_A\cap\,\supp A|_d)}
\,.
\endmultline
\tag E.2
$$
Next we use
Lemma 4.6 to bound
$$
\prod_{m=1}^N
\left|
{Z_m(\Int^{(0)}_m Y_A,h)
\over
Z_q^\prime(\Int^{(0)}_m Y_A,h)}
\right|
\leq
e^{a_q|\Int^{(0)} Y_A|
   +O(\epsilon)|\partial \Int^{(0)} Y_A|}
e^{2\gamma\tau N_\partial(\Int^{(0)} Y_A)}
\,.
\tag E.3
$$
Bounding $e_q-e_0\leq a_q+O(\epsilon)$,
observing that
$$
|\Int^{(0)} Y_A|+|\Int Y_A\cap\,\supp A|
=
|\Int Y_A|
\,,
\tag E.4
$$
and bounding
$
|\partial\Int^{(0)} Y_A|
\leq
|\partial\Int Y_A|+|\partial\supp A|
\leq
|\partial\Int Y_A|+2d|\supp A|
$,
we
get the bound
$$
\left|
K_q^\prime(Y_A)
\right|
\leq
C_A
e^{-(\tau-\gamma\tau-O(\epsilon))|Y_A|}
e^{2\gamma\tau N_\partial(\Int Y_A)}
e^{a_q|V(Y_A)|+O(\epsilon)|\supp A|}
\,.
\tag E.5
$$
Bounding now
$
N_\partial(\Int Y_A)
$
by
$C_1 |Y_A|$,
and observing that
$\prod_{Y\in Y_A}\chi_q(Y)\neq 0$
implies that
$a_q|V(Y_A)|\leq (\alpha+1+O(\epsilon))|Y_A|$
due to the bound (4.26), we finally get
$$
\left|
K_q^\prime(Y_A)
\right|
\leq
C_A
e^{-(\tau(1-(1+2C_1)\gamma)|Y_A|}
e^{(1+\alpha+O(\epsilon))|Y_A|}
e^{O(\epsilon)|\supp A|}
\,,
\tag E.6
$$
which implies the bound (5.15).

We are left with the proof of iii). By (E.1),
Lemma D.1 and the assumptions (3.9) and (3.25b),
$$
\multline
\left|
{d^k\over dh^k}
\rho(Y_A) e^{E_q(\overline{\supp Y_A})}
\right|
\leq
|k!|C_A C_0^{|k|}
\left(
|\overline{\supp Y_A}|
+|Y_A|
+2|\Int Y_A\cap\,\supp A|
\right)^{|k|}\times
\\
\times
e^{-(\tau-\gamma\tau)|Y_A|}
e^{2\gamma\tau N_\partial(\Int Y_A\cap \supp A)}
e^{(e_q-e_0)(|\supp Y_A|_d+|\Int Y_A\cap\,\supp A|_d)}
\,.
\endmultline
\tag E.7
$$
On the other hand,
$$
\multline
\left|
{d^k\over dh^k}
\prod_{m=1}^N
{Z_m(\Int^{(0)}_m Y_A,h)
\over
Z_q^\prime(\Int^{(0)}_m Y_A,h)}
\right|
\leq
\left(
(2C_0+O(\epsilon))
|\Int^{(0)}Y_A|
\right)^{|k|}
e^{a_q|\Int^{(0)} Y_A|}\times
\\
\times
e^{O(\epsilon)|\partial \Int^{(0)} Y_A|}
e^{2\gamma\tau N_\partial(\Int^{(0)} Y_A)}
\,
\endmultline
\tag E.8
$$
by
Lemma 4.7, while
$$
\left|
{d^k\over dh^k}
\prod_{Y\in Y_A}
\chi_q(Y)
\right|
\leq
\biggl(
\overline C_1
\sum_{Y\in Y_A}|V(Y)|
\biggr)^{|k|}
\,
\tag E.9
$$
by (D.6).

Combining the bounds (E.7) through (E.9),
and bounding
$$
\multline
\!\!\!\!
C_0\left(
        |\overline{\supp Y_A}|
        +|Y_A|
        +2|\Int Y_A\cap\,\supp A|
\right)
+(2C_0+O(\epsilon)) |\Int^{(0)}Y_A|
+\overline C_1
\sum_{Y\in Y_A}|V(Y)|\leq
\\
\leq
C_0 |\supp A|
+(2C_0+O(\epsilon))
\left(|Y_A|+|\Int Y_A|\right)
+\overline C_1
\sum_{Y\in Y_A}|V(Y)|
\leq|\supp A| e^{O(1)|Y_A|}
\,,
\endmultline
\tag E.10
$$
we may then continue as in the proof of (E.6) to
obtain the bound (5.16).
\qed

\bigskip
\subhead{Proof of Lemma 5.2}
\endsubhead
\bigskip

\noindent
We start
from the representation (5.4) and
use the assumptions (3.25) and (3.11) to bound
$$
\multline
\left|
\rho(Y_A)
\right|
\leq
C_A
e^{-\tau |Y_A|-E_0(\supp Y_A)}
e^{-E_0(\Int Y_A\cap\,\supp A)}
e^{-E_q(\Ext Y_A\cap\,\supp A)}\leq
\\
\leq
C_A
e^{\gamma\tau N_\partial(\overline{\supp Y_A})}
e^{-(\tau-O(\epsilon)) |Y_A|}
e^{O(\epsilon)|\supp A|}
e^{-f|\overline{\supp Y_A}|}
e^{-a_q|\Ext Y_A\cap\,\supp A|}
\,,
\endmultline
\tag E.11
$$
Lemma 4.6 to bound
$$
\left|
\prod_{m=1}^N
Z_m(\Int^{(0)}_m Y_A,h)
\right|
\leq
e^{-f|\Int^{(0)}Y_A|}
e^{O(\epsilon)(|Y_A|+|\supp A|)}
e^{\gamma\tau N_\partial(\Int^{(0)}Y_A)}
\,,
\tag E.12
$$
and the inequality (C.32)
in conjunction with
the estimate
$$
|\partial \Ext^{(0)}Y_A|
\leq
|\partial\supp A|
+|\partial V|
+|\partial V(Y_A)|
\leq
2d|\supp A|
+|\partial V|
+C_3|Y_A|
$$
to bound
$$
\multline
|Z_q(\Ext^{(0)}Y_A, h)|
\leq
e^{O(\tilde\epsilon)(|\partial V|+|\supp A|)}
e^{\gamma\tau N_\partial (\Ext^{(0)}Y_A)}
e^{-f|\Ext^{(0)}Y_A|}\times
\\
\times
e^{O(\tilde\epsilon)|Y_A|}
\max_{U \subset \Ext^{(0)}Y_A}
e^{-\frac{a_q}{4}|\Ext^{(0)}Y_A\setminus U|}
e^{-(\tau/4C_3)|\partial U| }
\,,
\endmultline
\tag E.13
$$
where $\tilde\epsilon$ is the constant introduced in
Lemma 5.2.
Combining the bounds (E.11) through (E.13) with (5.4),
we obtain
$$
\multline
\left|
Z_q(A|V,h)
\right|
\leq
C_A e^{O(\tilde\epsilon)|\supp A|}
e^{(\gamma\tau +O(\tilde\epsilon))|\partial V|}
e^{-f|V|}\times
\\
\times
\sum_{Y_A}
e^{-(\tau-1))|Y_A|}
\max_{U\subset\Ext^{(0)}Y_A}
e^{-\frac{a_q}{4}|\Ext Y_A\setminus U|}
e^{-(\tau/4C_3)|\partial U|}
\,,
\endmultline
\tag E.14
$$
where we used the bounds
$
{a_q}|\Ext\cap\,\supp A|
+({a_q}/4)|\Ext^{(0)} Y_A\setminus U|
\leq
({a_q}/4)|\Ext Y_A\setminus U|
$
and $N_\partial(V)\leq |\partial V|$.
Extracting
the  factor
$$
\align
&\max_{Y_A} e^{-\frac{\tau}4 |Y_A|}
\max_{U\subset\Ext^{(0)}Y_A}
e^{-\frac{a_q}{4}|\Ext Y_A\setminus U|}
e^{-(\tau/4C_3)|\partial U|}
\\
&\leq
\max_{Y_A}
e^{-(\tau/(4C_3)|\partial V(Y_A)|}
\max_{U\subset\Ext^{(0)}Y_A}
e^{-\frac{a_q}{4}|(V\setminus V(Y_A))\setminus U|}
e^{-(\tau/4C_3)|\partial U|}
\\
&\leq
\max_{S\subset V}
e^{-\frac{a_q}{4}|V\setminus S|}
e^{-(\tau/4C_3)|\partial S|}
\leq
\max
\left\{
e^{-\frac{a_q}{4}|(V|}\,,\,
e^{-(\tau/4C_3)|\partial V|}
\right\}
\endalign
$$
from the right hand side of (E.14),
we are left with a sum
$
\sum_{Y_A} e^{-(3\tau/4-1)|Y_A|}
$
which we bound as follows
$$
\sum_{Y_A} e^{-(3\tau/4-1)|Y_A|}
\leq
\sum_{n=0}^\infty
{1\over n!}
\biggl[
\sum_{Y:V(Y)\cap\,\supp A\neq\emptyset}
e^{-(3\tau/4-1)|Y|}
\biggr]^n
\leq
e^{O(\tilde\epsilon)|\supp A|}
\,.
$$
Putting everything together,
we obtain the bound i) of Lemma 5.2.

\medskip

In order to prove ii), we generalize
(E.11) through (E.13) to derivatives.
In (E.11), these derivatives produce
an extra
factor,
$$
\align
C_A |k|!
\left(
        C_0|\overline{\supp Y_A}|
        +C_0|\supp A\setminus\supp Y_A|
\right)^{|k|}
&\leq
C_A |k|!
\left(
        2C_0|{\supp A}|
        +C_0|Y_A|
\right)^{|k|}
\\
&\leq
C_A |k|!
\left(C_0|{\supp A}|\right)^{|k|}
e^{O(1)|Y_A|},
\endalign
$$
while in (E.12) and (E.13), they produce
factors
$$
|k|!
\left(
        (C_0+O(\epsilon))|\Int^{(0)} Y_A|
\right)^{|k|}
$$
and
$$
|k|!
\left(
        (C_0+O(\epsilon))|\Ext^{(0)} Y_A|
\right)^{|k|}
\,.
$$
On the right  hand side of (E.14),
this leads to an extra factor
$$
C_A |k|!
\left(
        (C_0+O(\epsilon)) |{V}|
\right)^{|k|}
e^{O(1)|Y_A|}
\,.
$$
Observing that the sum
$
\sum_{Y_A} e^{-(3\tau/4-O(1))|Y_A|}
$
can be bounded by
$
e^{O(\tilde\epsilon)|\supp A|}
$
as well, we obtain Lemma 5.2 ii).
\qed

\tenpoint
%


\enddocument
\end